\documentclass{aa}  
\usepackage{graphicx}
\usepackage{color}
\definecolor{red}{rgb}{1,0,0}
\definecolor{blue}{rgb}{0,0,1}

\usepackage{rotating}

\usepackage{natbib}
\usepackage{txfonts}
\begin{document}
 \title{Resolving asymmetries along the pulsation cycle \\ of the Mira star \object{X Hya}
   \thanks{Based on observations obtained with the ESO VLTI/ATs telescopes under the program ID 084.D-0326. } }

   \author{X. Haubois\inst{1,2,3}
          \and       
          M. Wittkowski\inst{4}
         \and    
         G. Perrin\inst{2}
         \and    
         P. Kervella\inst{5,2}
        \and
    A. M\'{e}rand\inst{1}
     \and
         E. Thi\'{e}baut\inst{6}         
   \and
          S.T. Ridgway\inst{7}
         \and \\
        M. Ireland\inst{8}
         \and
        M. Scholz\inst{3,9}
                         }

    \institute{European Southern Observatory (ESO), Alonso de Cordova 3107, Casilla 19001, Vitacura, Santiago 19, Chile, \email{xhaubois@eso.org}
\and LESIA, Observatoire de Paris, CNRS UMR 8109, UPMC, Universit\'e Paris Diderot, PSL Research University, 5 place Jules Janssen, 92195 Meudon, France
\and  Sydney Institute for Astronomy, School of Physics, University of Sydney, Sydney, NSW 2006, Australia 
\and European Southern Observatory (ESO), Karl-Schwarzschild-Str. 2, D-85748 Garching bei M\"{u}nchen, Germany
\and Unidad Mixta Internacional Franco-Chilena de Astronom\'{i}a (UMI 3386), CNRS/INSU, France \& Departamento de Astronom\'{i}a, Universidad de Chile, Camino El Observatorio 1515, Las Condes, Santiago, Chile.
   \and Centre de Recherche Astrophysique de Lyon, CNRS/UMR 5574, 69561 Saint Genis Laval, France 
 \and Kitt Peak National Observatory, National Optical Astronomy Observatories,  P.O. Box 26732,  Tucson, AZ 85726-6732 U.S.A. 
 \and Research School of Astronomy \& Astrophysics, Australian National University, Canberra, ACT 2611, Australia 
 \and Zentrum f{\"u}r Astronomie der Universit{\"a}t Heidelberg (ZAH), Institut f{\"u}r Theoretische Astrophysik, Albert-Ueberle-Str. 2, 69120 Heidelberg, Germany
}
   \date{Received; accepted }

 
  \abstract
   {The mass-loss process in Mira stars probably occurs in an asymmetric way where dust can form in inhomogeneous circumstellar molecular clumps. Following asymmetries along the pulsation cycle can give us clues about these mass-loss processes.}
   {We imaged the Mira star \object{X Hya} and its environnement at different epochs to follow the evolution of the morphology in the continuum and in the molecular bands.}
   {We observed X Hya with AMBER in J-H-K at low resolution at two epochs. We modelled squared visibilities with geometrical and physical models. We also present imaging reconstruction results obtained with MiRA and based on the physical \textit{a priori} images.}
   {We report on the angular scale change of X Hya between the two epochs. 1D CODEX profiles allowed us to understand and model the spectral variation of squared visibilities and constrain the stellar parameters. Reconstructed model-dependent images enabled us to reproduce closure phase signals and the azimuthal dependence of squared visibilities. They show evidence for material inhomogeneities located in the immediate environment of the star. }
   {}


   \maketitle
%

\section{Introduction}



When reaching the tip of the asymptotic giant branch (AGB), low-to-intermediate-mass stars begin a Mira phase before evolving to the planetary nebulae and finally white dwarf stages. During this phase, their pulsation period of a few hundred days is associated with a several-magnitude photometric variability in the infrared \citep{1993A&AS...97..729L} and a mass-loss rate up to $10^{-4} M_{\sun}$/year \citep{2008A&A...487..645R}. While this phenomenon is well explained by the $\kappa$-mechanism, the way the radial oscillations interact with the stellar outer atmosphere and the way mass is dispersed into the interstellar medium (ISM) is not well understood. A molecular layer or MOLsphere extending to about 1 stellar radius from the photosphere has been systematically observed \citep[e.g.][]{,2002ApJ...579..446M,2004A&A...426..279P,2009A&A...496L...1L,2008A&A...479L..21W,2011A&A...532L...7W}. The hypothesis derived from spectroscopic H-band data \citep{1984ApJS...56....1H} assumes free-falling material disassociated by the pulsation shock-front and recombining in the post-shock zone. For oxygen-rich stars like the one we study in this paper, one scenario predicts the formation of Fe-free silicate grains that allow a strong enough radiative pressure to drive a wind provided the grain size is about 1 micron, which is a typical size of grains observed in the ISM \citep{2008A&A...491L...1H}.



Another piece of the Mira star puzzle are the bright asymmetries. \cite{2006ApJ...652..650R},  \cite{2009ApJ...707..632L}, \cite{2011A&A...532L...7W} and \cite{2015MNRAS.446.3277C} showed that asymmetries are common in Mira atmospheres and can be modelled (though not uniquely) as bright features containing up to several percent of the total flux. A number of phenomena have been proposed to explain these observations such as variable extinction resulting from condensation of dust grains in asymmetric regions, shock features or inhomogeneities in the molecular layers. Moreover, since protoplanetary nebulae (PPN) are also known to present asymmetric morphologies \citep{2007AJ....134.2200S}, we hope to understand the connection between the AGB and PPN structures by studying the morphology of the Mira circumstellar envelopes which are likely expelled with some degree of asymmetry. 

The goal of our study is therefore to characterize these asymmetries and establish a link with the mass-loss process. Near-IR spectro-interferometry is ideally suited for this study, since it delivers spectral and spatial information on both the stellar photosphere and the atmosphere or envelope. We here investigate the variation in the morphology of the Mira star \object{X Hya} based on two-epoch AMBER measurements.
Previous interferometric observations of \object{X Hya} were made with the IOTA interferometer \citep{2006ApJ...652..650R} in H band. They found that the object showed no sign of asymmetry up to a maximum baseline of 24.5 meters for a visual pulsation phase of 0.5. Medium-resolution K-band AMBER data were also obtained on \object{X Hya} at phase 0.7 \citep{2011A&A...532L...7W}.  Asymmetries were detected, more prominently in $\rm H_{2}O$ bands, and were interpreted as clumps or inhomogeneities in this molecular layer. 

We here aim at imaging these asymmetries and observe their evolution along the pulsation cycle.
In Sect.~\ref{obs}, we present the observations. In Sect.~\ref{modfit}, we show the results of the parametric model fitting analysis as well as radiative transfer modelling. In Sect.~\ref{ima}, we show the results of the image reconstruction approach. Before concluding, we discuss these results with similar observations reported in the literature in Sect.~\ref{discuss}.


\section{Observation and data reduction}
\label{obs}

\begin{figure*}
 \centering
 \includegraphics[width=5.5in]{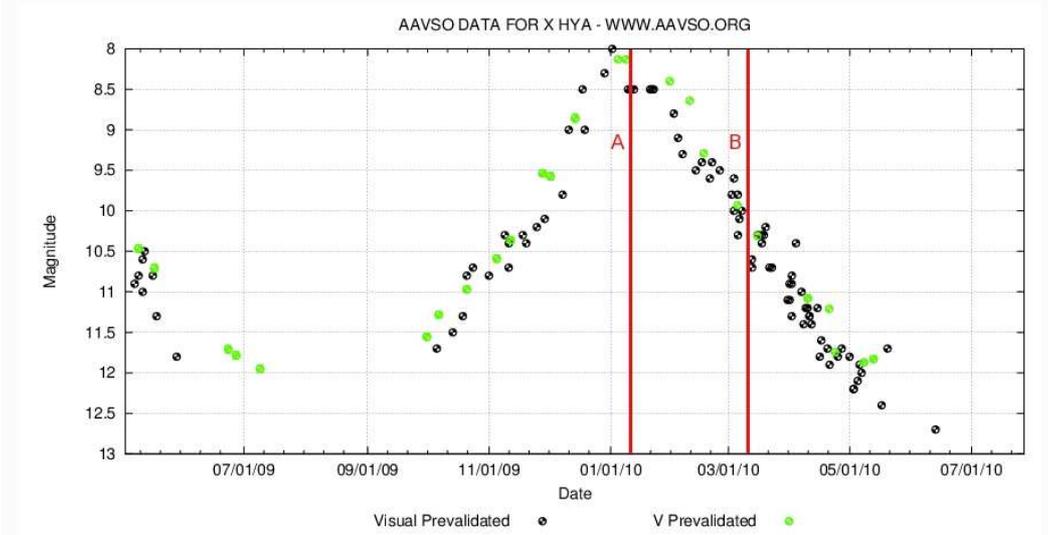}
\caption{AAVSO $V$-band light curve of \object{X Hya}. The period of its brightness variations has been estimated to 301 days \citep{2008MNRAS.386..313W}. The two observation epochs are marked by red vertical lines and correspond to visual phases 0.0 and 0.2.}
    \label{log_obs}
 \end{figure*}

\begin{figure*}
 \centering
 \includegraphics[width=3.2in]{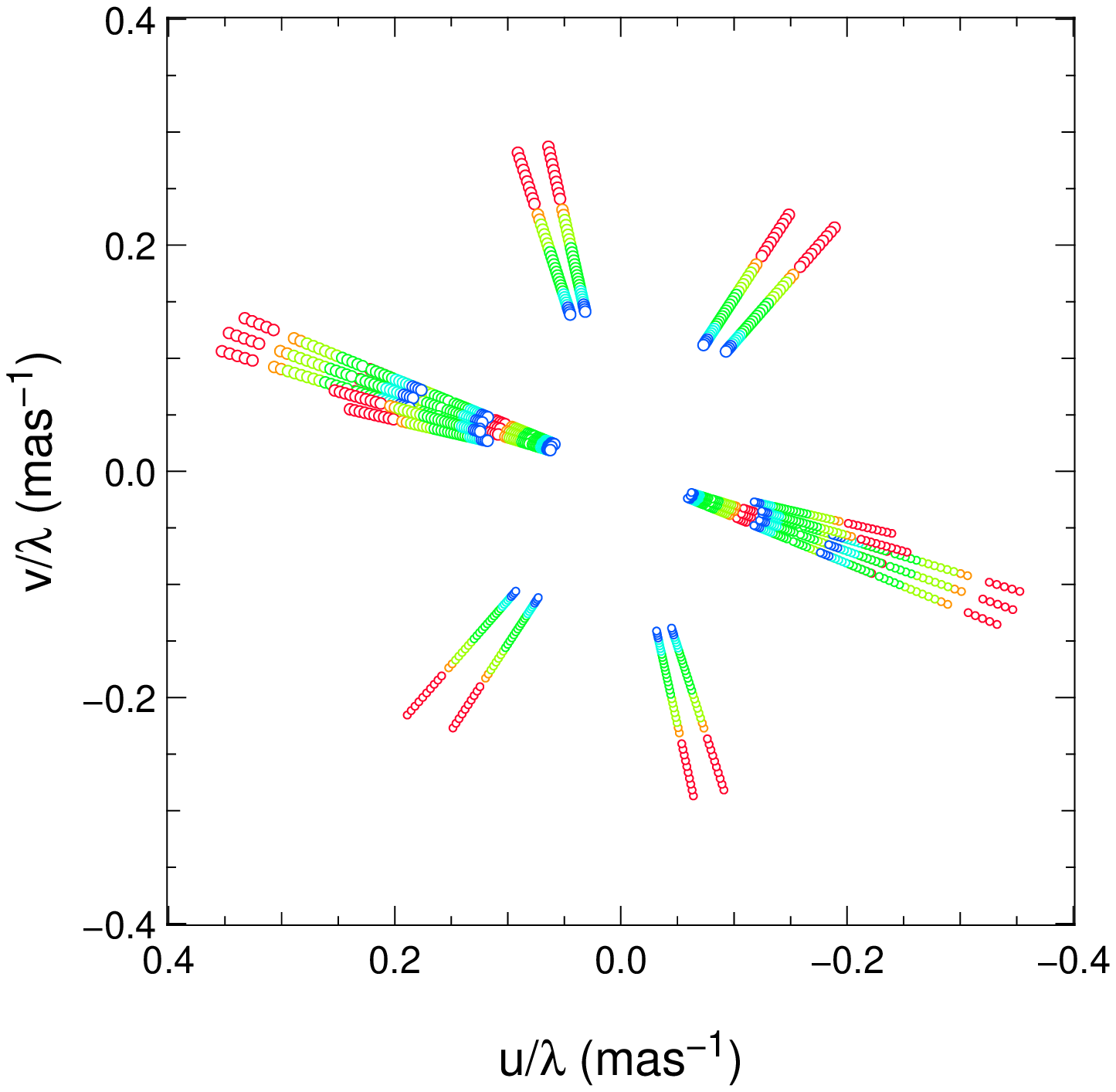}
  \includegraphics[width=3.2in]{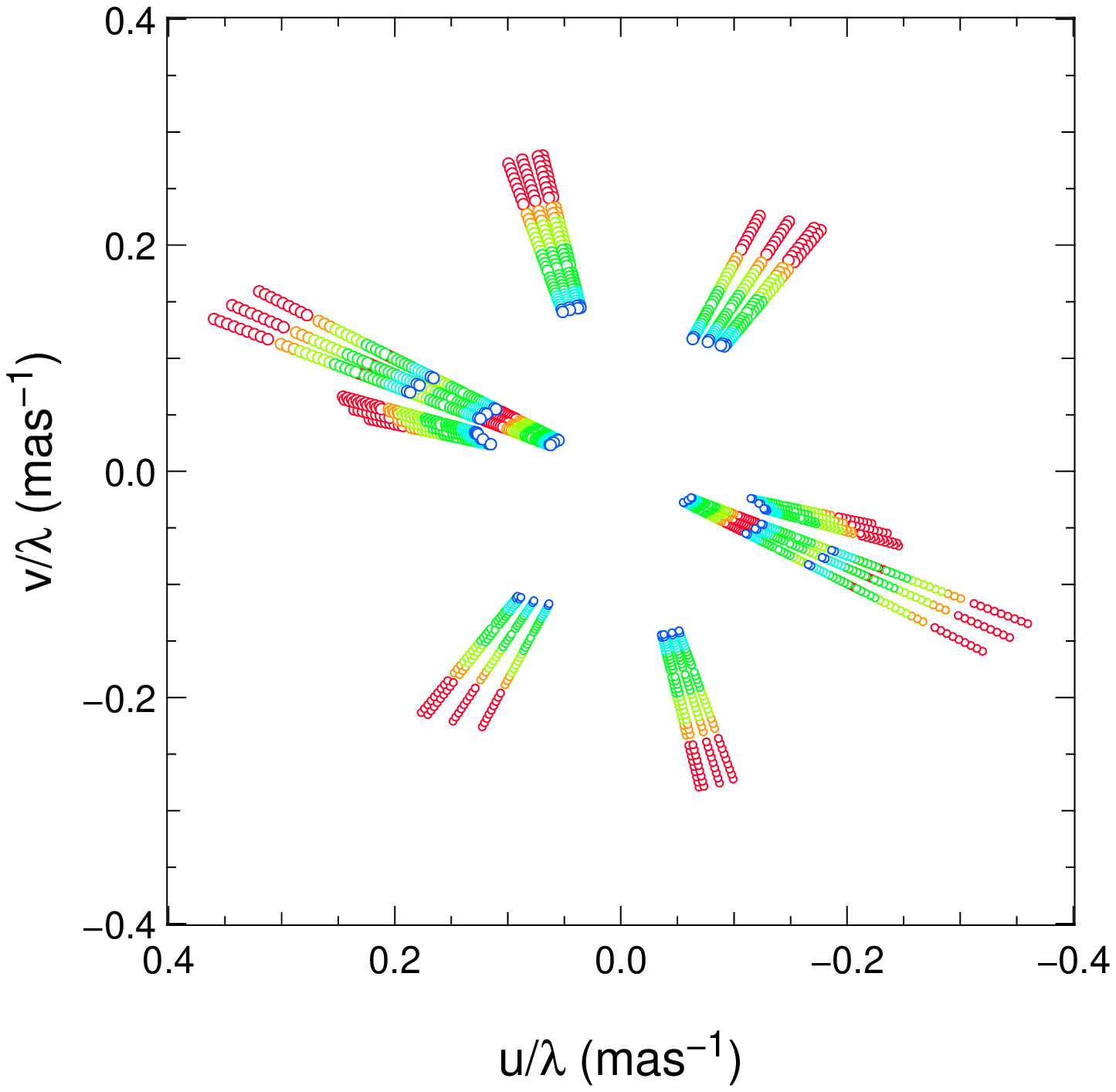}
\caption{Uv coverage of the observations of the two epochs: left epoch A, right epoch B. Colours indicate the six spectral bands explicated in Table~\ref{tab:bands} (1: red, 2: orange, 3: light green, 4: dark green, 5: turquoise, 6: blue). }
       \label{uvcov}
 \end{figure*}

We observed the Mira star \object{X Hya} with AMBER+FINITO \citep{2007A&A...464....1P} at two epochs. In Fig.~\ref{log_obs}, these observation epochs are marked in the brightness variation cycle taken from the AAVSO \footnote{http://www.aavso.org/} database.  They correspond to visual phases of 0.0 and 0.2. X Hya is a M7e Mira star with a pulsation period of 301 days, it is located at a distance of $\sim 440$ pc \citep{2008MNRAS.386..313W}.  
A multi-wavelength observation is required to distinguish the molecular layer from the stellar photosphere. We thus performed a multi-wavelength J-H-K observation (covering wavelengths from 1.1 $\mu$m to 2.4 $\mu$m) at low spectral resolution with AMBER. A log of the observations is presented in Table~\ref{tab:Logob}. The U-V plane we covered with our measurements was very similar at the two epochs as shown in Fig.~\ref{uvcov}.

\begin{table*}[!h]
\centering
\begin{tabular}{lccccccl}
\hline
Date/MJD & Configuration & Calibrator  & Coherence time$^{1}$ (ms) & Fried parameter$^{1}$ (cm) & Frame selection  & Integration time (s) $/$\\
&&&&&&calibrated point\\
\hline
Epoch A (Visual Phase: 0.0) &&& \\
\hline
09-01-2010/55205   &  D0-H0-K0  & HR3802 & 5.3 $\pm$ 1.2 & 2.0 $\pm$0.4 & 20$\%$  & 48.5
\\
10-01-2010/55206    &D0-G1-H0 & HR3802 &  3.2 $\pm$ 0.8 & 2.2 $\pm$ 0.5 &40$\%$ & 97.0
\\
\hline
Epoch B (Visual Phase: 0.2)&&& &\\
\hline
16-03-2010/55271    & D0-H0-K0  & HR3802& 16.0  $\pm$ 6.5 & 3.0 $\pm$ 1.2  & 35$\%$ & 85.0
\\
&D0-G1-H0 & HR3749&  16.0  $\pm$ 6.5 & 3.0 $\pm$ 1.2 &35$\%$  & 85.0
\\
\hline
\end{tabular}
\caption{Log of the observations. $^{1}$ As an indication on the atmospheric conditions, coherence times and Fried parameters were taken from the DIMM measurements in the visible and are expressed as $\mu \pm 1 \sigma$, where $\mu$ is the average and $\sigma$ is the standard deviation computed over each half-night of our observations.}
\label{tab:Logob}
\end{table*}



Reduced squared visibilities and closure phases were obtained with the amdlib package \citep{2007A&A...464...29T,2009A&A...502..705C}.
To optimise the stability of the transfer function, we adjusted the frame selection with a threshold of 15 microns for the piston value. The percentage of the frames left after a selection based on the signal-to-noise ratio (S/N), and consequently the total integration time per calibrated point, varied for each run because of different atmospheric conditions. These information are provided in Table~\ref{tab:Logob}.


The spectral calibration was based on a fit of telluric features and on the positions of flux minima between absorption bands (see Fig.~\ref{graph_CODEX}). We compared these positions between the observed spectra of the calibrators and a synthetic spectrum. This led us to multiply the wavelength table of our data by a factor of 1.04 to match the positions of the observed spectra with those of the synthetic spectrum. Interferometric calibrators were observed and their measurements bracketed those of X Hya.

\section{Model fitting of the squared visibilities}
\label{modfit}

\begin{figure*}[h!]
 \centering
\includegraphics[width=3.5in]{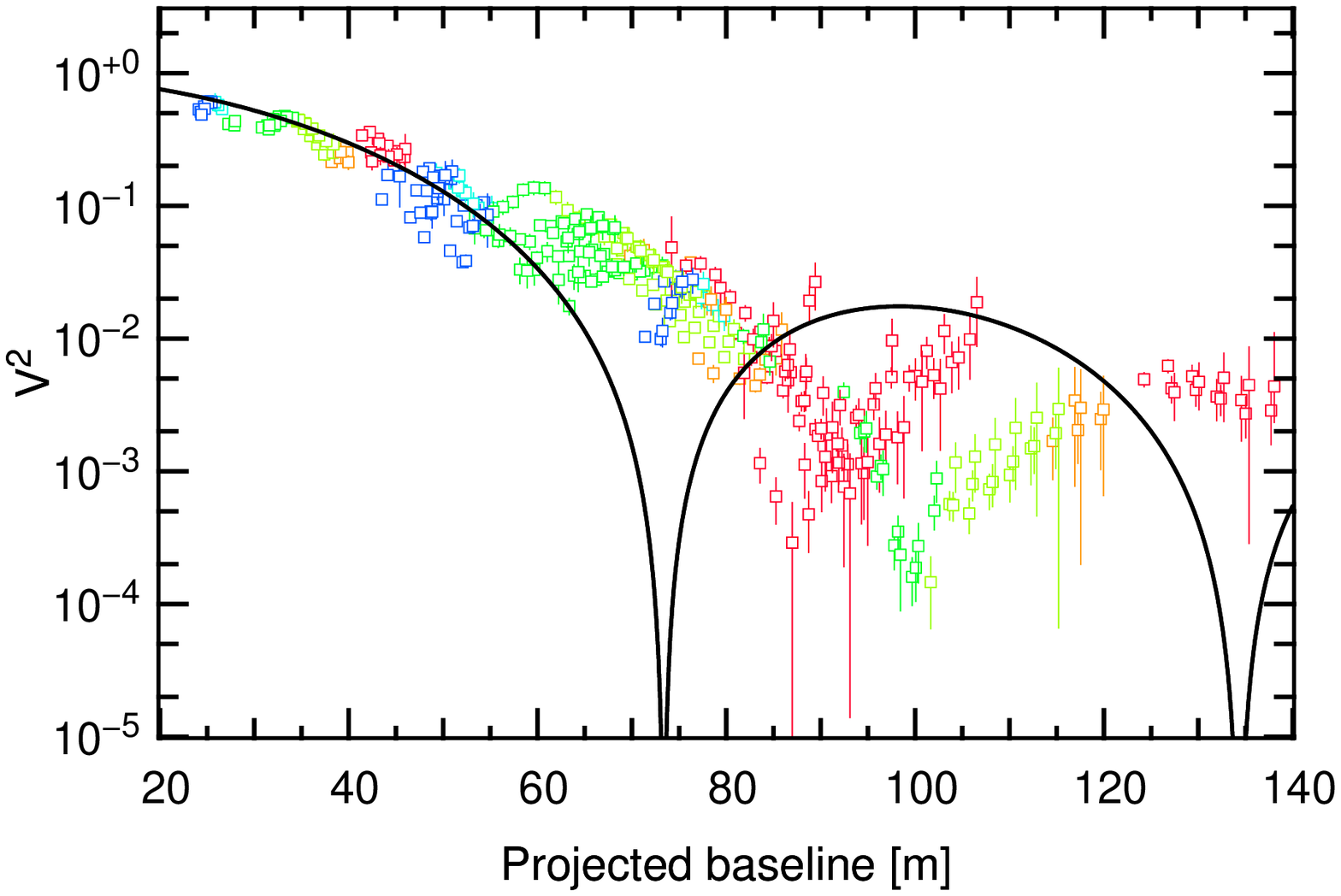}
\includegraphics[width=3.5in]{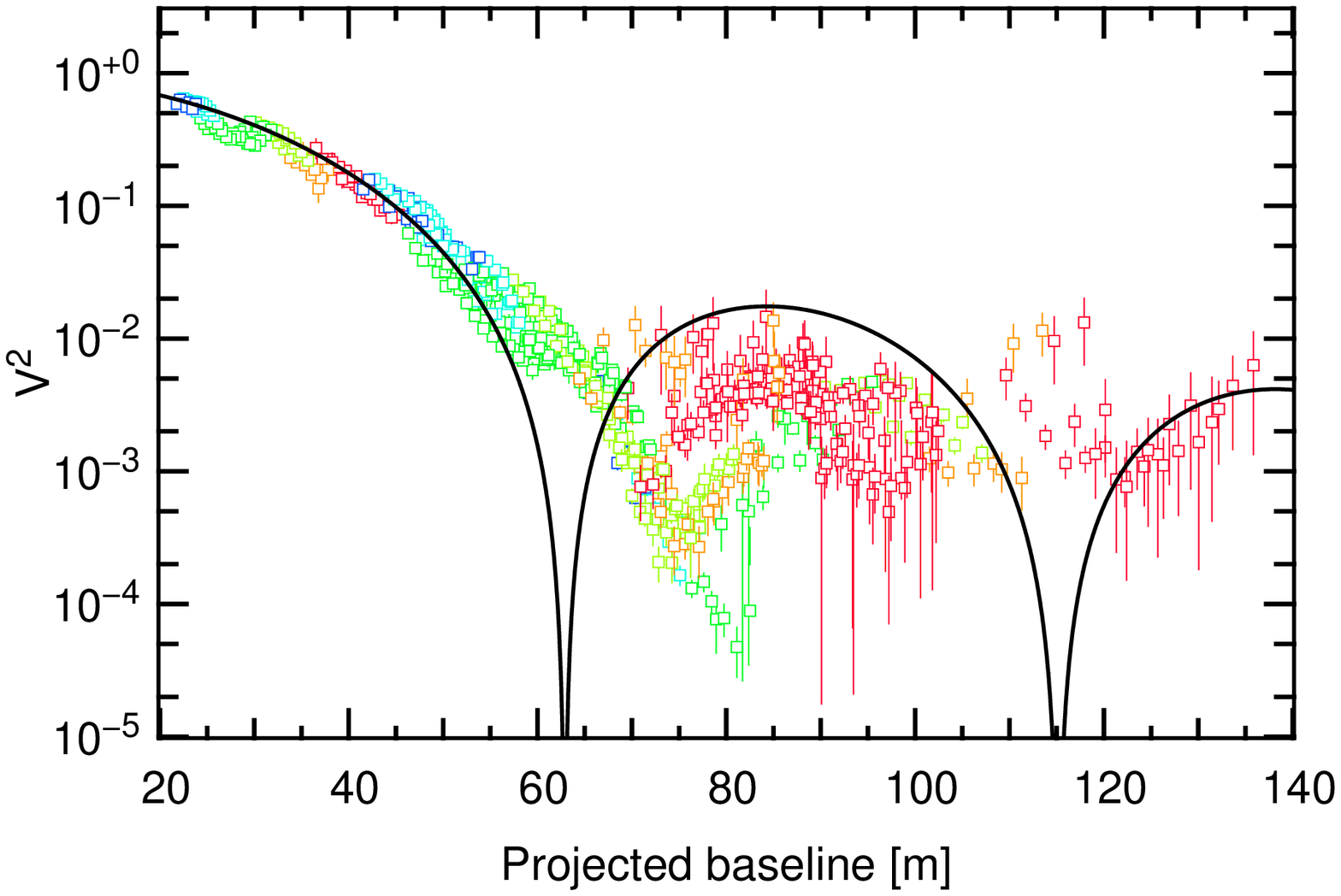}
\caption{Squared visibilities data at epoch A (left) and epoch B (right). Colours indicate the six different spectral bands as in Fig.~\ref{uvcov}. The solid line represents a 6 mas (epoch A) and a 7 mas (epoch B) uniform disks at 1.75 micron (average wavelength of the covered spectral intervals).
}
       \label{v2data}
 \end{figure*}

The datasets we obtained at each epoch are rich and complex. Even at short baselines, squared visibilities ($V^{2}$) show a significant departure from a simple geometry such as an uniform disk model (Fig.~\ref{v2data}). To understand the spectral and morphological signatures on the $V^{2}$ individually, we first divided the data into six spectral bands, as shown in Table~\ref{tab:bands}. This division is based on an analysis of the visibility variation with the wavelength (see Fig.~\ref{graph_CODEX}) which is coherent with a previous spectro-interferometric study of X Hya \citep{2011A&A...532L...7W} and also based on the molecular absorption bands shown in Fig.~4 of \cite{2000A&AS..146..217L}. Although we labelled these bands "Continuum", "CO", and "$H_{2}O$" based on the dominating "species", they may also be contaminated by other molecular and atomic features that we cannot distinguish with a low spectral resolution. In this context, the label "Continuum" does not exclude some molecular contamination that may be present at these wavelengths, in particular in the case of the Continuum (1) band, which overlaps with a part of an H$_{2}$O absorption band. Our choice represents a compromise between the location of molecular absorption
bands and a relevant subdivision of the data to show their spectral variation.

In the following, we present a progressive modelling approach of the squared visibilities in the six spectral bands, which are representative of the centro-symmetric intensity distribution of the object. Before using a complex dynamical model, we performed a first standard modelling with the uniform disk function which delivers a basic reference size estimate.

\begin{table}[h!]
\centering
\begin{tabular}{cccc}
\hline
Spectral band & Wavelength range & Contribution \\  
number  & & \\\hline
1 & $1.1-1.4 \,\rm\mu m  $  & Continuum (1)  & \\ 
2 & $1.4-1.5 \,\rm\mu m  $ &  $\rm H_{2}O$ (1)  & \\ 
3 & $1.5-1.7 \,\rm\mu m  $ & CO (1)  & \\  
4 & $1.7-2.1 \,\rm\mu m$ &  $\rm H_{2}O$ (2) & \\  
5 & $2.1-2.25 \,\rm\mu m $ & Continuum (2) & \\  
6 & $2.25-2.4 \,\rm\mu m $ & CO (2)  & \\ \hline
\end{tabular}
\caption{\label{tab:bands} Division of the covered spectral range in six spectral bands.}
\end{table}

\subsection{Uniform disk model}
We first modelled a uniform disk (UD) to obtain a first estimate of the angular size of the object in the different bands and at the different position angles (PA) we observed.  We only considered squared visibility data whose spatial frequencies belong to the first lobe so as not to bias this estimate with any contamination from limb darkening or structures smaller than the stellar diameter. An average of the measurements over the PAs is shown in Table~\ref{tab:DU}.

\begin{table}[h!]
\centering
\begin{tabular}{ccccc}
\hline
Spectral  & $ \phi_{UD} $ (mas)  & $ \phi_{UD} $ (mas) & $\Delta(\phi_{UD}) / \phi_{UD}$ & $ \phi_{UD} $ (mas)$^{1}$\\  
band  & $\Phi =0.0$ & $\Phi =0.2$  & & $\Phi =0.7$  \\
number  & (epoch A) &  (epoch B)  & &\\ \hline
1 & 5.94 $\pm$ 0.15 \,\rm  & 7.41 $\pm$ 0.40 \,\rm  &  0.25 $\pm$ 0.05 & -   \\ 
2  &  6.90 $\pm$ 0.13 \,\rm & 7.85  $\pm$ 0.49 \,\rm   & 0.14  $\pm$ 0.05 & -  \\ 
3 &  6.03 $\pm$ 0.59 \,\rm & 7.19 $\pm$ 0.85 \,\rm  & 0.20  $\pm$ 0.13 & -  \\  
4 & 6.52 $\pm$ 0.23 \,\rm &   8.24 $\pm$ 0.86 \,\rm  & 0.27  $\pm$ 0.09 & 8.3 $\pm$ 0.1 \\  
5  & 6.20 $\pm$ 0.35 \,\rm & 7.02 $\pm$ 0.86 \,\rm  & 0.14  $\pm$ 0.10  & 6.2 $\pm$ 0.1 \\  
6   & 6.80 $\pm$ 0.42 \,\rm &  7.54 $\pm$  0.79 \,\rm  & 0.11  $\pm$ 0.09 & 9.0 $\pm$ 0.1 \\ \hline
\end{tabular}
\caption{\label{tab:DU} Fitting results of uniform disk models performed in the first lobe of the squared visibilities. Diameter values were averaged over the PAs. Results are expressed as $\mu \pm 1 \sigma$, where $\mu$ is the average and $\sigma$ is the standard deviation computed over all the PAs. The fourth column represents the relative variation of UD diameter between epochs A and B, relatively to epoch A.  $^{1}$The last column reports measurements from \cite{2011A&A...532L...7W}. }
\end{table}

This first basic modelling description contains rich information. The UD diameter systematically increases between the two epochs in every spectral band. This agrees with the visual pulsation cycle seen in Fig.~\ref{log_obs}: as the envelope expands, the temperature decreases and corresponds to a dimmer visual magnitude.

VINCI \citep{2000SPIE.4006...31K} measurements of X Hya were reduced and analysed for comparison. They were taken in $K$ band on 2002-01-27, corresponding to a phase of 0.25 which was close to the phase of the second epoch of our AMBER measurements (0.2).  The UD diameter we derived is 7.55 $\pm$ 0.70 mas which is compatible with our results for band 5 (2.1-2.25 $\mu$m) and 6 (2.25 -2.4 $\mu$m) with the overlap of error bars.

The UD diameters reported in \cite{2011A&A...532L...7W} that were obtained for phase $\Phi=0.7$ (see last column of Table~\ref{tab:DU}) are also compatible with our measurements for bands 4 and 5 (\rm $H_{2}O$ (2) and continuum). This is not the case for band 6 (CO (2)), however, for which Wittkowski et al. found a high value of 9.0 $\pm$ 0.1 mas.


Linear limb-darkened disk diameters were also computed to model the $V^{2}$ data up to the second lobe \citep[see Eq.~5 with $B = 0$ in][]{2009A&A...508..923H}. Results are presented in Table~\ref{tab:DA}. As judged by the high reduced $\chi^{2}$ values (whose maximum was 100) and the extreme values found for the linear coefficient (forced to be between -1 and 1), the addition of limb darkening did not improve the modelling compared to the UD model. This suggests that $V^{2}$ data can only be reproduced by a more complex intensity distribution.


\begin{table*}[h!]
\centering
\begin{tabular}{ccccc}
\hline
Spectral band  &  $ \phi_{LD}$ &  $ A_{LD}$  & $ \phi_{LD}$ &  $ A_{LD}$\\  
number  & $\Phi =0.0$ & $\Phi =0.0$ & $\Phi =0.2$ & $\Phi =0.2$ \\
  &  (Epoch A) &  (Epoch A) & (Epoch B) &  (Epoch B)  \\ \hline
  
1 & 5.73  $\pm$ 0.04 \,\rm mas  &  1.00 $\pm$ 0.11 \,\rm &  8.47 $\pm$ 0.08 \,\rm mas & 1.00 $\pm$ 0.12\,\rm  \\ \hline
2  &8.56  $\pm$ 0.28  \,\rm mas & 1.00 $\pm$ 0.12 \,\rm & 6.88  $\pm$ 0.07 \,\rm mas  & 1.00 $\pm$ 0.10 \,\rm \\  \hline
3 & 5.67 $\pm$ 0.05  \,\rm mas  & 1.00 $\pm$ 0.11 \,\rm  & 6.96  $\pm$ 0.02 \,\rm mas  & 1.00 $\pm$ 0.05 \,\rm  \\  \hline
4&  6.37  $\pm$ 0.34 \,\rm mas & 1.00 $\pm$ 0.11 \,\rm &  10.58 $\pm$ 0.10 \,\rm mas & 0.58 $\pm$ 0.06\,\rm \\  \hline
5 & 6.00 $\pm$ 0.37 \,\rm mas &  1.00 $\pm$ 0.10 \,\rm & 7.19 $\pm$ 0.06 \,\rm mas & -1.00 $\pm$ 0.10 \,\rm  \\  \hline
6  &  6.46 $\pm$ 0.36 \,\rm mas & -1.0 $\pm$ 1.16 \,\rm & 7.58  $\pm$ 0.05   \,\rm mas & 1.00  $\pm$ 0.12 \,\rm \\ \hline
\end{tabular}
\caption{\label{tab:DA} Fitting results of linear limb-darkened disk models. These measurements were made over all PAs. The LD coefficient was forced to be between -1 and 1. Reduced chi2 values are between 2  and 100.}
\end{table*}


\subsection{Radiative transfer modelling}
\begin{figure*}
 \centering
   \includegraphics[width=3.5in]{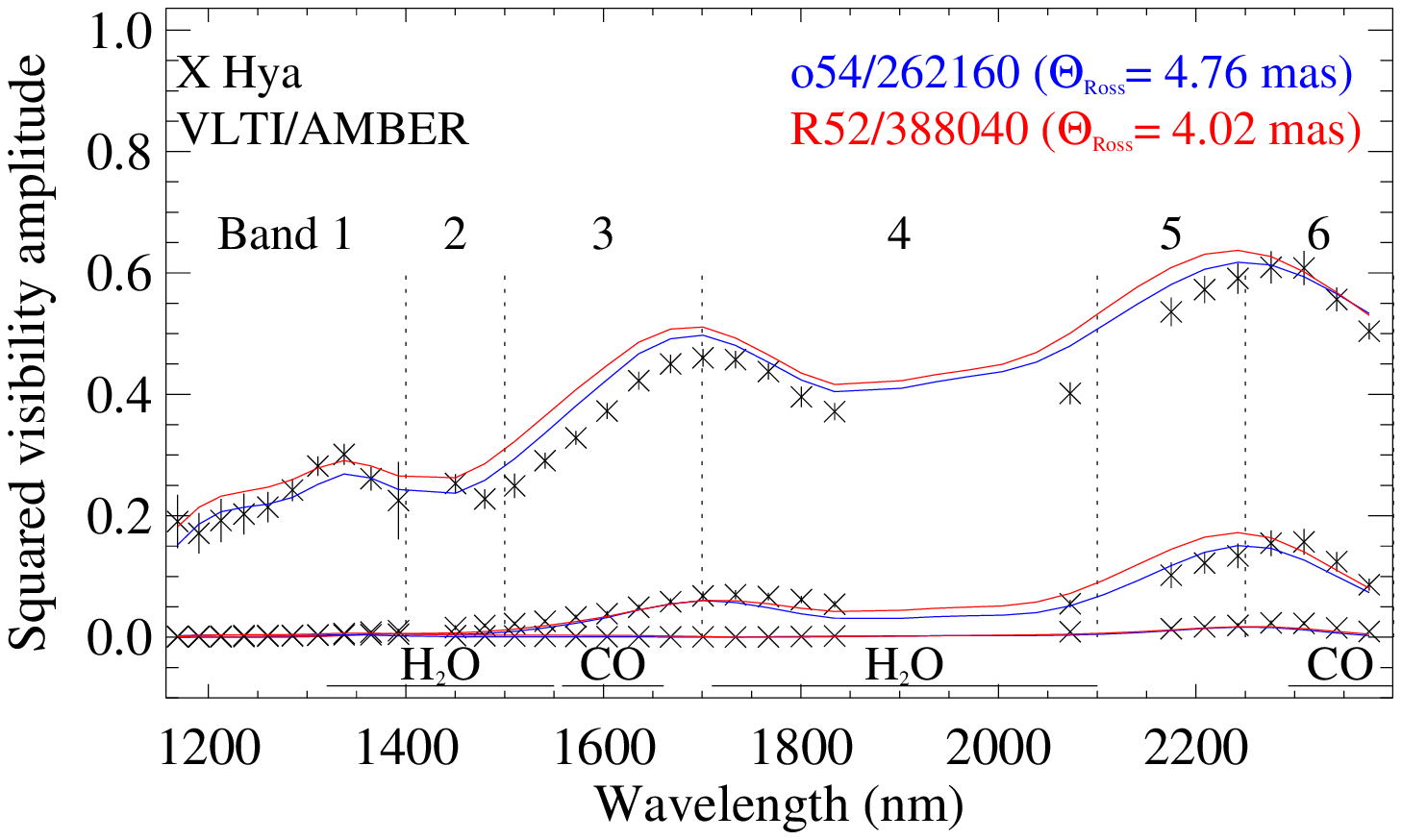}  
 \includegraphics[width=3.5in]{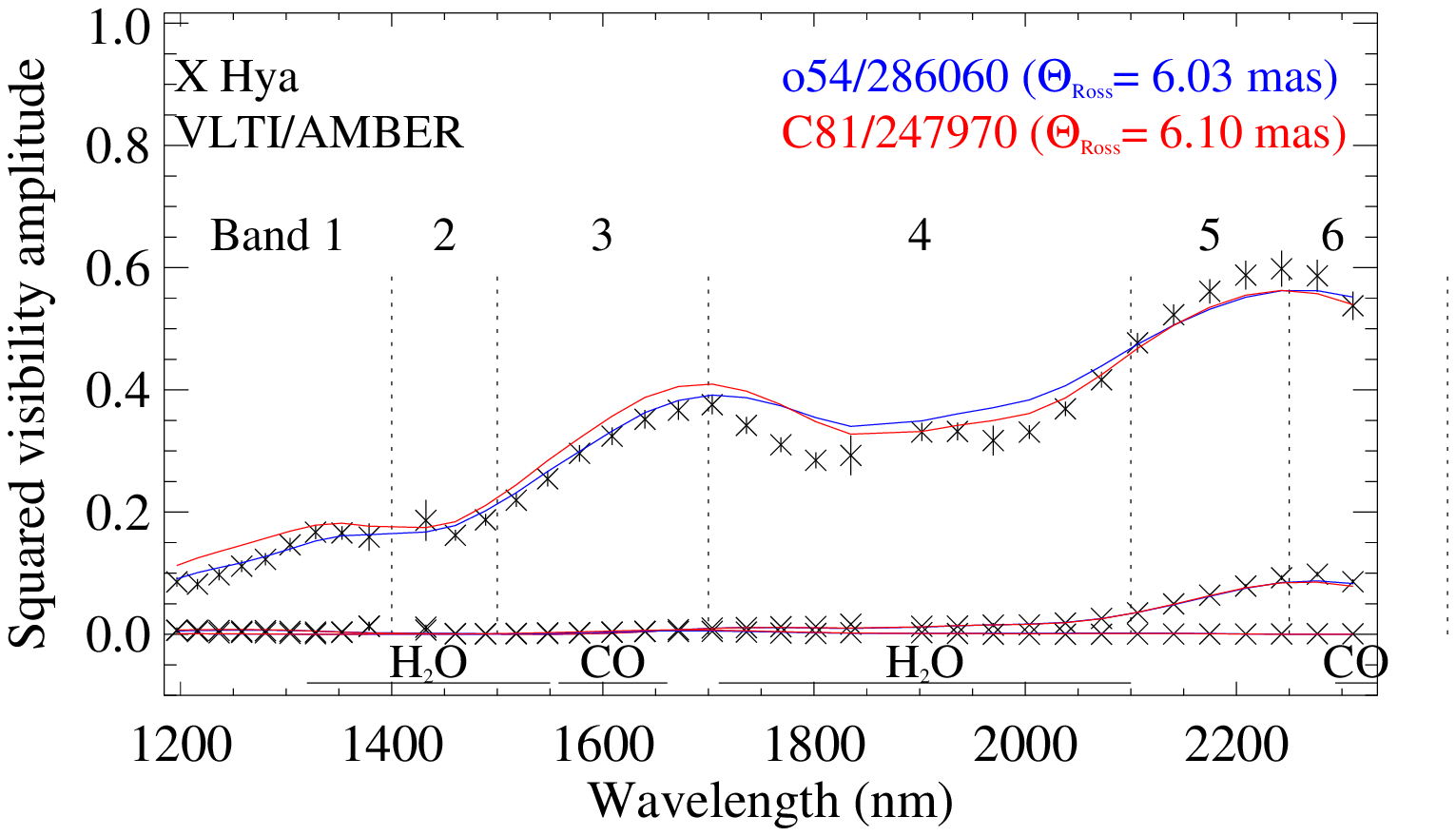} 
   \includegraphics[width=3.5in]{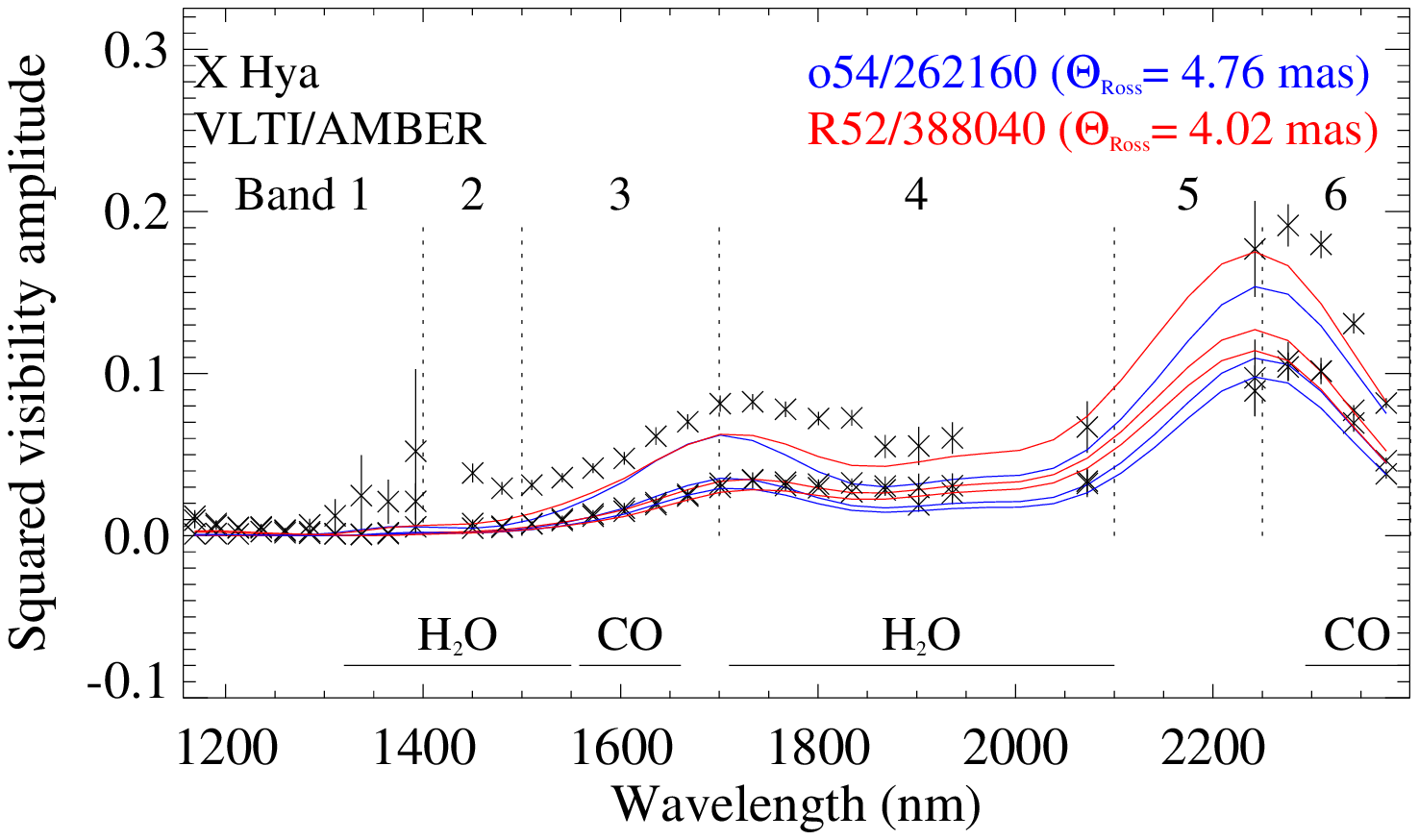}  
  \includegraphics[width=3.5in]{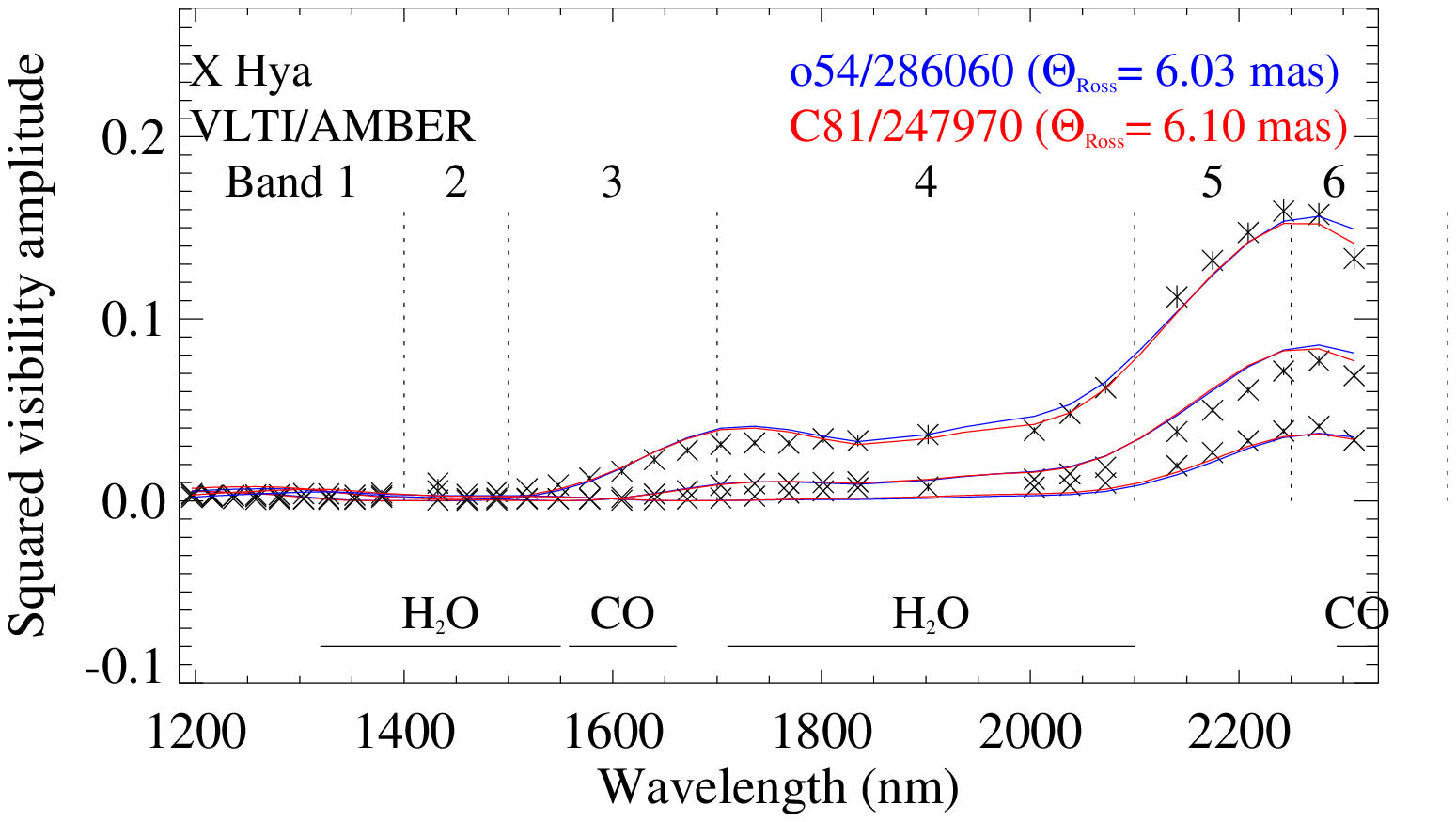}     
\caption{Modeling of the squared visibility data (in black) with R52 (red) and o54 (blue) CODEX models.
Our division in 6 spectral bands is indicated as well as the molecular absorption bands from \cite{2000A&AS..146..217L}. Each panel shows three visibility spectra corresponding to the three baselines of one triplet: D0-H0, H0-K0, and D0-K0 (upper panels) and D0-G1, D0-H0, and G1-H0 (lower panels). The baseline ground lengths for D0-H0/H0-K0/D0-K0 and D0-G1/D0-H0/G1-H0 are respectively 64m/32m/96m and 72m/64m/72m. The left panels represent Epoch A whereas the right ones represent Epoch B. The position angle variation of the configuration between the two epochs is less than 1 degree for D0-H0-K0 and less than 5 degrees for D0-G1-H0. 
}
       \label{graph_CODEX} 
 \end{figure*}

To better understand the spectral dependence of this complex dataset, we used a grid of dynamic model atmosphere series computed with the CODEX code \citep{2008MNRAS.391.1994I,2011MNRAS.418..114I}. These models develop a self-excited pulsation mechanism and sample opacities over 4300 wavelengths running from 200 nm to 50 $\mu$m, with 1 nm sampling between 200 nm and 3 $\mu$m. For more details about CODEX and the grid of stellar parameters for which the model series were computed, we refer to \cite{2011MNRAS.418..114I}. For each set of stellar parameters, these models produce radial profiles of the intensity across the chosen spectral band that we turned into $V^{2}$ by a Fourier transform. These latter were finally fitted to our data by adjusting the Rosseland diameter and the pulsation phase. Figure~\ref{graph_CODEX} shows $V^{2}$-fitted curves for two triplets of baselines at two epochs. At the difference of the simple geometric models, these physical models that take into account $\rm H_{2}O$ and CO opacities do reproduce the variation of $V^{2}$ with wavelength fairly well. The results of the model parameter fitting are summarised in Table~\ref{tab:CODEX}.

\begin{table}[h!]
\centering
\begin{tabular}{ccc}
\hline
Models &  Epoch A, $\Phi =0.0$  & Epoch B, $\Phi =0.2$   \\  
\hline
Best R52 model & 388040 & 388040   \\
$\theta_{Ross}$ (mas)  & 4.02 & 4.93  \\
$\chi_{red}^{2}$ & 8.1 & 21.9  \\
$\Phi_{CODEX}$& 0.7  & 0.7  \\
\hline
Best o54 model & 262160 & 286060  \\
$\theta_{Ross} (mas)$  & 4.76  & 6.03 \\
$\chi_{red}^{2}$ &  7.4 & 19. \\
$\Phi_{CODEX}$& 0.7  & 0.4  \\
\hline
\end{tabular}
\caption{\label{tab:CODEX} Results of the fitting of the CODEX radiative transfer models. }
\end{table}


%
%
%
%

As for the previous parametric modelling, the Rosseland diameters found with the model fitting all increase from epoch A to epoch B. Relatively to epoch A, the increase in diameter from epoch A to epoch B was 23 \% and 27 \% for the R52 and o54 CODEX series, respectively. Compared to the same modelling for $\Phi =0.7$ reported in Table 3 of \cite{2011A&A...532L...7W}, we globally find a poorer fit by the attempted models to the data as can be seen from the values of the reduced $\chi^{2}$. The R52 model fits the data almost equally well as the o54 series. These two models differ from each other in the parent star parameters (o54: M = 1.1 M$_\odot$, L = 5400 L$_\odot$, R = 216 R$_\odot$, P = 330 days; R52: M = 1.1 M$_\odot$, L=5200 L$_\odot$, R= 209 R$_\odot$, P= 307 days). The fact that the CODEX series were computed for a limited parameter space could also explain the lack of a better fit to our data. Moreover, the phases fitted by the CODEX models (0.7 and 0.4) do not match the visual phases (0.0 and 0.2, respectively). This discrepancy in the phases might be explained by the degeneracy in the CODEX models with the parameters and the pulsation phases: the same atmosphere might be equally modeled by two different sets of parameters and phases. On the other hand, CODEX models are 1D and do not take into account non-radial features. The presence of such non-radial features in some azimuth ranges causes the visibility curve to depart from a pure radial profile and would explain why the phase determination is not accurate. Among the possible origins for non-radial intensity distributions we can cite : surface spots, non-radial pulsation, an asymmetric shock front, an inhomogeneous molecular layer and dust environment, presence of a companion, etc. In the following, we present an imaging work that aims at unveiling the non-radial and asymmetric properties in the morphology of the object.


\section{Imaging}
\label{ima}

\begin{figure*}[h!]
 \centering
 \includegraphics[width=3.5in]{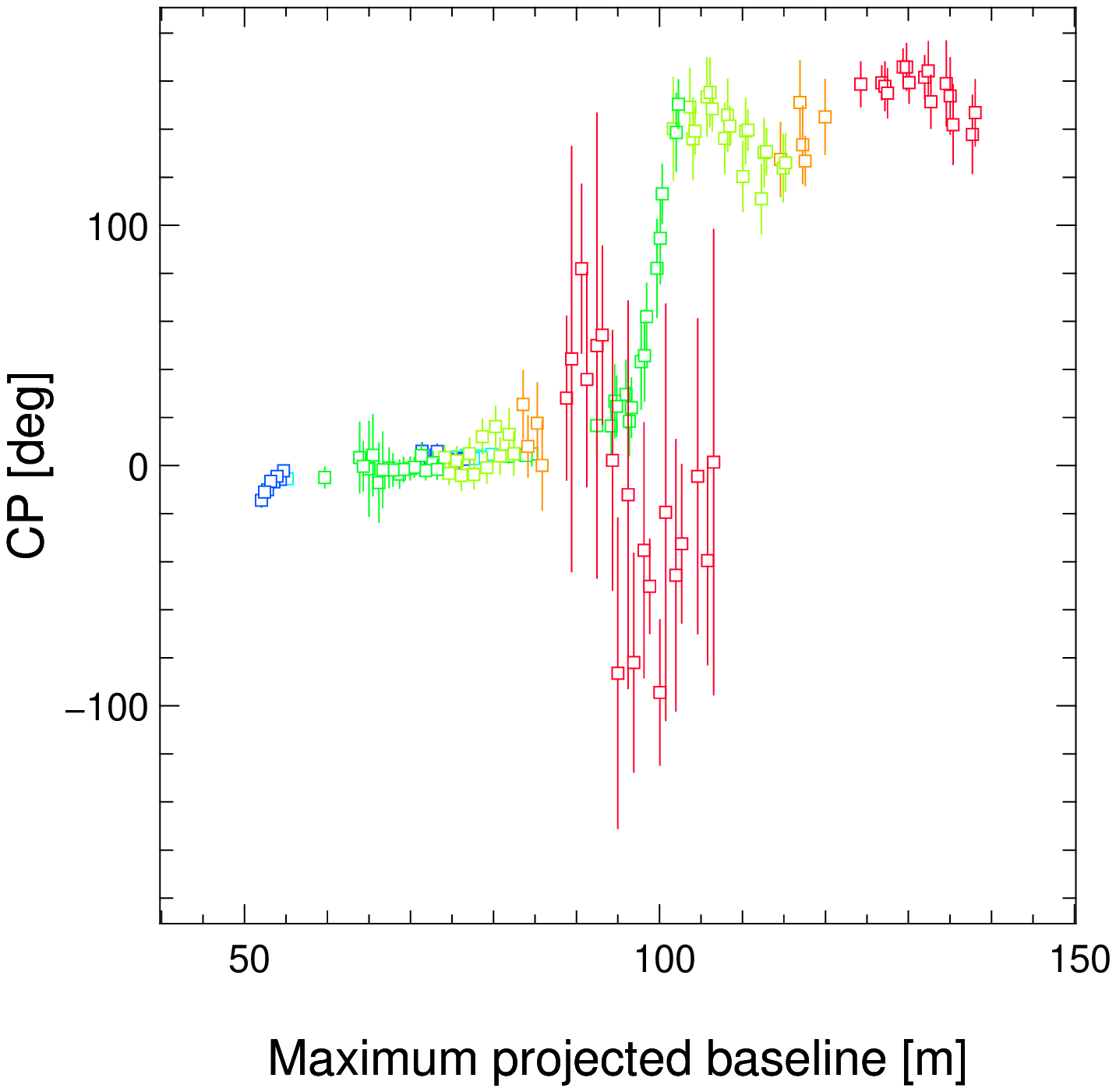} 
  \includegraphics[width=3.5in]{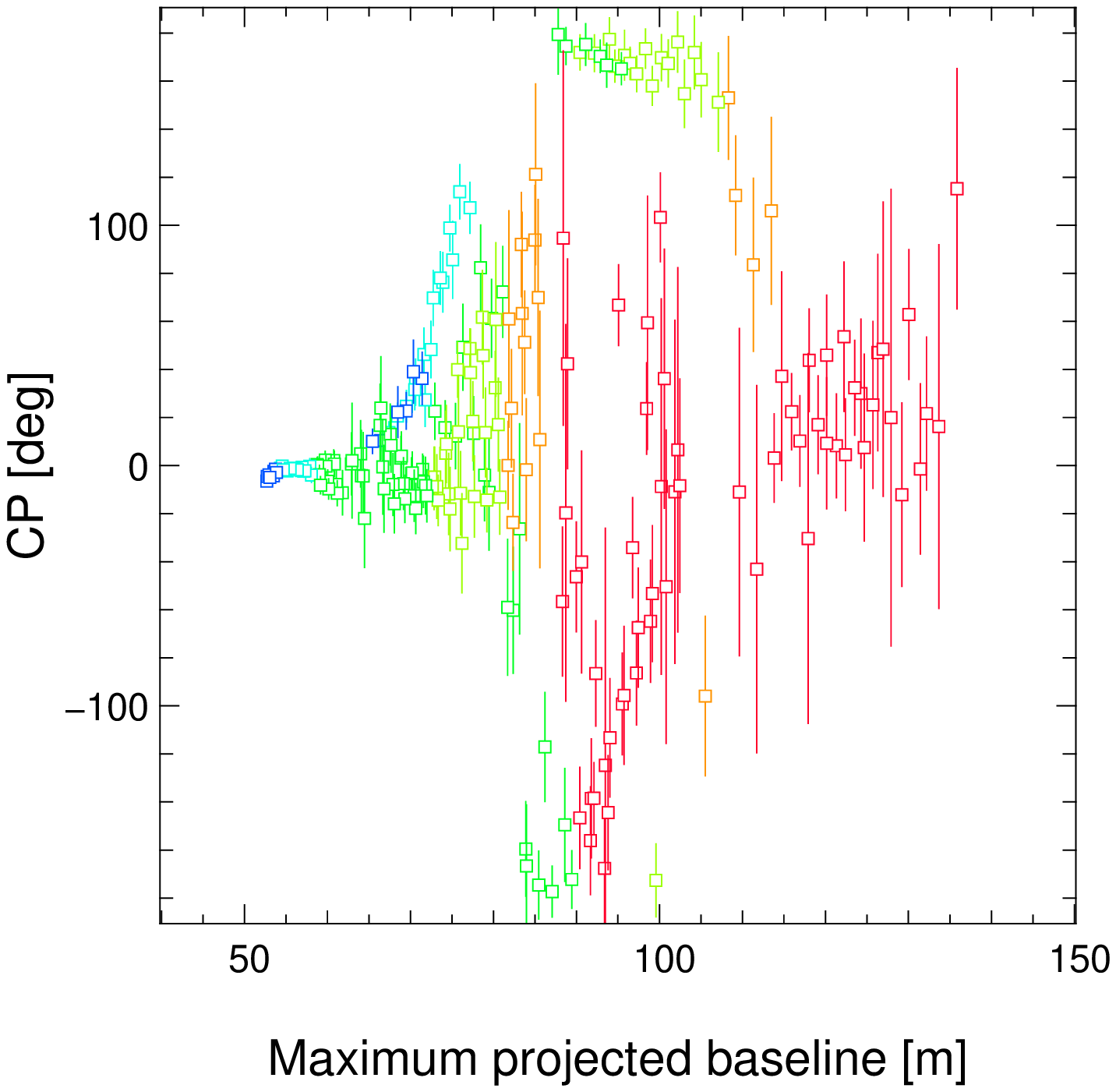}  
\caption{Closure phases in degrees at epochs A (left) and B (right). Colours indicate the six different spectral bands with the same coding as in Fig.~\ref{uvcov}.
}
       \label{cpdata} 
 \end{figure*}

In the previous sections, we only took into account the $V^{2}$ data, which means that the object was thought to be centro-symmetric. However, since AMBER recombines the beams from three telescopes, we do have a phase information called closure phase that is a measurement of the asymmetry in the brightness distribution of the object. Figure~\ref{cpdata} shows that closure phase departs significantly from the 0 modulo $180\degr$ values which means that X Hya was not centro-symmetric in most of the spectral bands.

The closure phase signal of epochs A and B exhibits a noisy pattern at the spectral band 1, which is located around 100m of maximum projected baseline (second visibility lobe). This might be due to a combination of poor atmospheric conditions and low object visibilities at these spatial frequencies. The object can be considered centro-symmetric below $\sim$80 meters which corresponds to $\sim$5.5 mas at 2.15 $\mu$m. For epoch B, closure phases show a much higher dispersion and higher uncertainties. The maximum baseline where the closure phases depart from the null value lies between 55 and 60 meters, which indicates a higher degree of asymmetry than for epoch A. These results agree with the $H$-band measurements presented in \cite{2006ApJ...652..650R} that reported that the closure phase signal of X Hya was consistent with the null value up to a maximum baseline of 24.5 meters (observations at $\Phi =0.5$). 

\begin{figure*}[h!]
 \centering
  \includegraphics[width=3.5in]{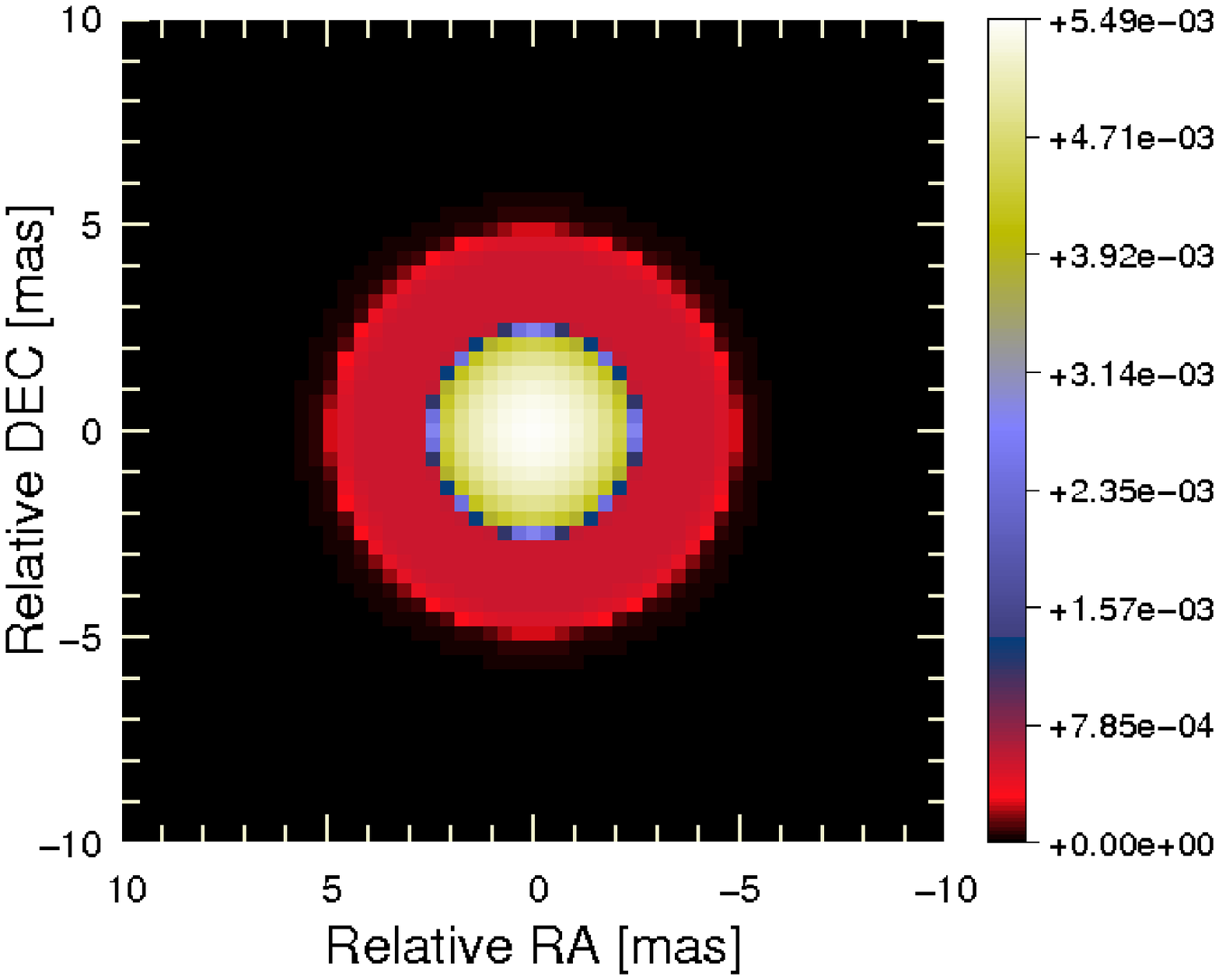}  
\caption{Image built from the CODEX intensity profile R52/388040 at 1.9 micron. This image was used as a prior and initial guess for the image reconstruction of band 4 datasets ($\rm H_{2}O (2)$).}
       \label{prior} 
 \end{figure*}

To reconstruct the spatial intensity distribution out of such a complex dataset of $V^{2}$ and closure phases,  it is mandatory to use image reconstruction algorithms such as MiRA \citep{2008SPIE.7013E..43T}. However, given our limited uv coverage in individual bands, we cannot aim at reconstructing a model-independent image. Nevertheless, we carried out an image reconstruction approach based on our parametric modelling results. We therefore used 1D intensity profiles from the CODEX models that we turned into initial guess and \textit{a priori} image for the reconstruction. 
Several regularisation techniques were used including the total variation (TV) regularisation recommended in benchmarking studies for image reconstruction with MiRA \citep{2011A&A...533A..64R}. The best results were obtained with the TV and quadratic regularisations towards an \textit{a priori} object image derived from the R52/388040 model intensity profile (Fig.~\ref{prior}).

For the quadratic regularisation, the reconstructed images were obtained by adjusting the data and by regularising towards this prior image. This was done by minimising the following criterion: 

\begin{equation}
 J =  \sum_{x,y} [ J_{data}(x,y)+ \mu f_{prior} (x,y)]
\end{equation}
with
\begin{equation}
 f_{prior} = \sum_{x,y} [I(x,y) - P(x,y)]^2     
\end{equation}

where I(x,y) is the reconstructed image and P(x,y) is the prior image, x and y are the coordinates, and $\mu$, also called hyperparameter, sets the weight between data ($V^2$ and closure phase) fitting, represented by the criterion $J_{data}$,  and regularisation towards the prior image, the criterion called $f_{prior}$.

\begin{figure*}[!h]
 \centering   
 \includegraphics[width=3.2in]{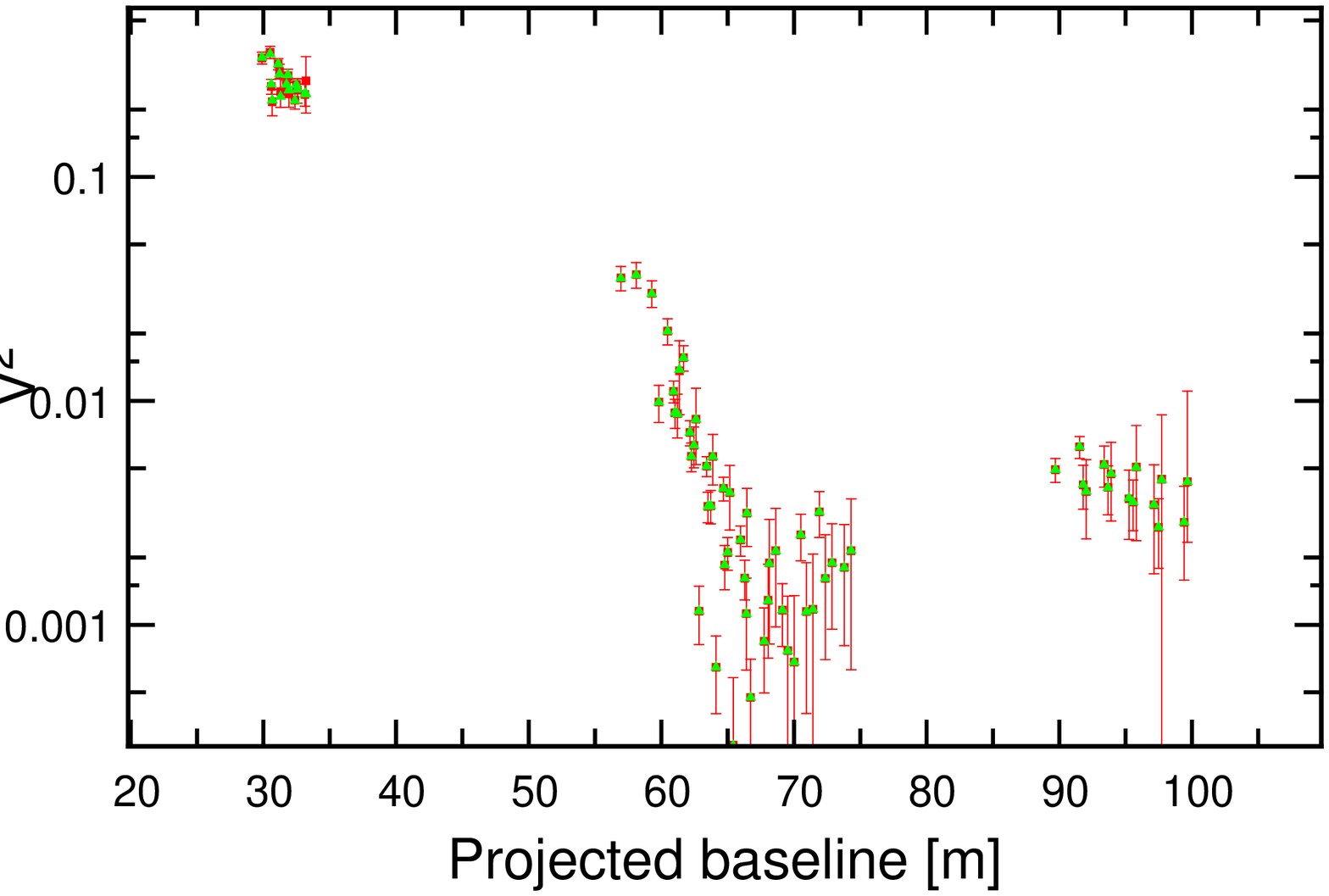}
 \hspace{0.1in}
   \includegraphics[width=3.2in]{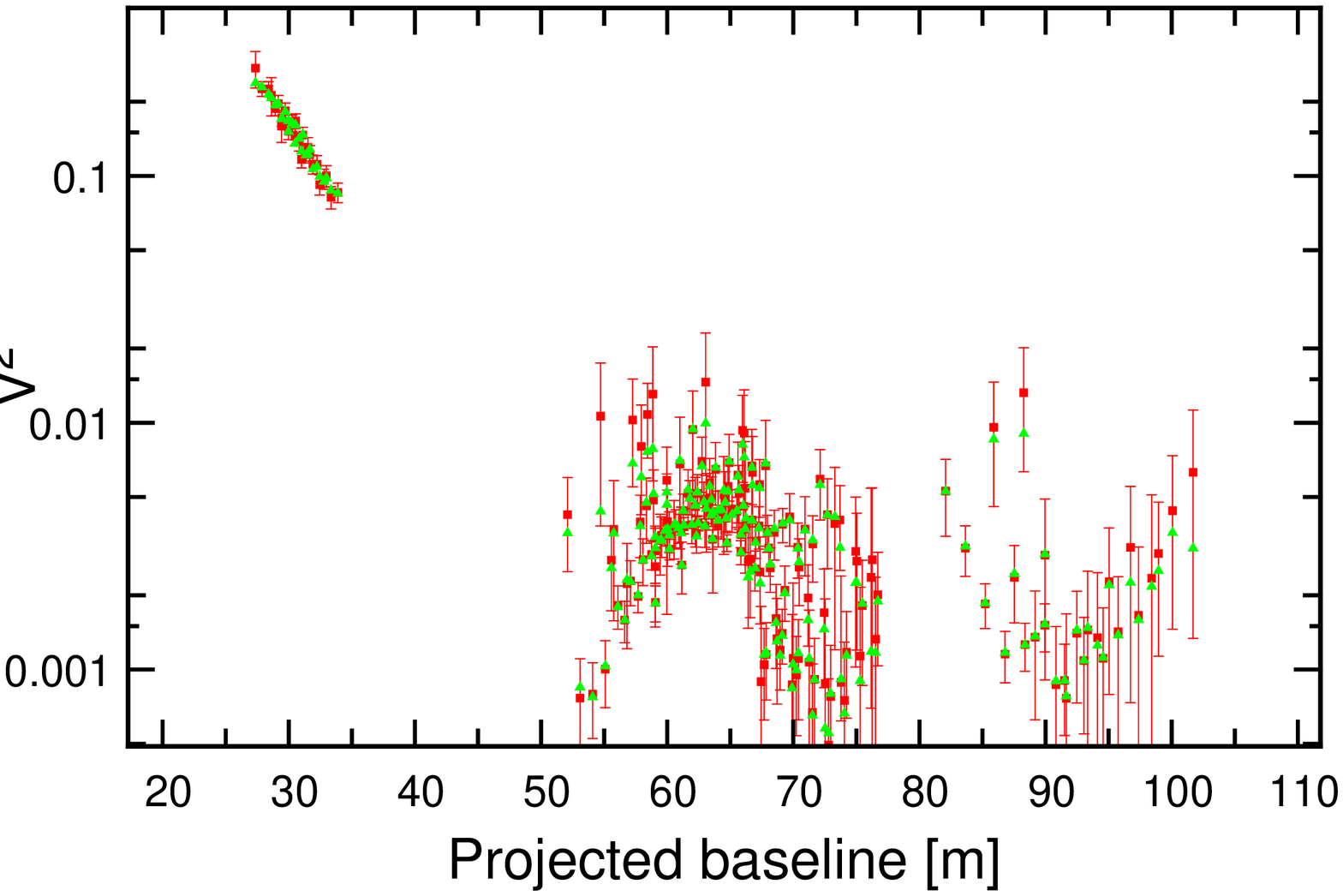}
  \vspace{0.1in}
  \includegraphics[width=3.2in]{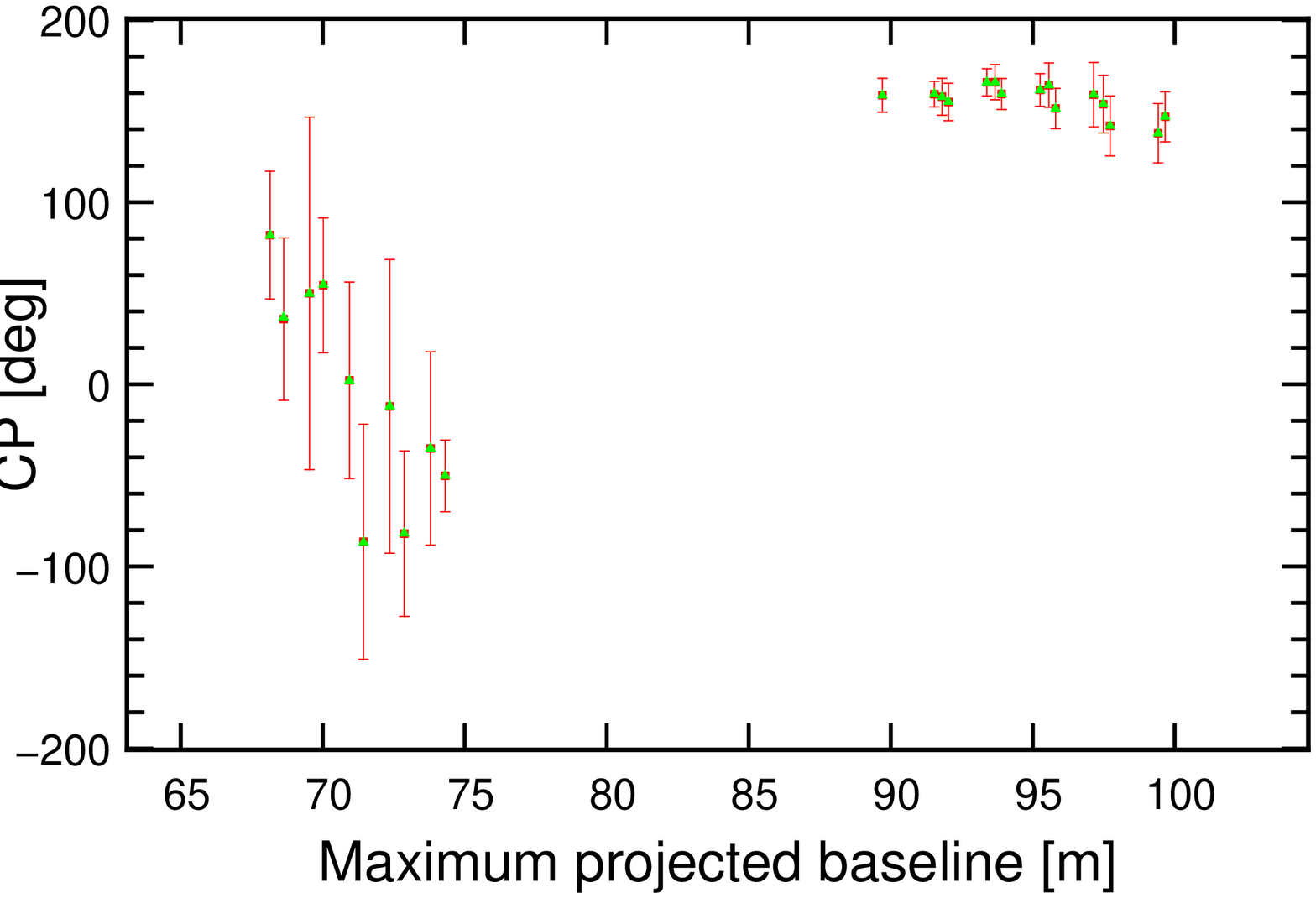}  \hspace{0.1in}
  \includegraphics[width=3.2in]{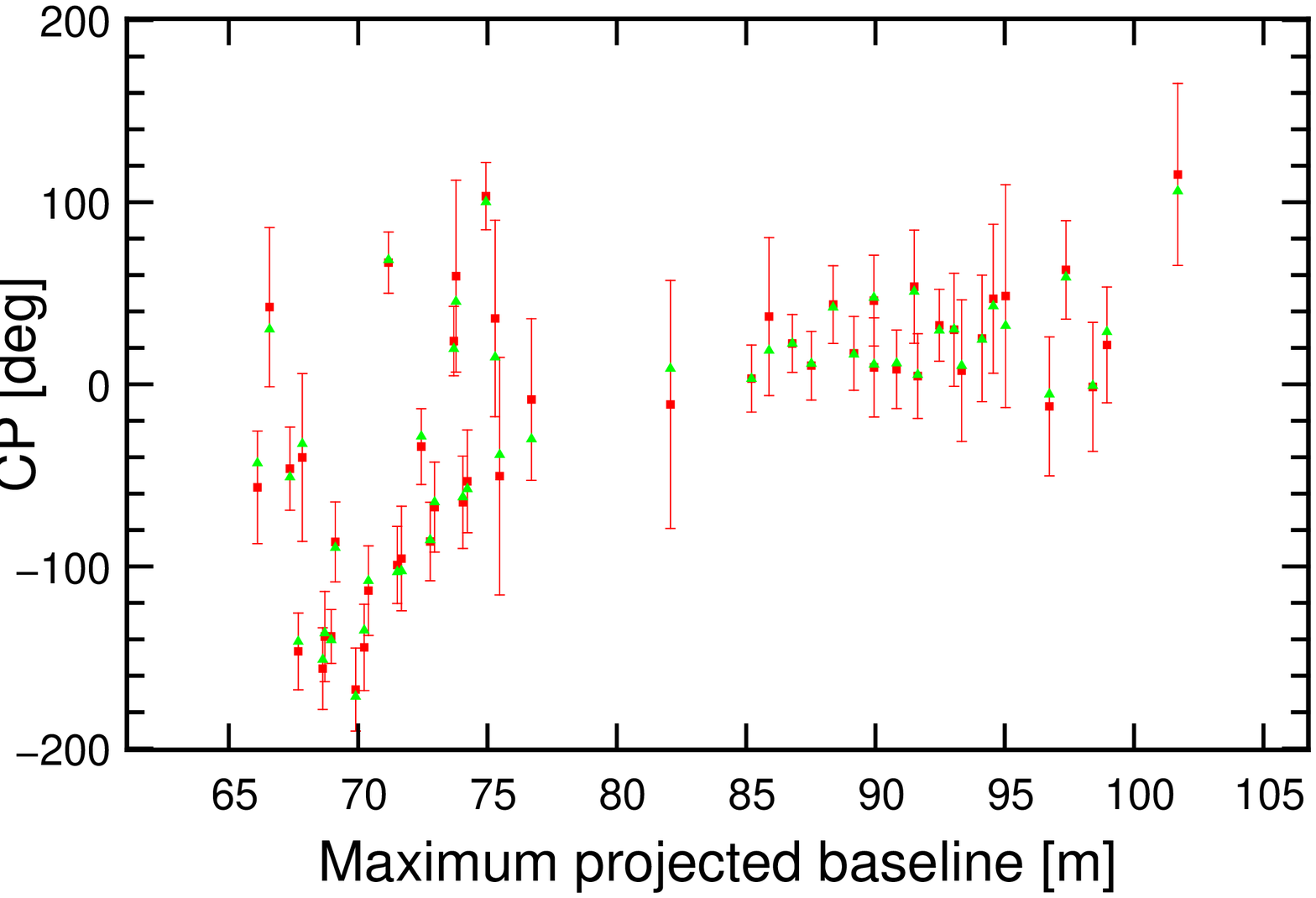}
\caption{Squared visibilities (Upper panels) and closure phases (lower panels) data points and their fit by the image reconstruction process are plotted in red and  green respectively, for Epoch A (left) and Epoch B (right). This dataset corresponds to band 1, i.e. Continuum (1).}
       \label{band1}
 \end{figure*}

\begin{figure*}[!h]
 \centering   
 \includegraphics[width=3.2in]{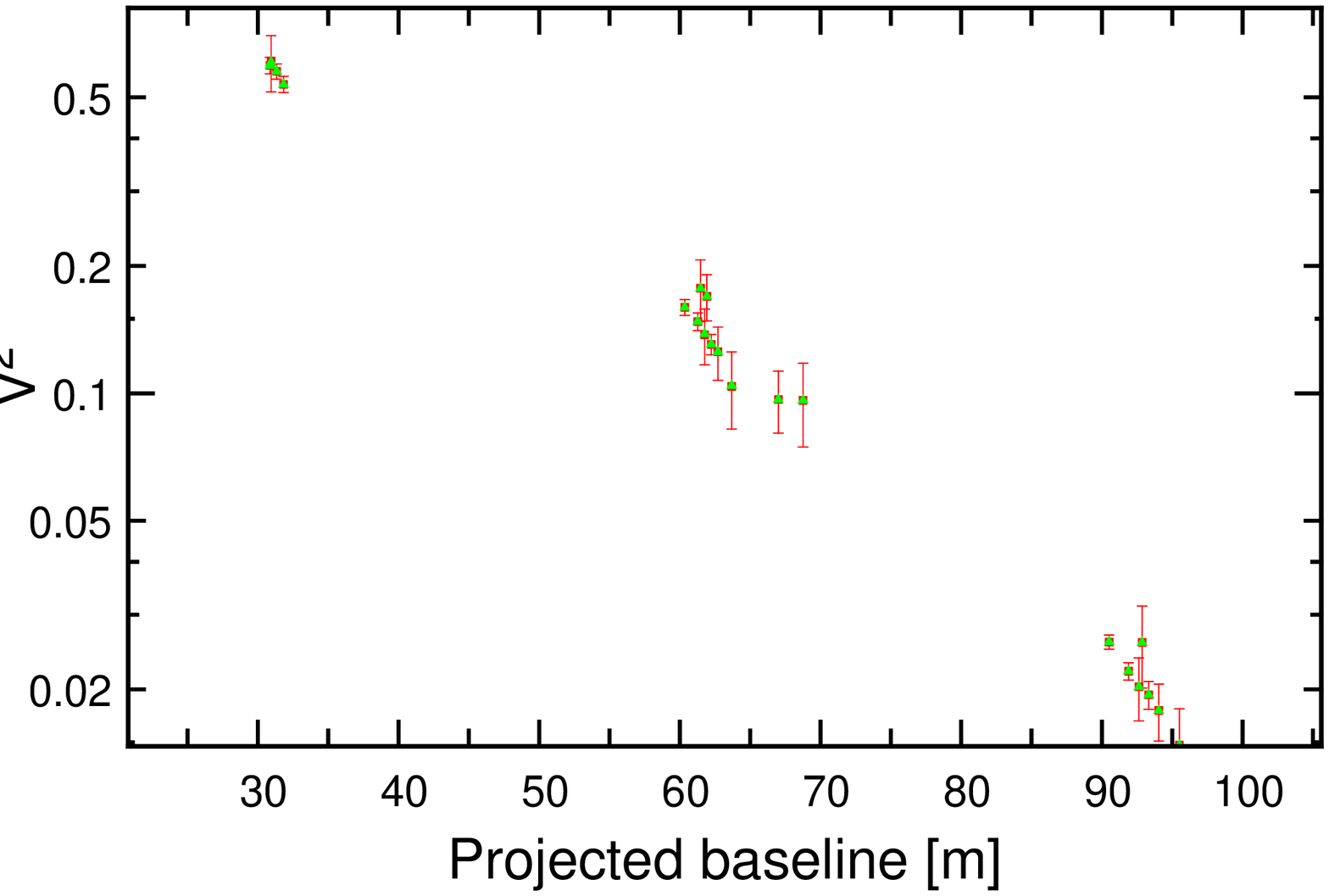}
 \hspace{0.1in}
   \includegraphics[width=3.2in]{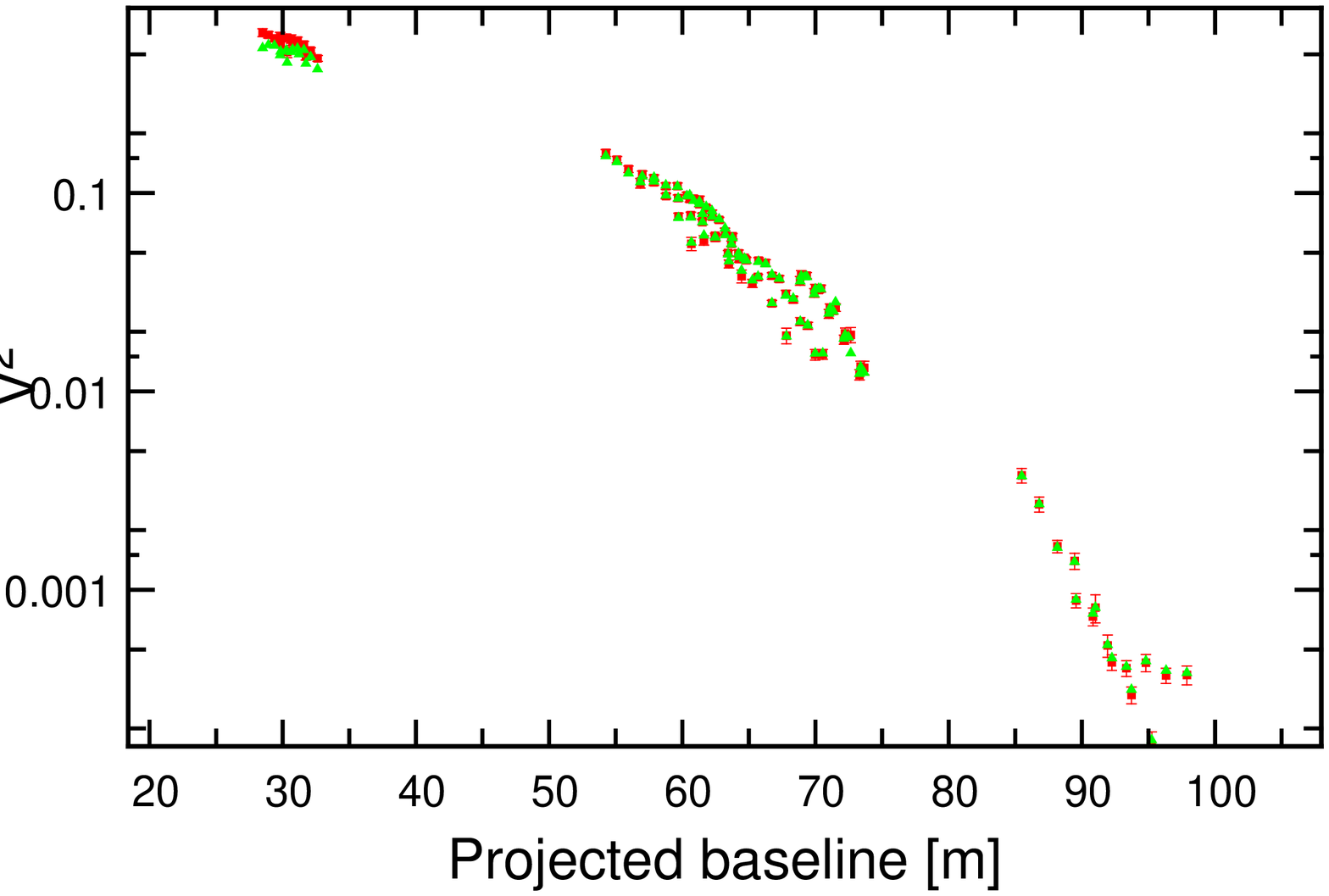}
  \vspace{0.1in}
  \includegraphics[width=3.2in]{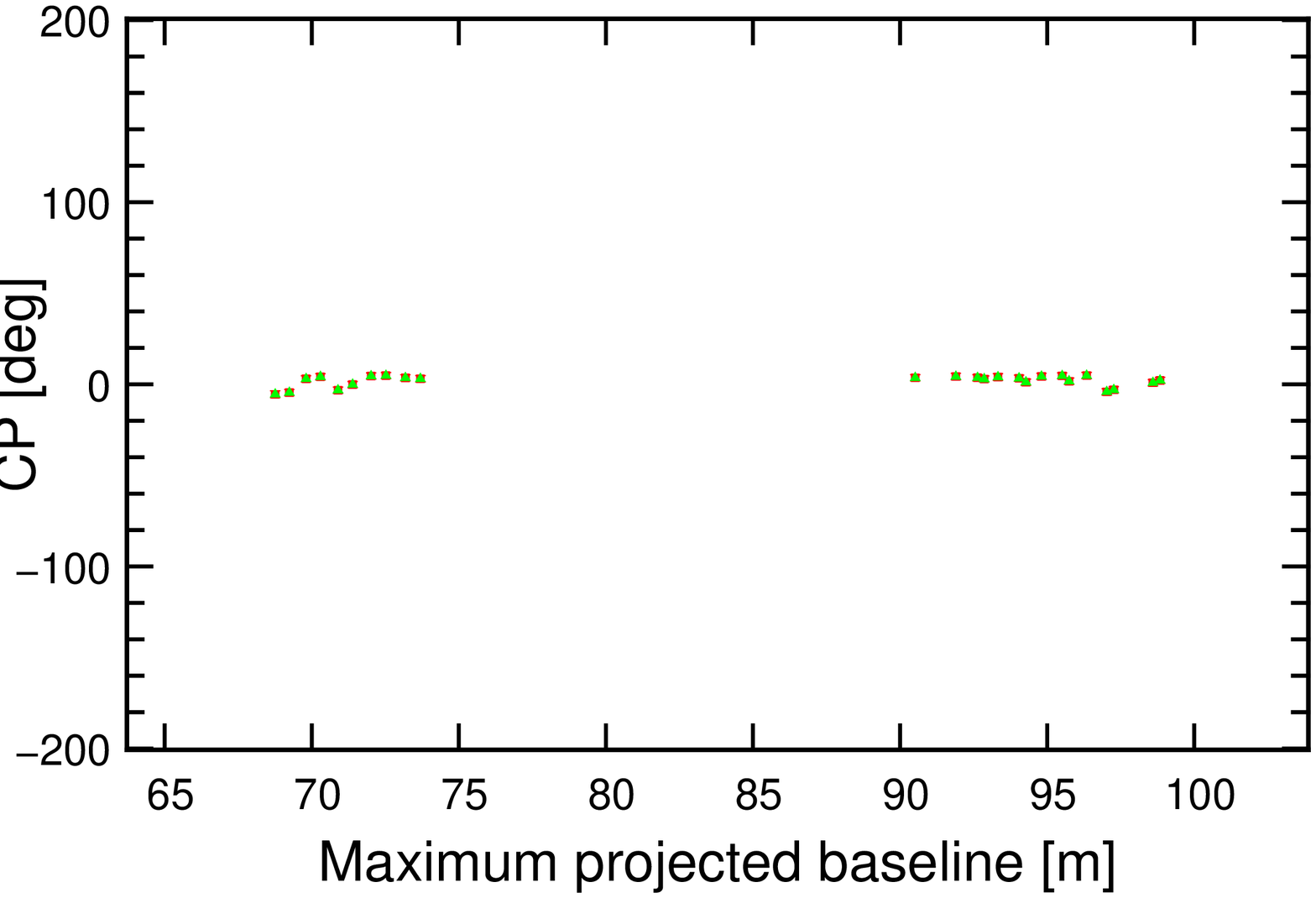}\hspace{0.1in}
  \includegraphics[width=3.2in]{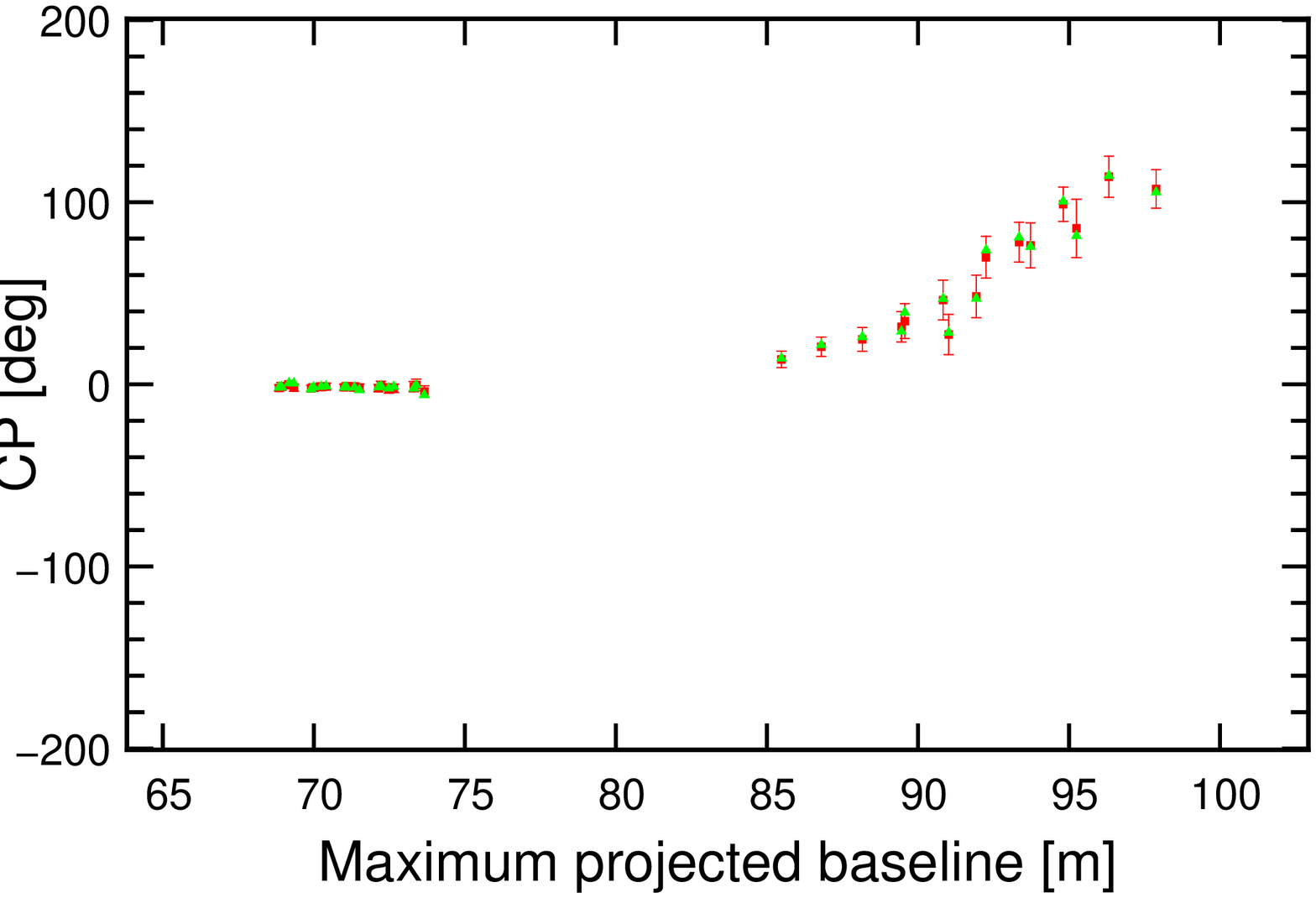}
\caption{Same as Fig.~\ref{band1} for band 5, Continuum (2).}
       \label{band5}
 \end{figure*}

\begin{figure*}[!h]
 \centering   
 \includegraphics[width=3.3in]{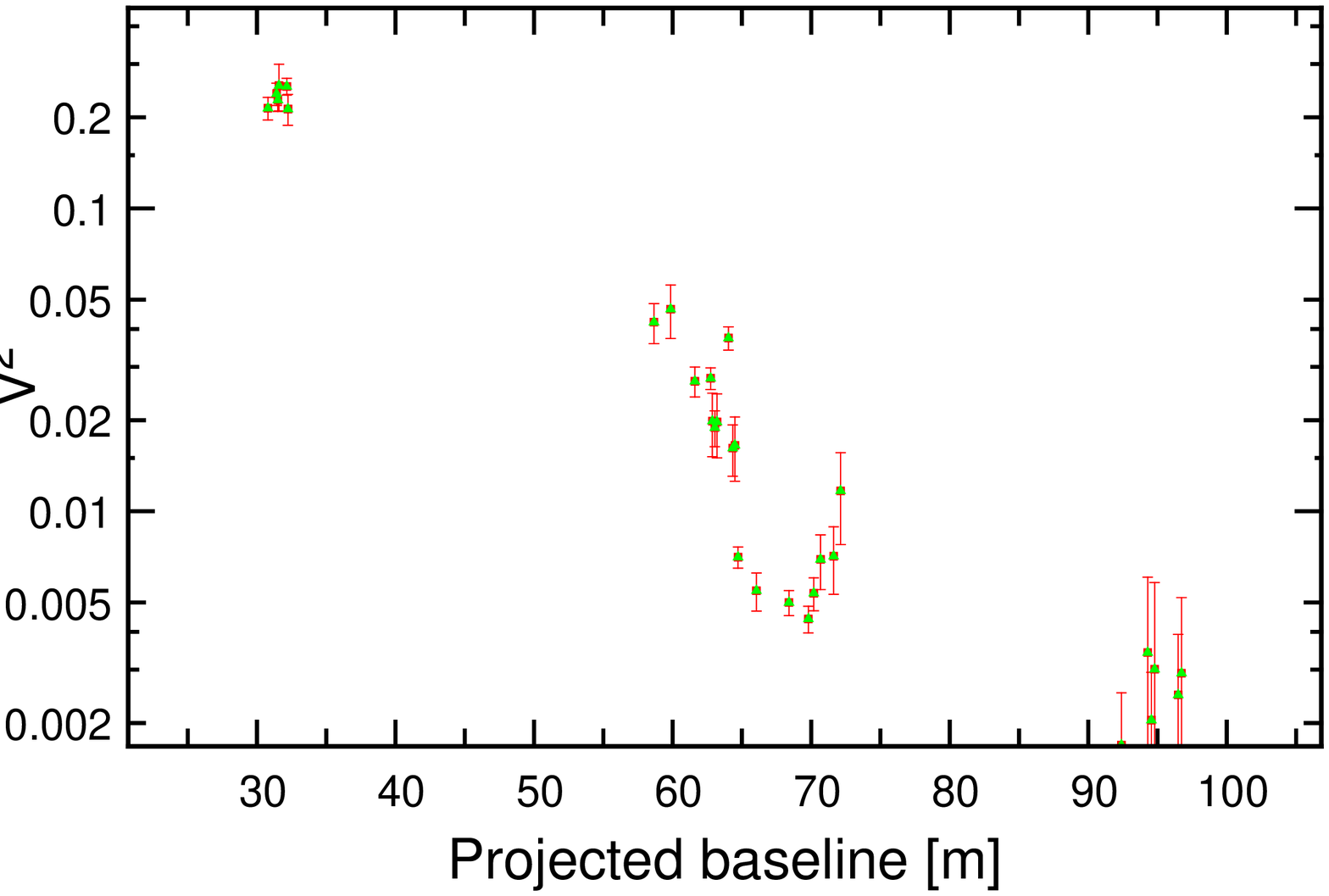}
 \hspace{0.1in}
   \includegraphics[width=3.3in]{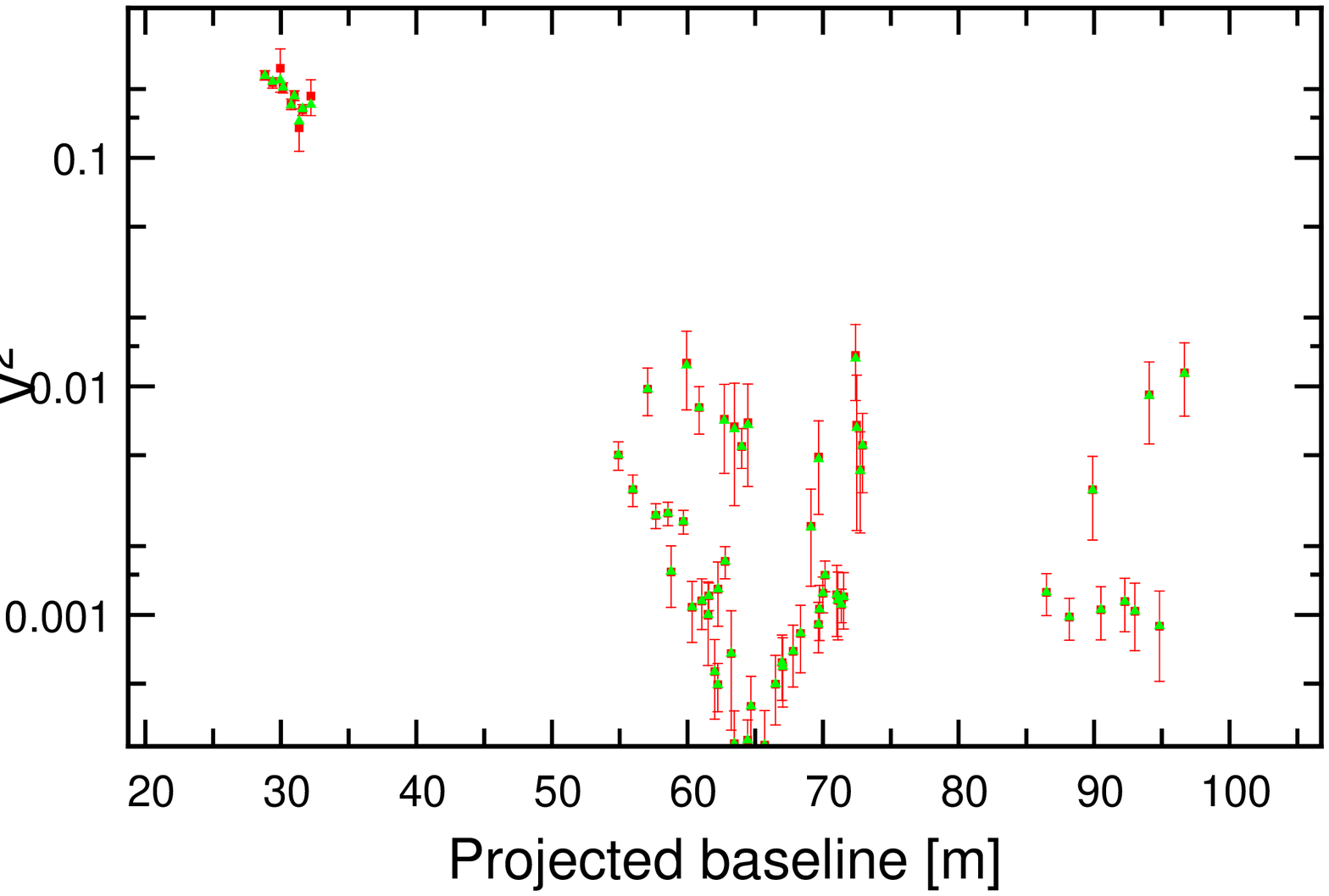}
  \vspace{0.1in}
  \includegraphics[width=3.3in]{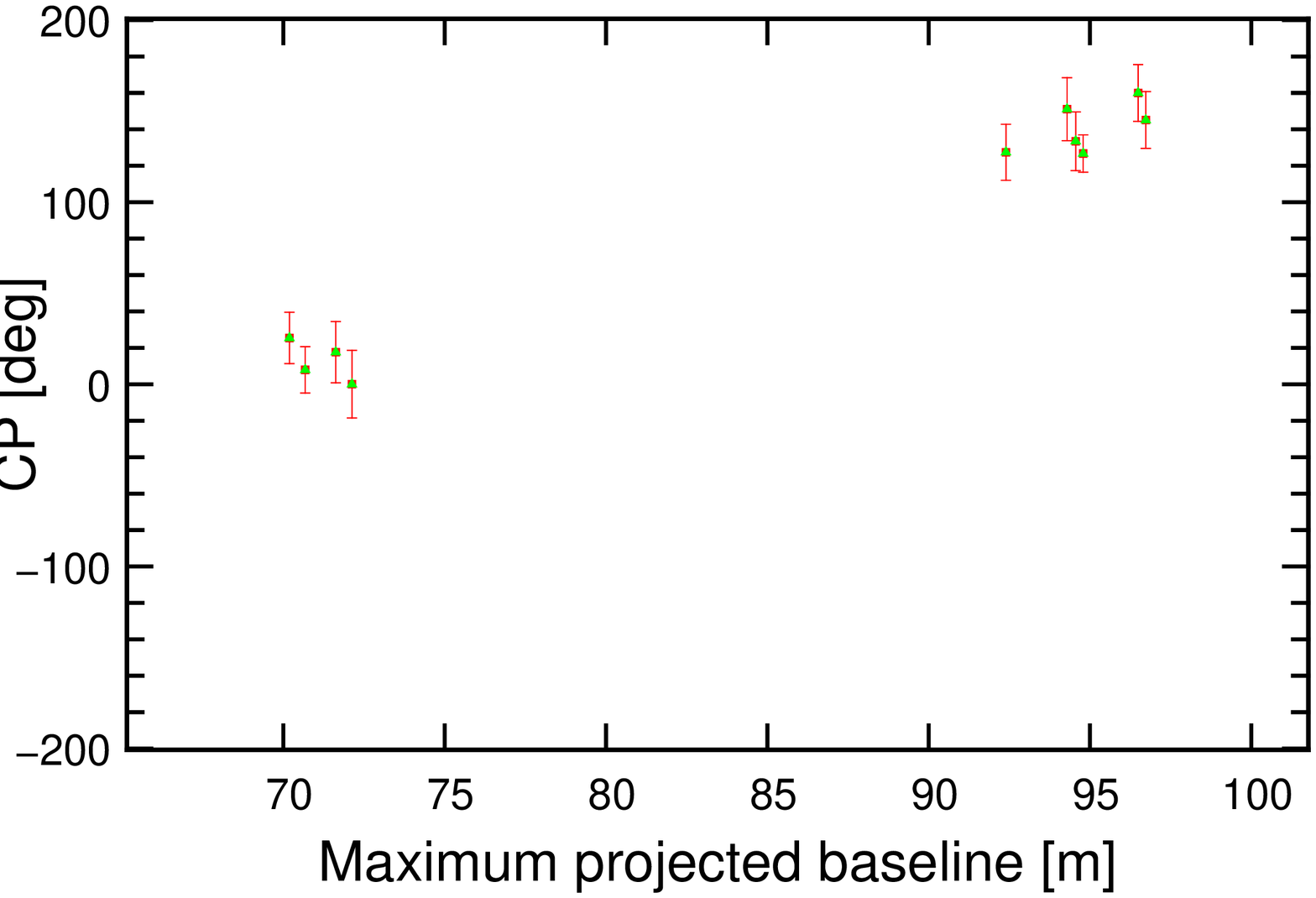}\hspace{0.1in}
  \includegraphics[width=3.3in]{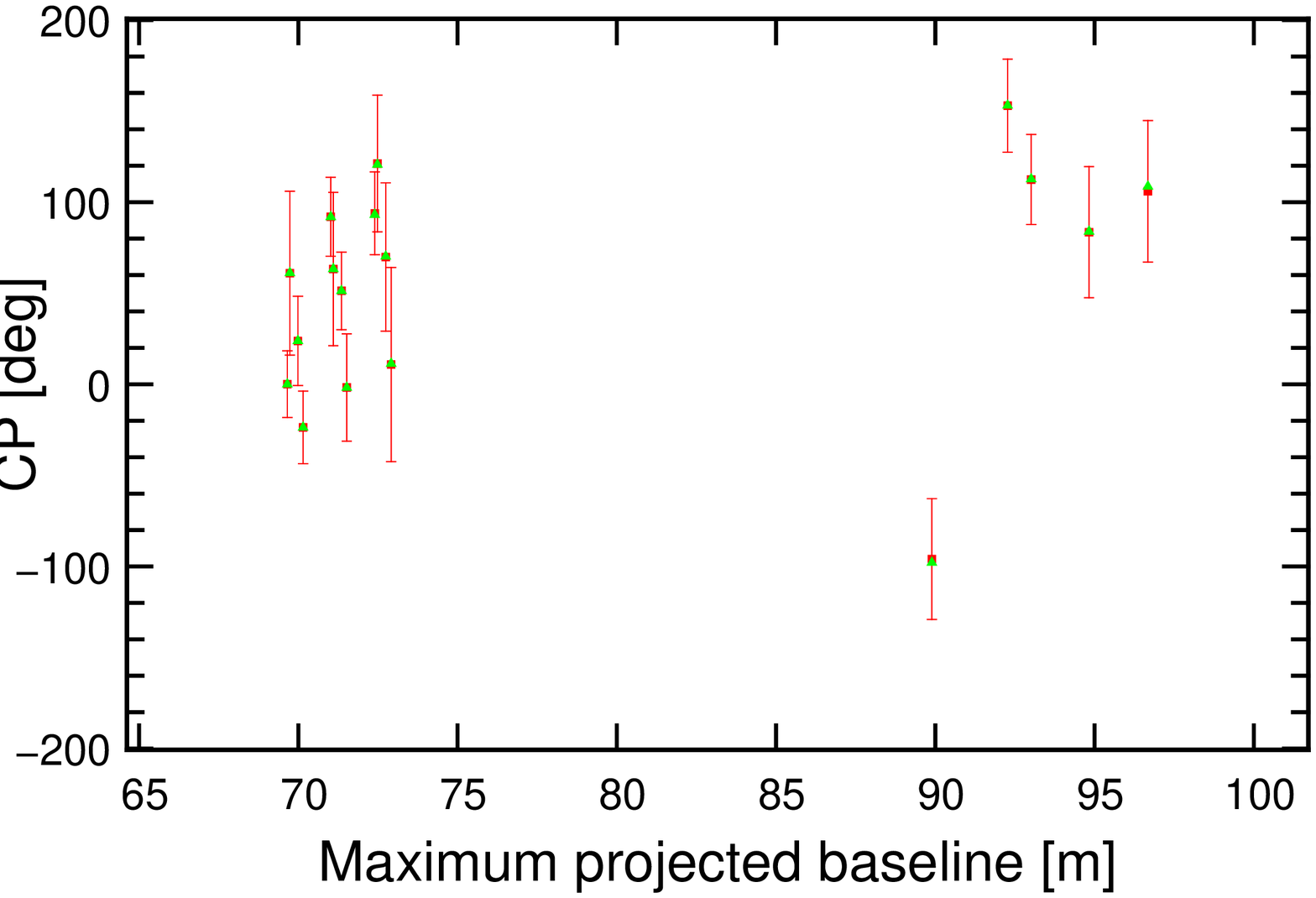}
\caption{Same as Fig.~\ref{band1} for band 2, $\rm H_{2}O$ (1).}
       \label{band2}
 \end{figure*}

\begin{figure*}[!h]
 \centering
 \includegraphics[width=3.3in]{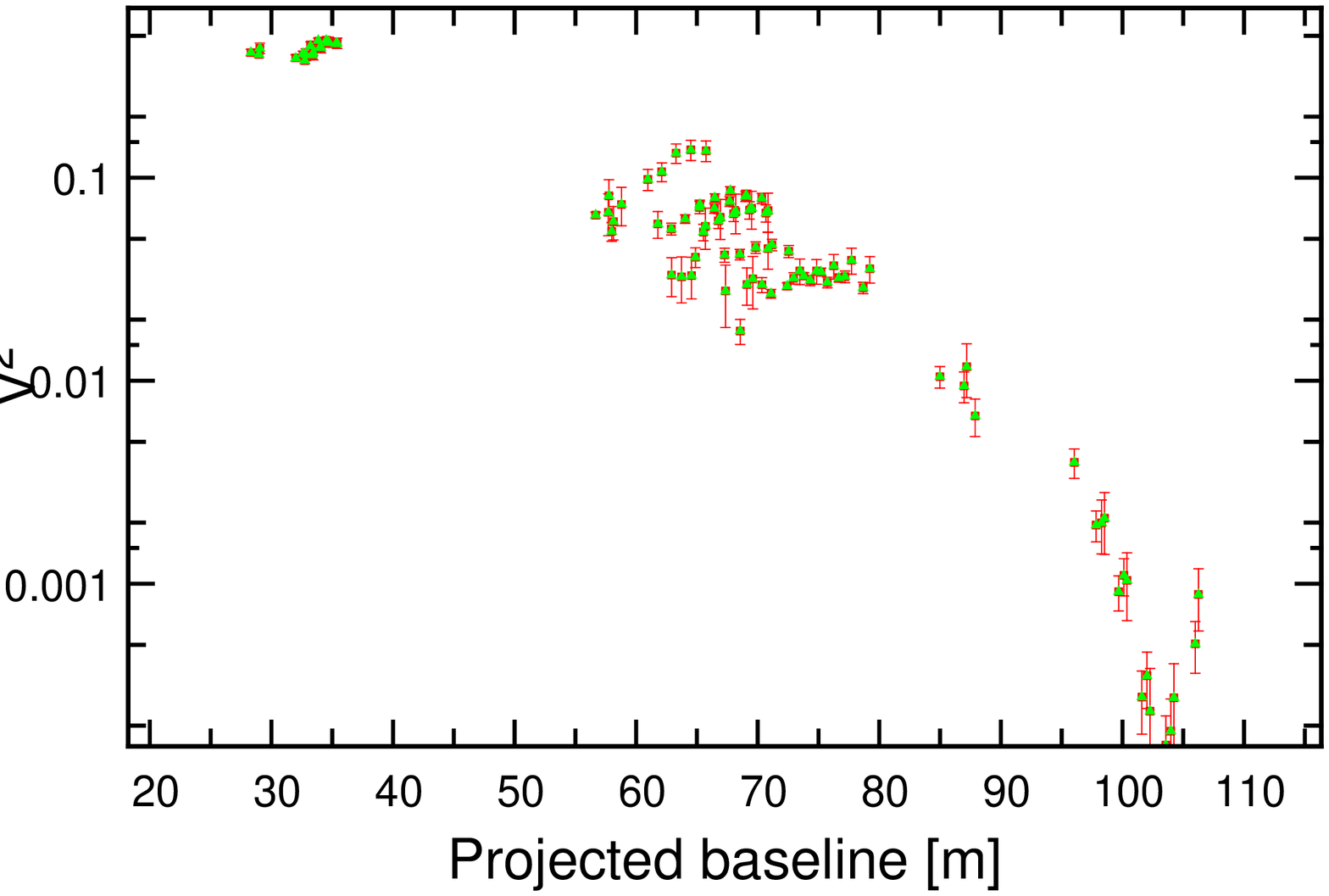} 
        \hspace{.1in}
         \includegraphics[width=3.3in]{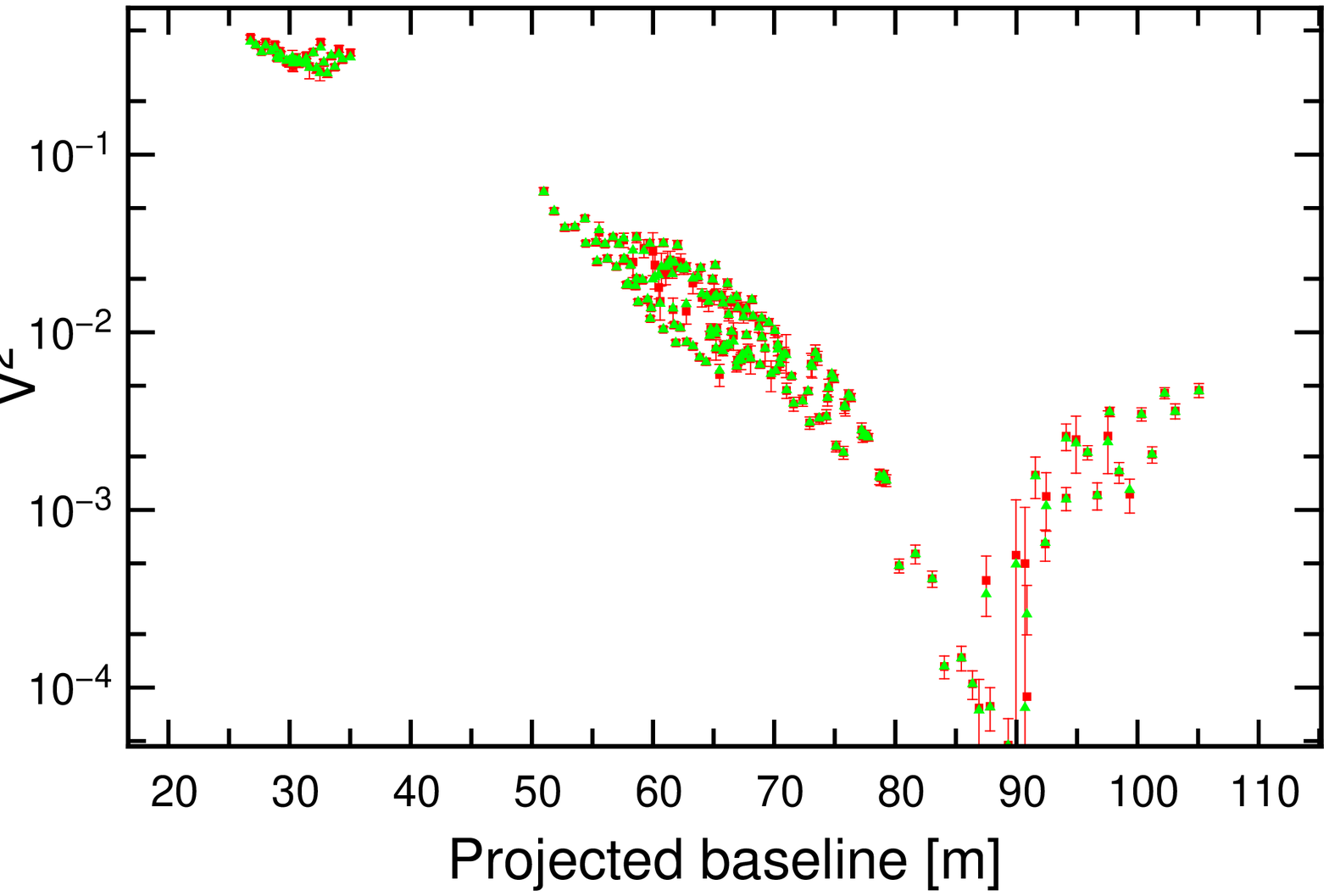}  
          \vspace{0.1in}
         \includegraphics[width=3.3in]{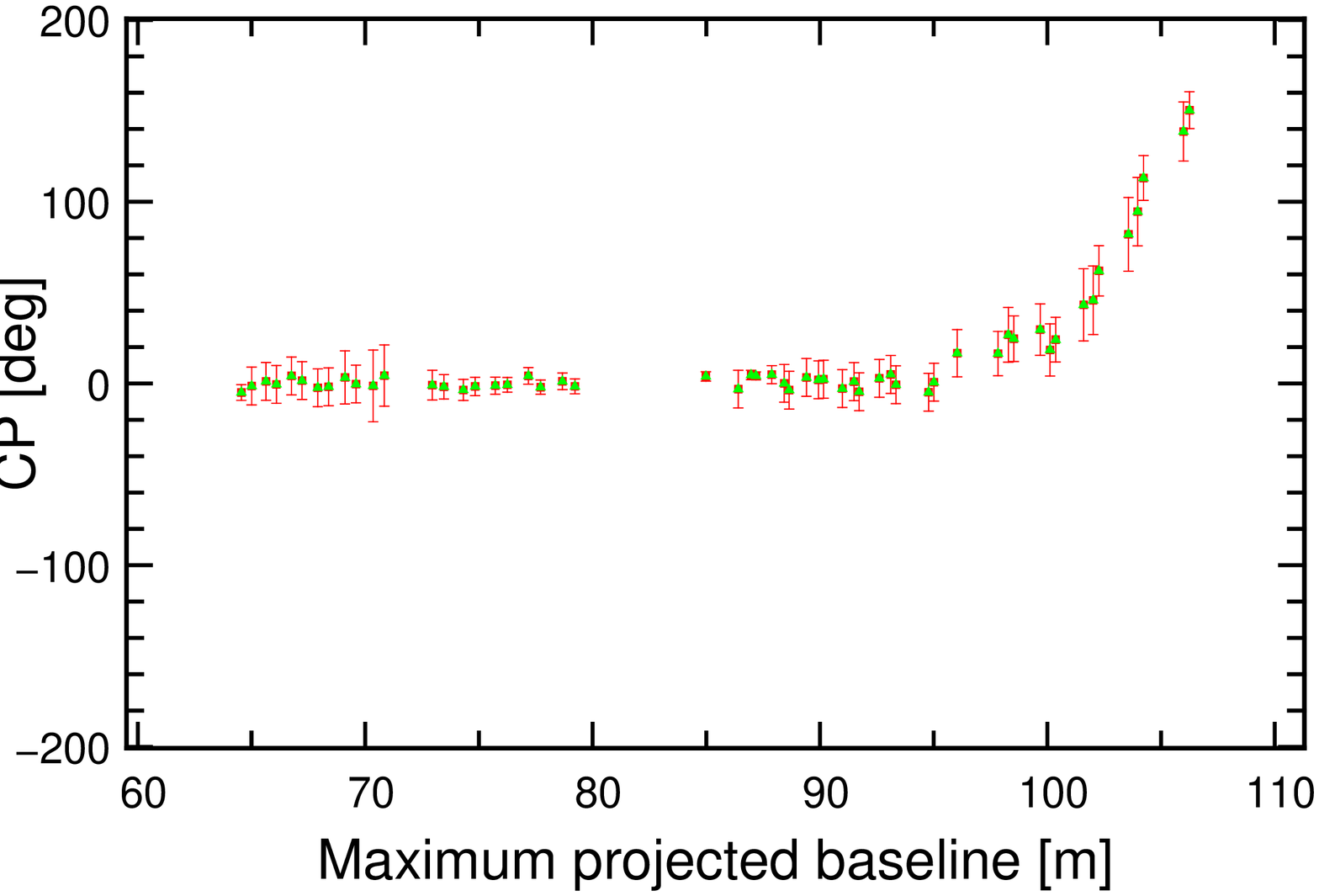} 
       \hspace{.1in}
  \includegraphics[width=3.3in]{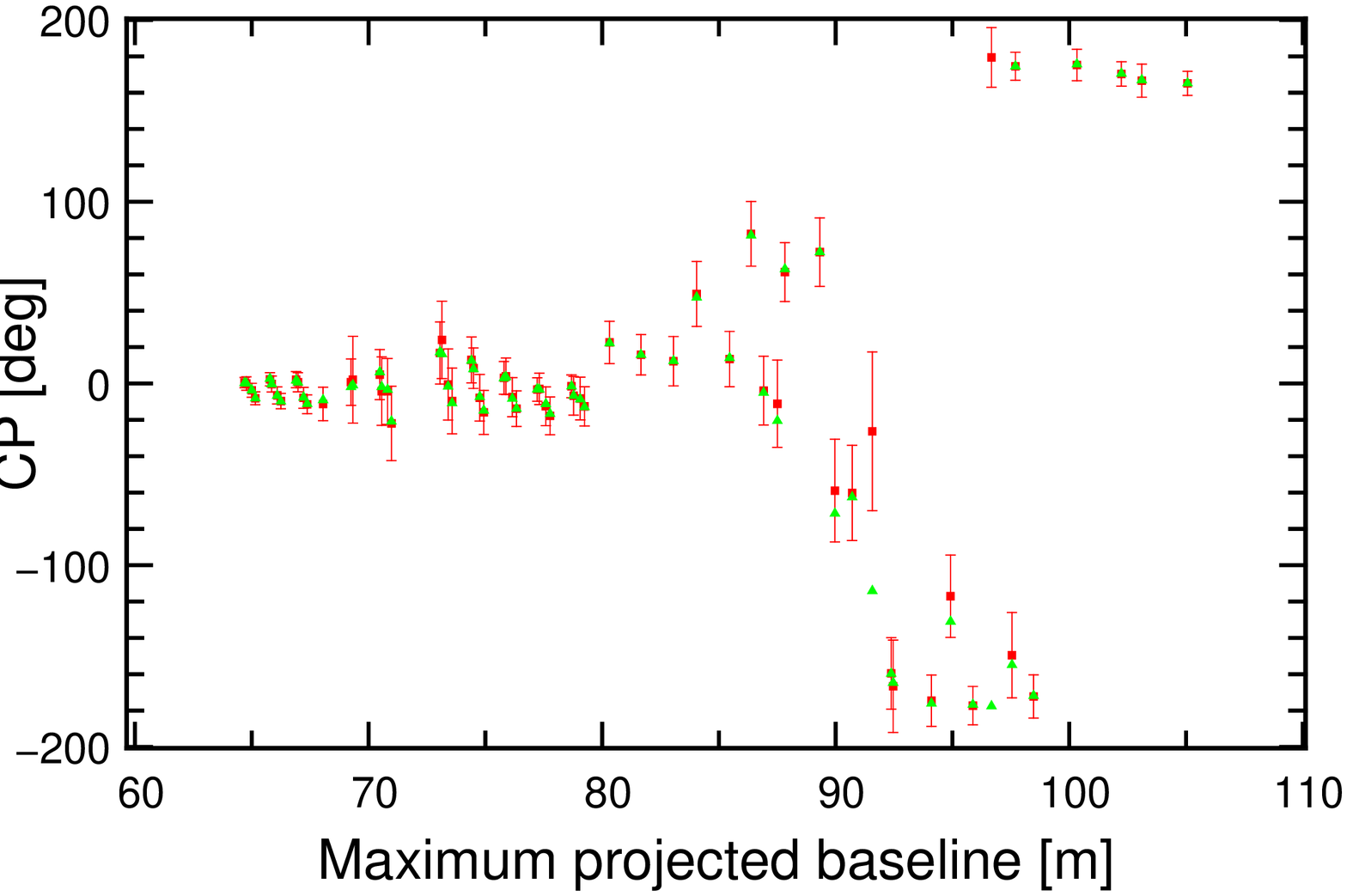} 
  \caption{Same as Fig.~\ref{band1} for band 4, $\rm H_{2}O$ (2).}
       \label{band4}
 \end{figure*}

\begin{figure*}[!h]
 \centering   
   \includegraphics[width=3.3in]{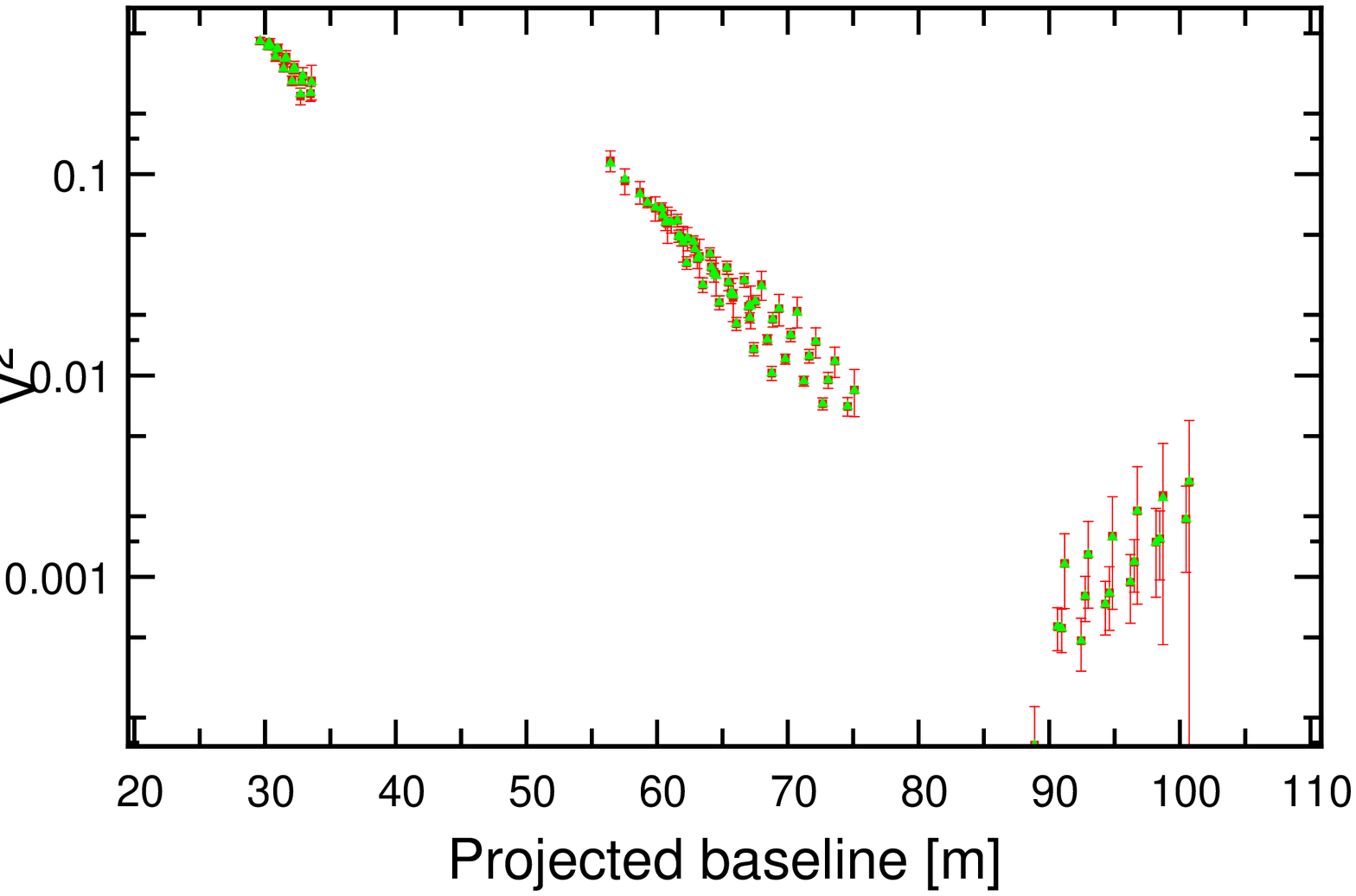}
 \hspace{0.1in}
   \includegraphics[width=3.3in]{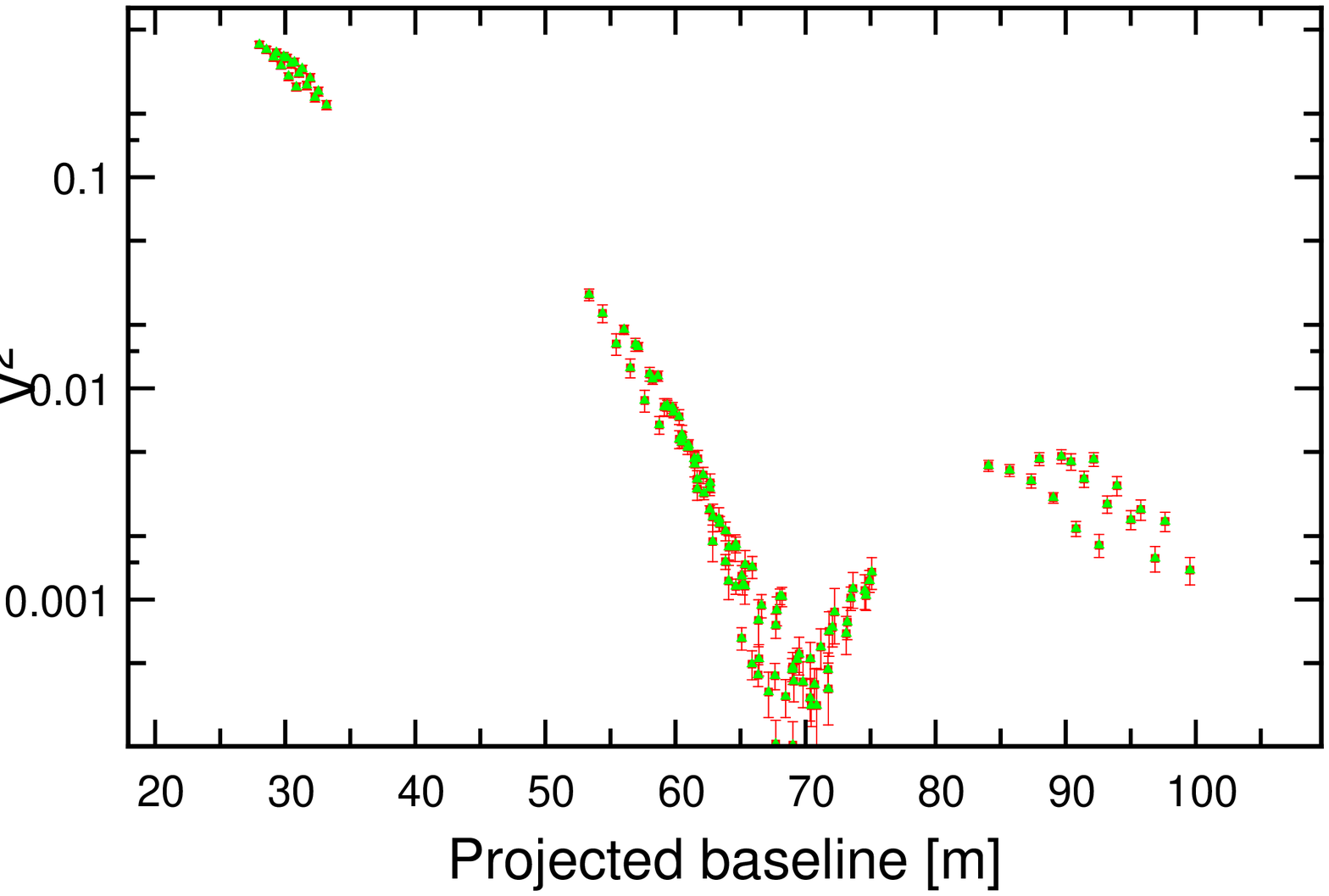}
  \vspace{0.1in}
  \includegraphics[width=3.3in]{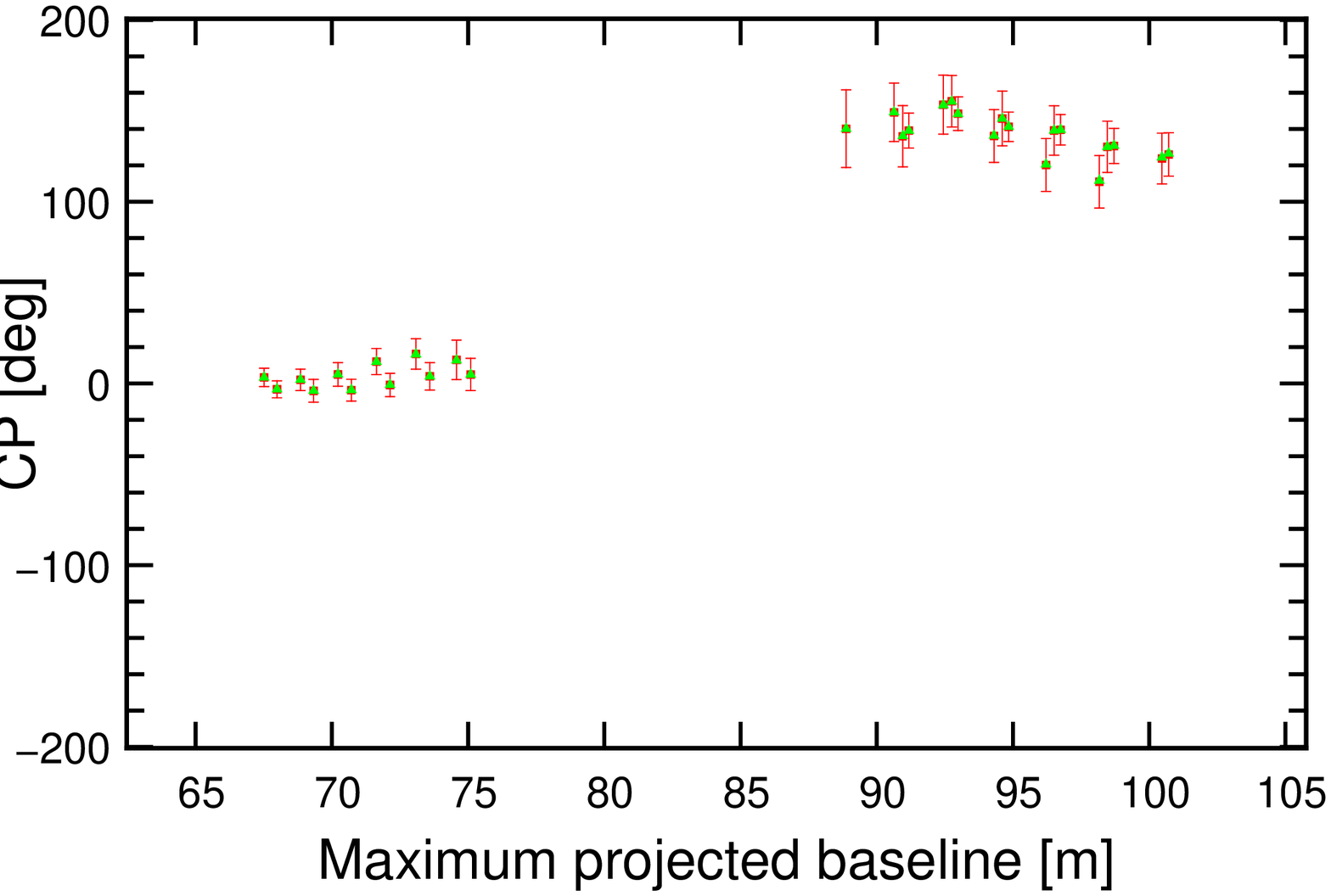}
  \hspace{0.1in}
  \includegraphics[width=3.3in]{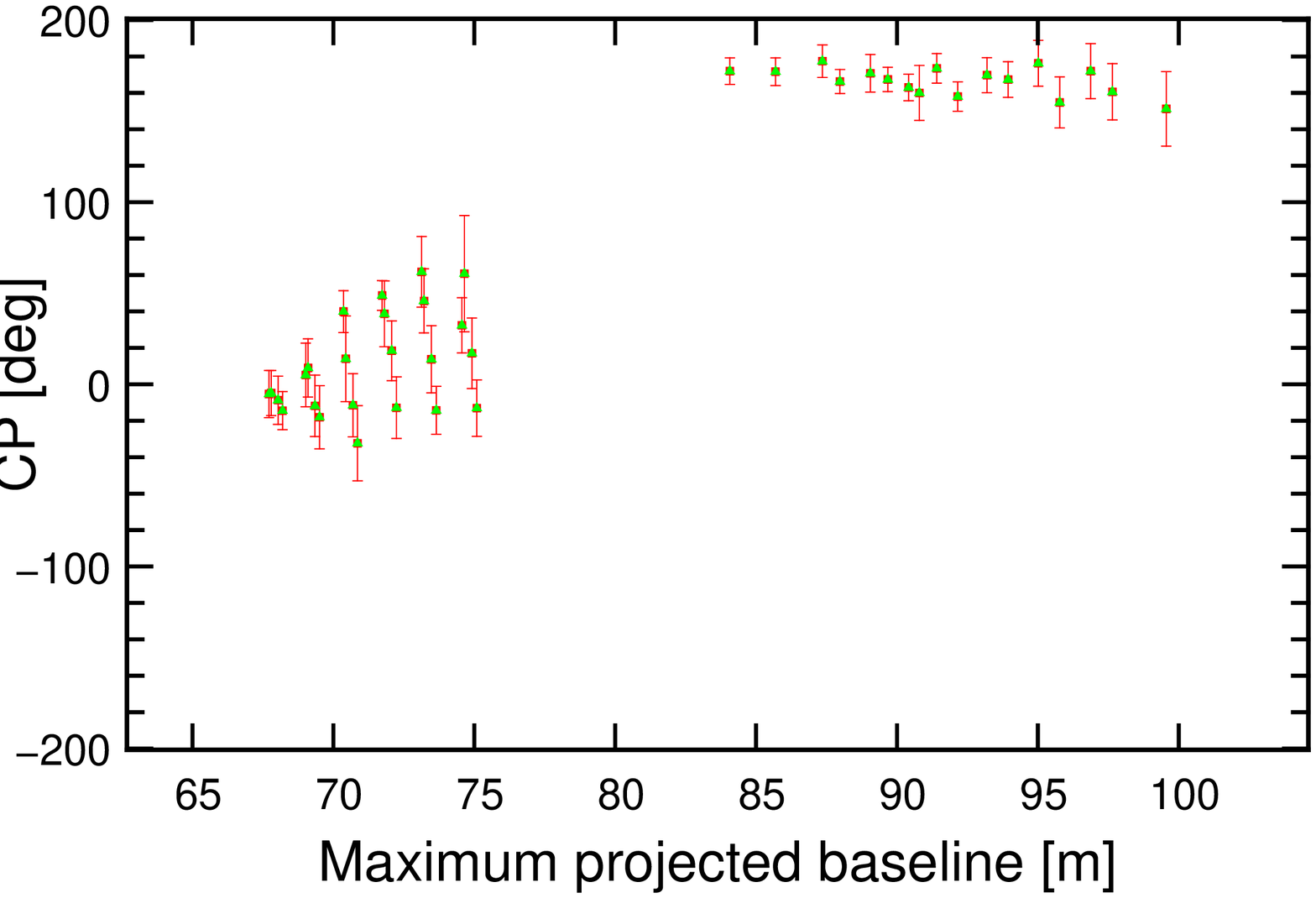}
\caption{Same as Fig.~\ref{band1} for band 3, $CO$ (1).}
       \label{band3}
 \end{figure*}

\begin{figure*}[!h]
 \centering   
 \includegraphics[width=3.3in]{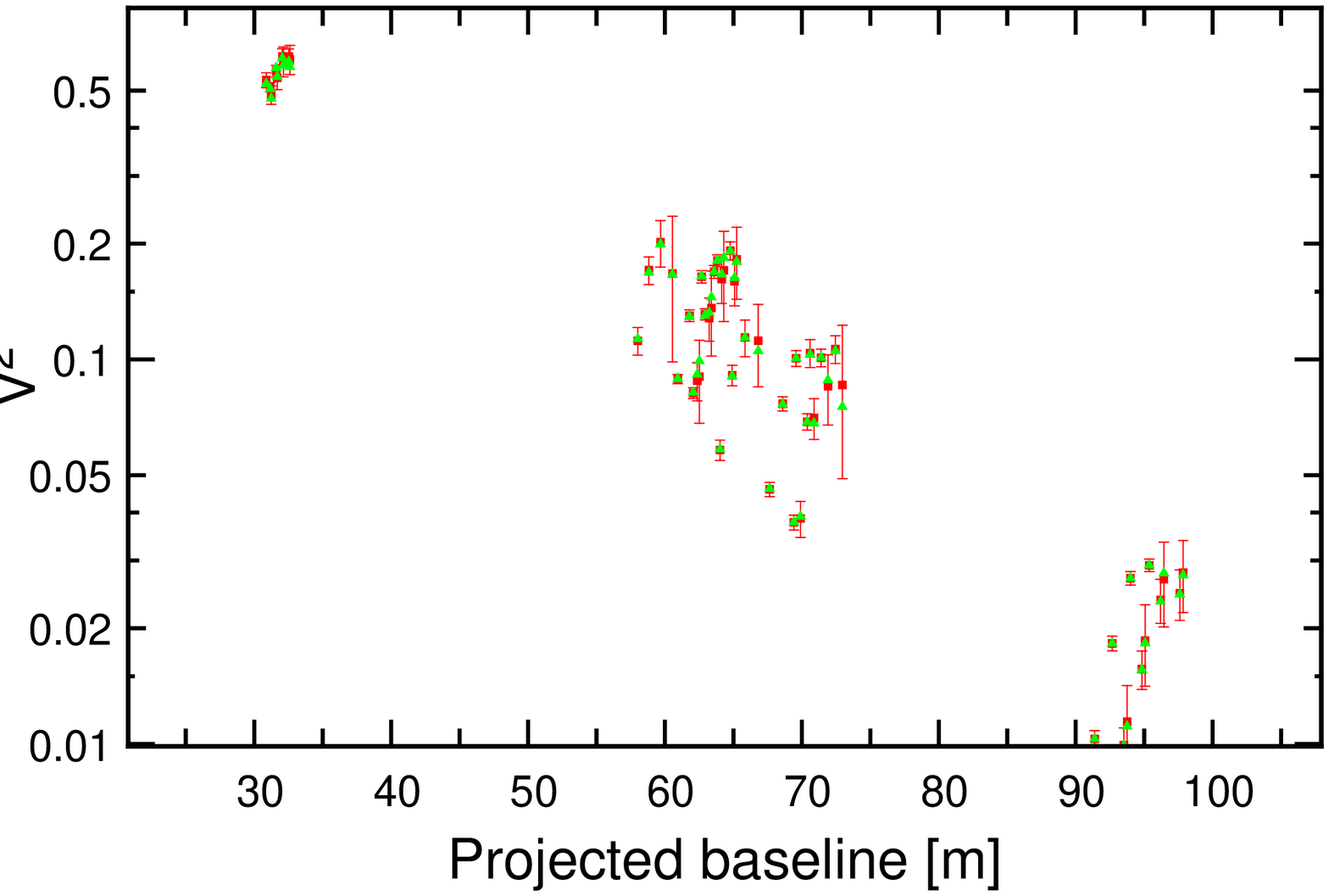}
 \hspace{0.1in}
   \includegraphics[width=3.3in]{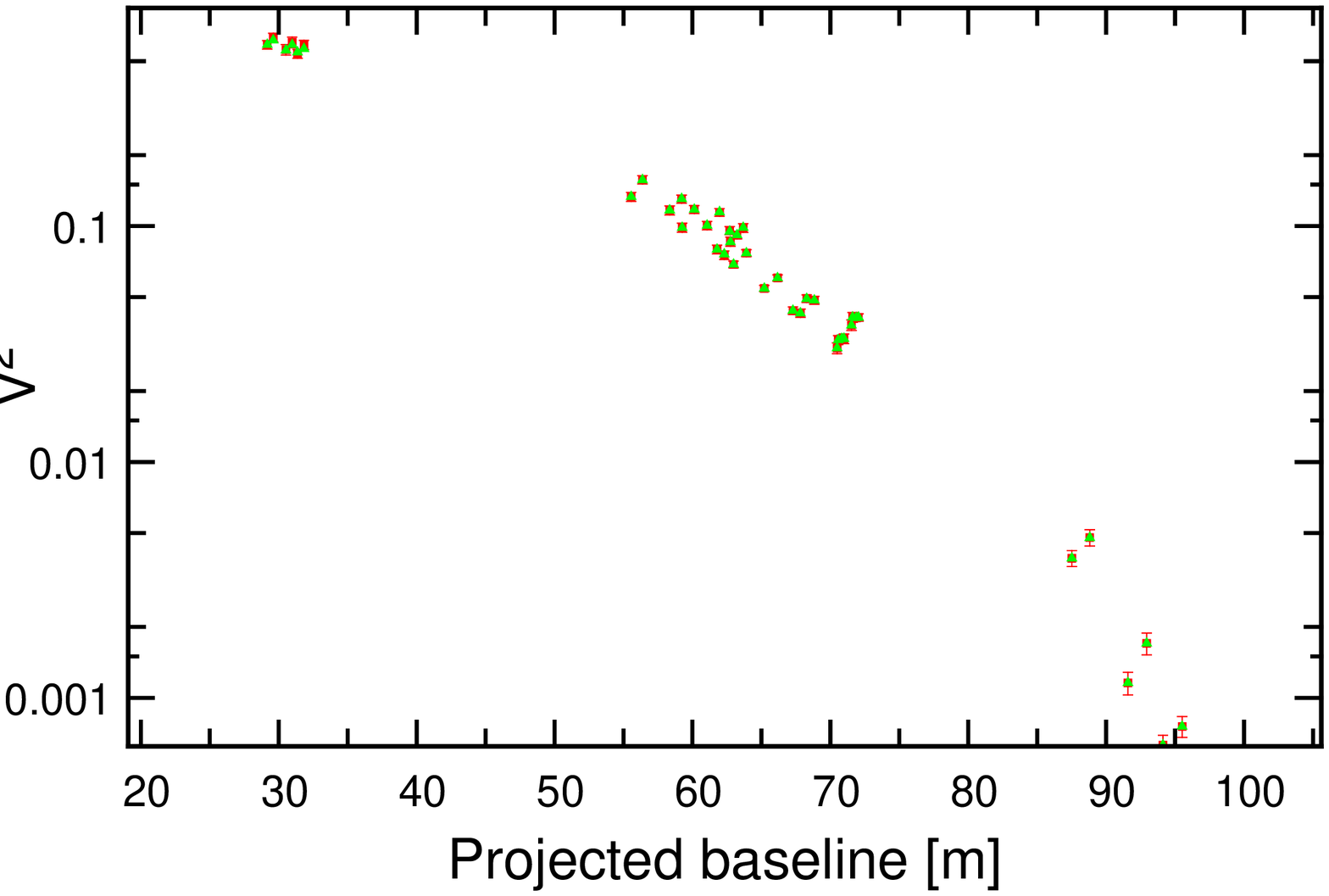}
  \vspace{0.1in}
  \includegraphics[width=3.3in]{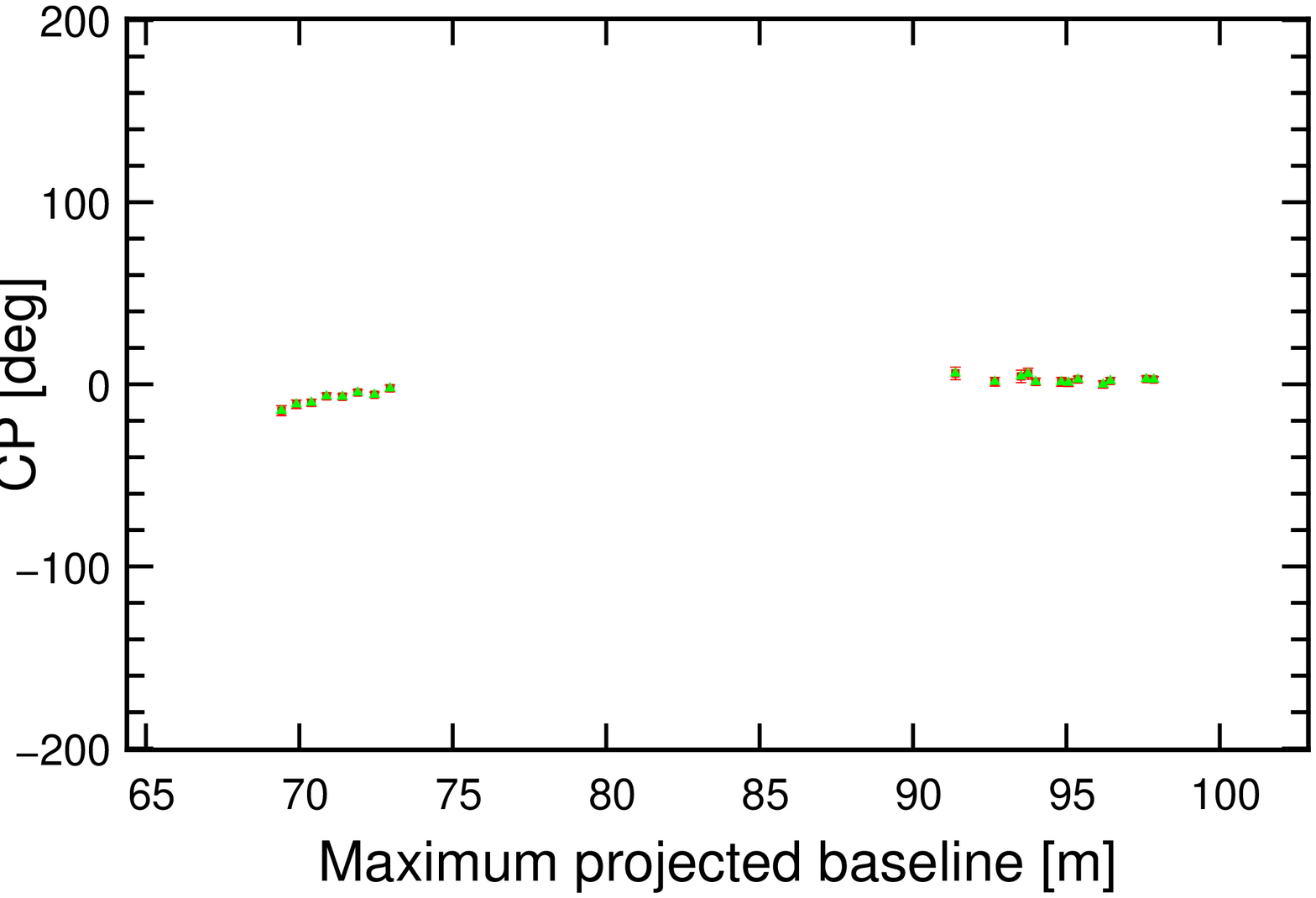}
  \hspace{0.1in}
  \includegraphics[width=3.3in]{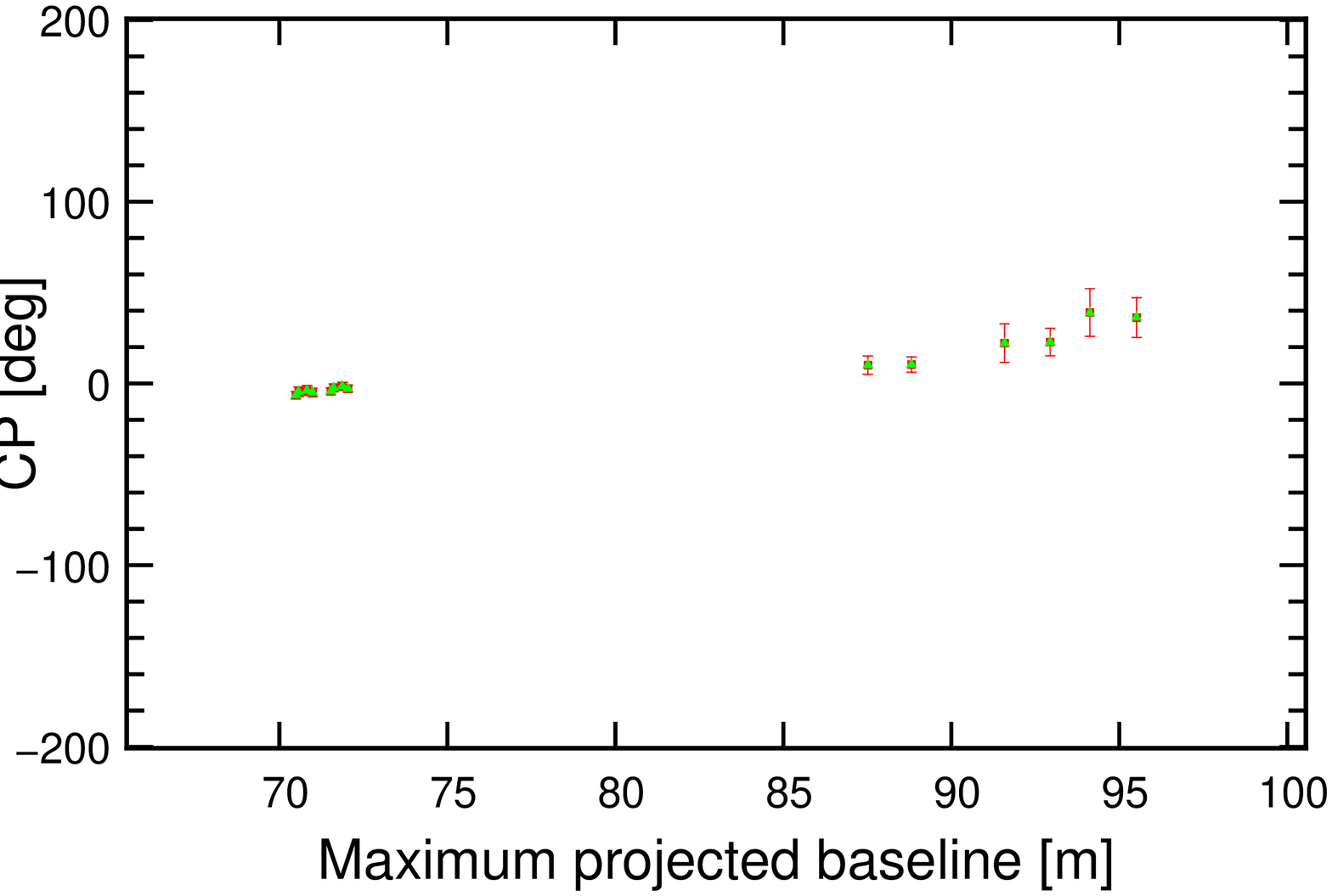}
\caption{Same as Fig.~\ref{band1} for band 6, $CO$ (2).}
       \label{band6}
 \end{figure*}

For the TV regularisation, the \textit{a priori} object image was used as a guess image to guide the data fitting. As shown in  Figs.~\ref{band1},~\ref{band5},~\ref{band2},~\ref{band4},~\ref{band3}, and \ref{band6}, data are convincingly fitted after the image reconstruction process.  Table~\ref{tab:MIRA} summarises the reduced $\chi^{2}$ values as well as hyperparameter values and functions used in MiRA to reconstruct images. 

\begin{table}[h!]
\centering
\begin{tabular}{ccc}
\hline
Spectral band  &  Regularisation / $\mu$ / $\chi^{2}_{red} $   & Regularisation / $\mu$ / $\chi^{2}_{red}$   \\  
number  & $\Phi =0.0$   & $\Phi =0.2$  \\
  &  (epoch A)  & (epoch B)   \\ \hline
1 & Quadratic / 1e5 / 3.6e-2  & Quadratic / 1e6  / 0.3 \\ 
2 & Quadratic / 1e3 / 1.2e-3  &Quadratic / 1e5 /  4.2e-2   \\ 
3 & TV / 1e2 / 0.25  & Quadratic / 1e5 / 1.8e-2    \\ 
4 & Quadratic / 1e4 / 1.2e-2  & TV / 1e3 / 1.4    \\ 
5 &  Quadratic / 5e3 / 1e-3  & Quadratic / 1e3 / 1.1    \\ 
6 &   TV / 1e2 / 0.4  & TV / 1e2 / 0.5     \\ \hline
\end{tabular}
\caption{\label{tab:MIRA} Regularisation functions and hyperparameters used in MiRA to produce reconstructed images presented in Figs.~\ref{images1}, \ref{images2} and \ref{images3}. Reduced $\chi^{2}$ values are specified and represent the value of the minimised criterion per datum at the end of the reconstruction process.}
\end{table}

%

\begin{figure*}[p]
\includegraphics[width=3.5in]{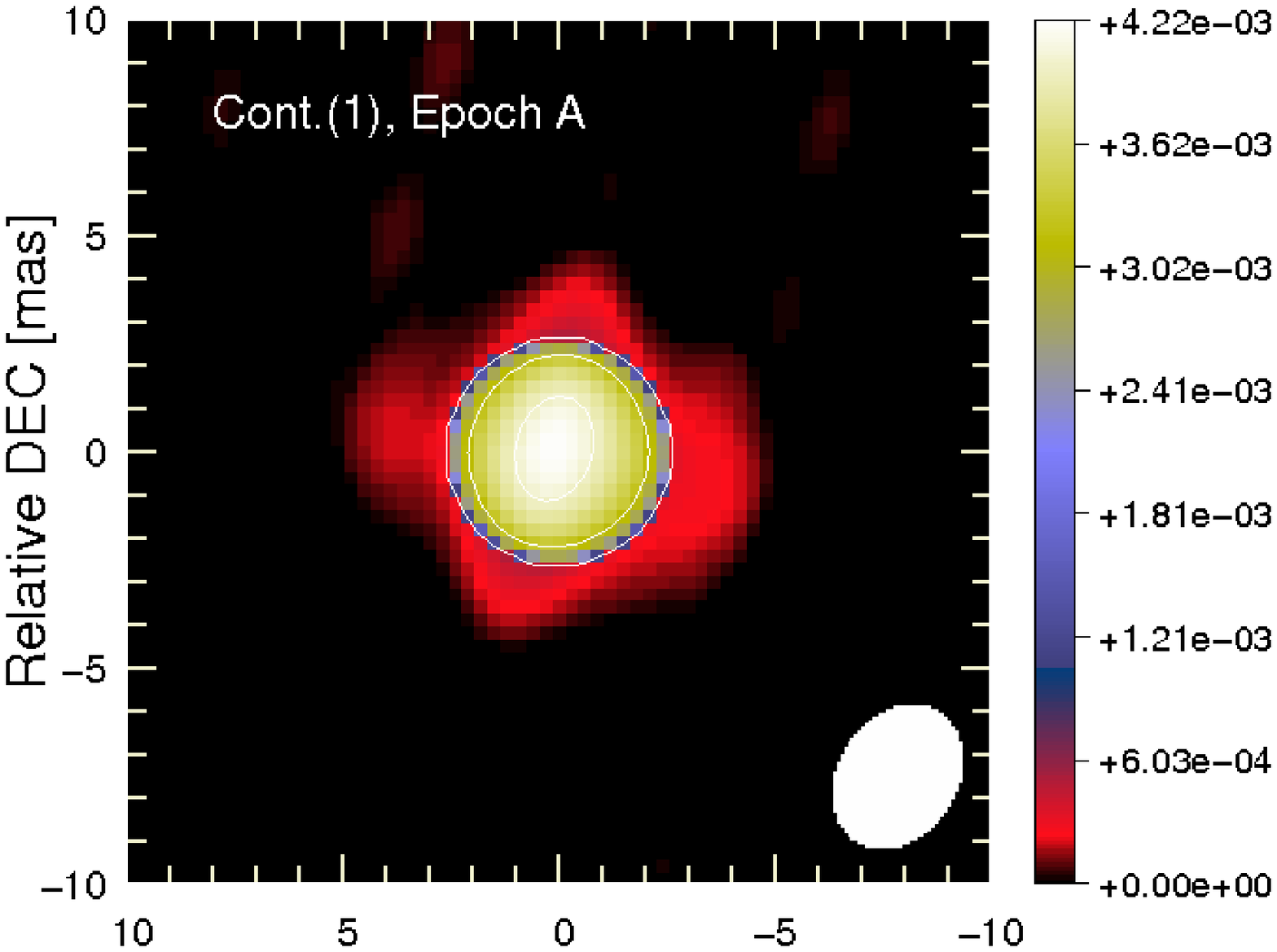}
\includegraphics[width=3.5in]{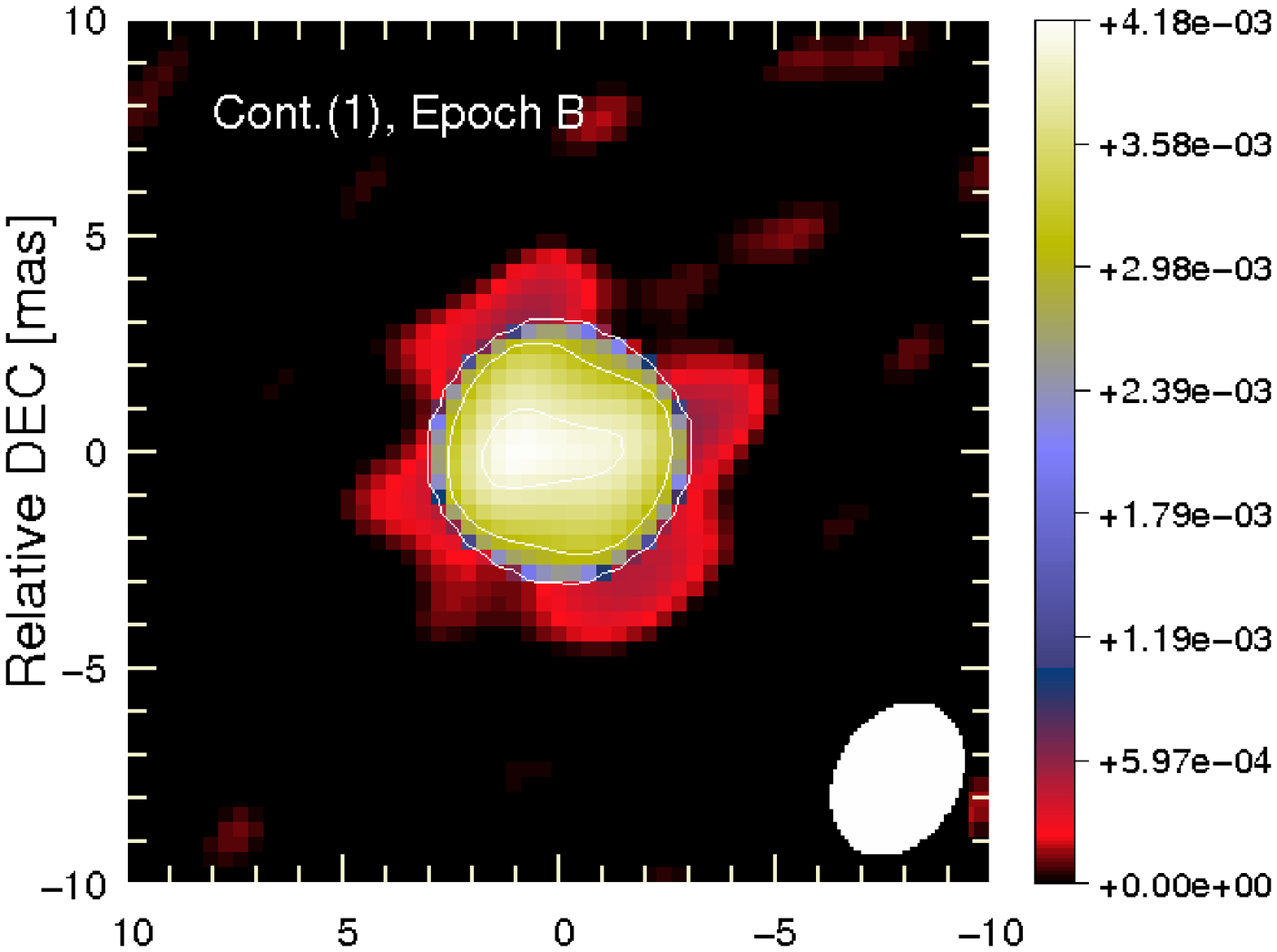}
 \includegraphics[width=3.5in]{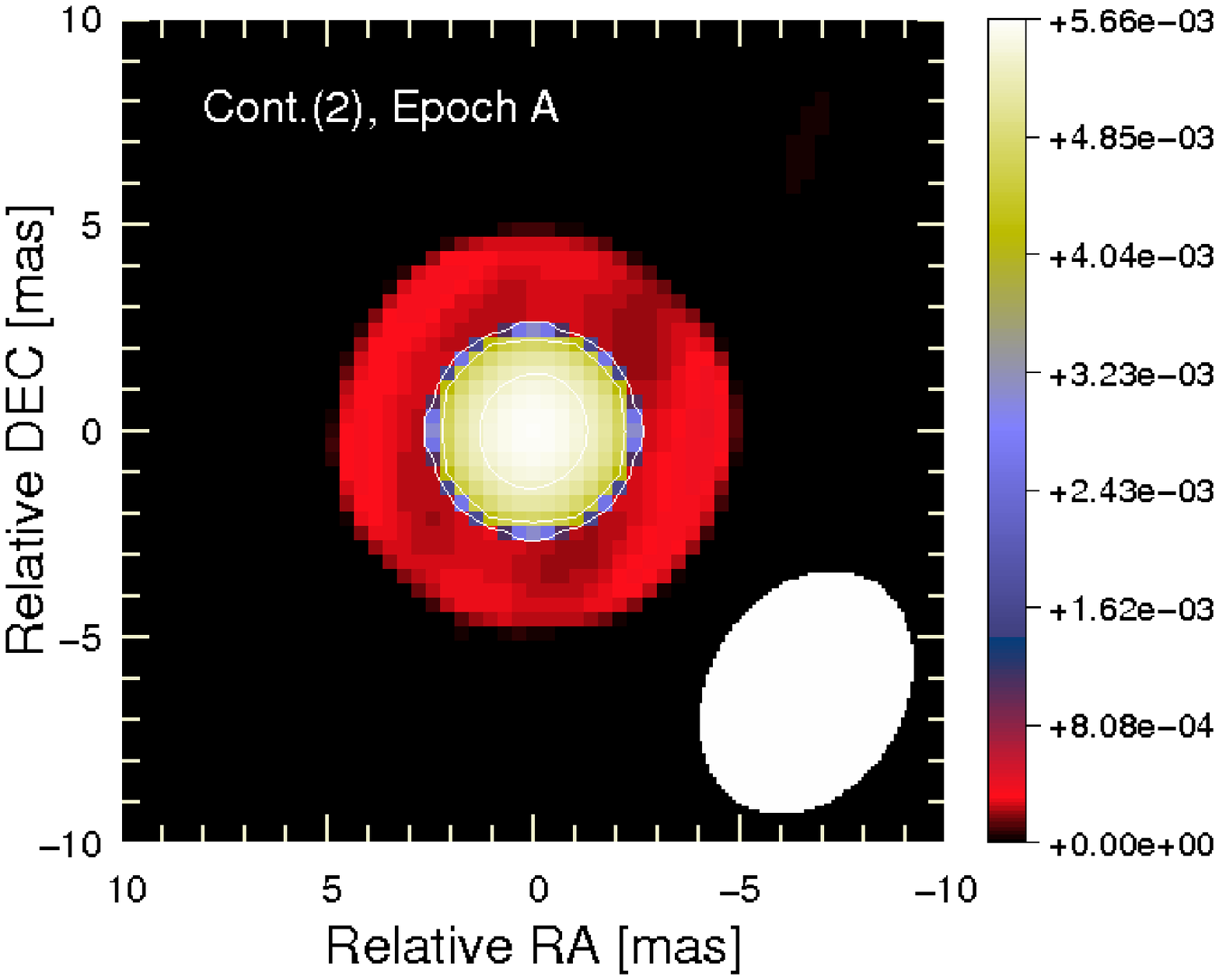}
  \hspace{0.2in} 
\includegraphics[width=3.5in]{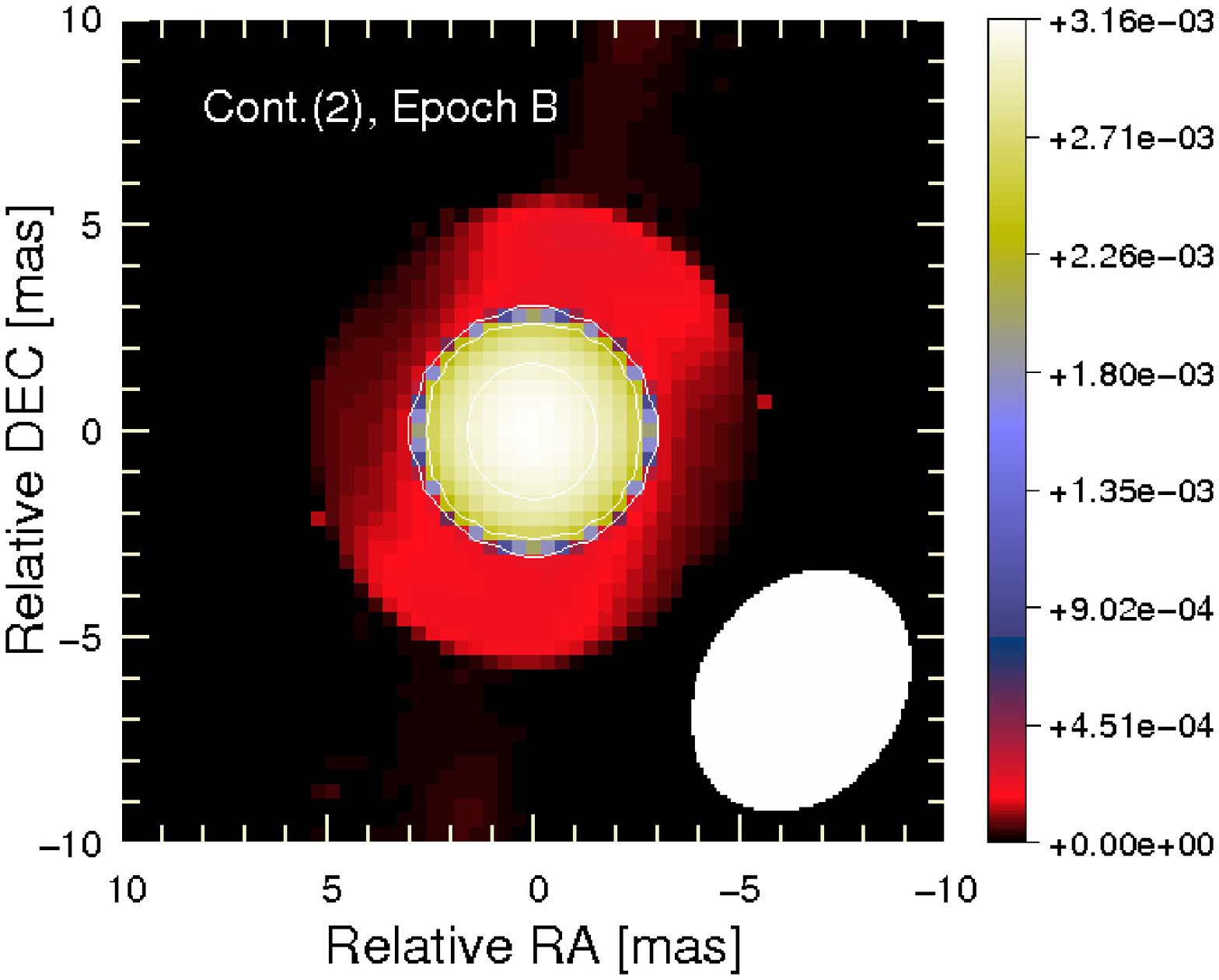}
\caption{Reconstructed images with MiRA for spectral bands 1 and 5 corresponding to the continuum bands at epochs A (left column) and B (right column). The interferometric beam size is displayed in the bottom right corner of each image. White contours represent iso-intensity levels of 25\%, 75\% and 95\% of the maximum intensity of the image.}
     \label{images1}
 \end{figure*}

\begin{figure*}[p]
\includegraphics[width=3.5in]{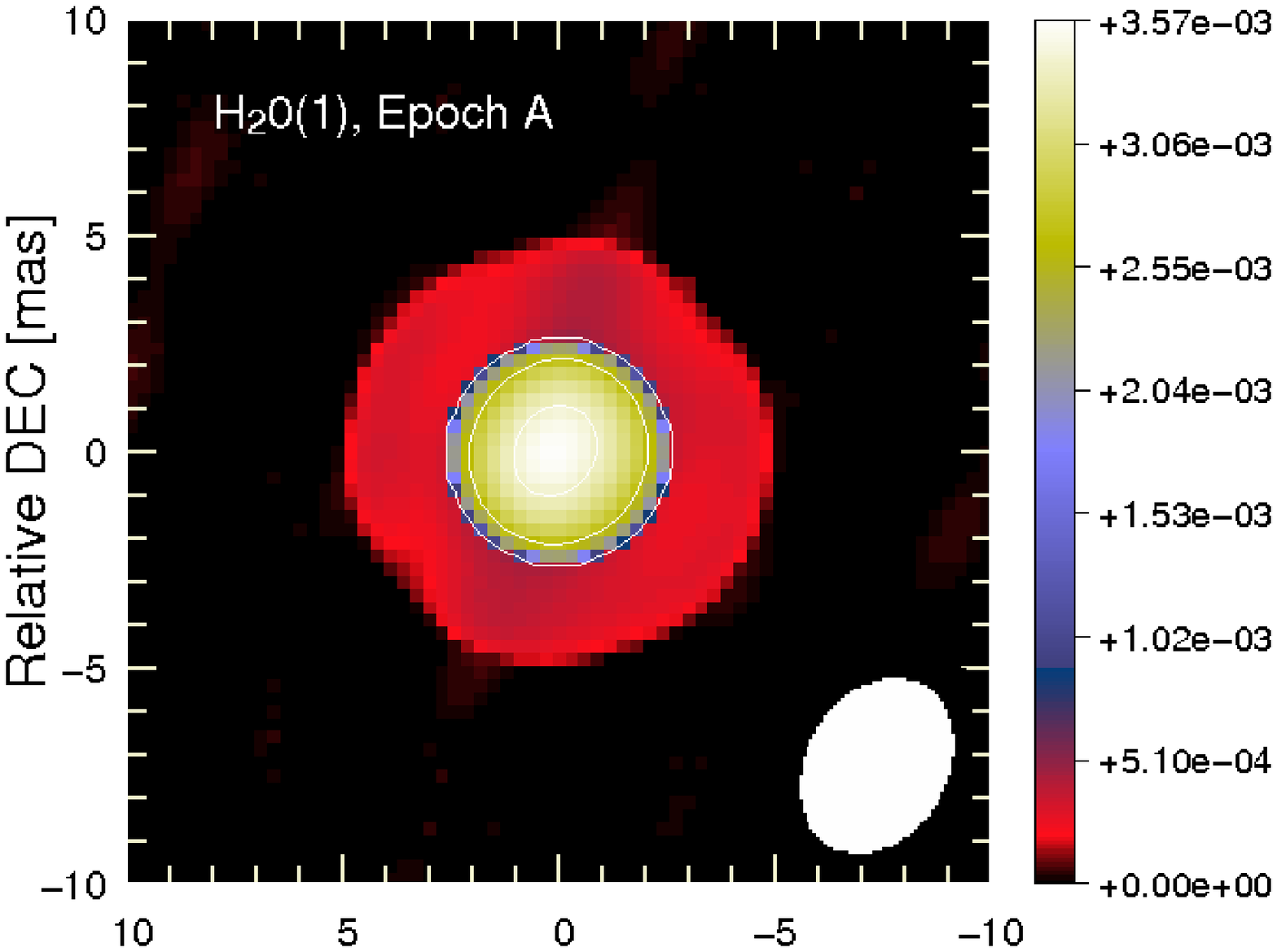}
\includegraphics[width=3.5in]{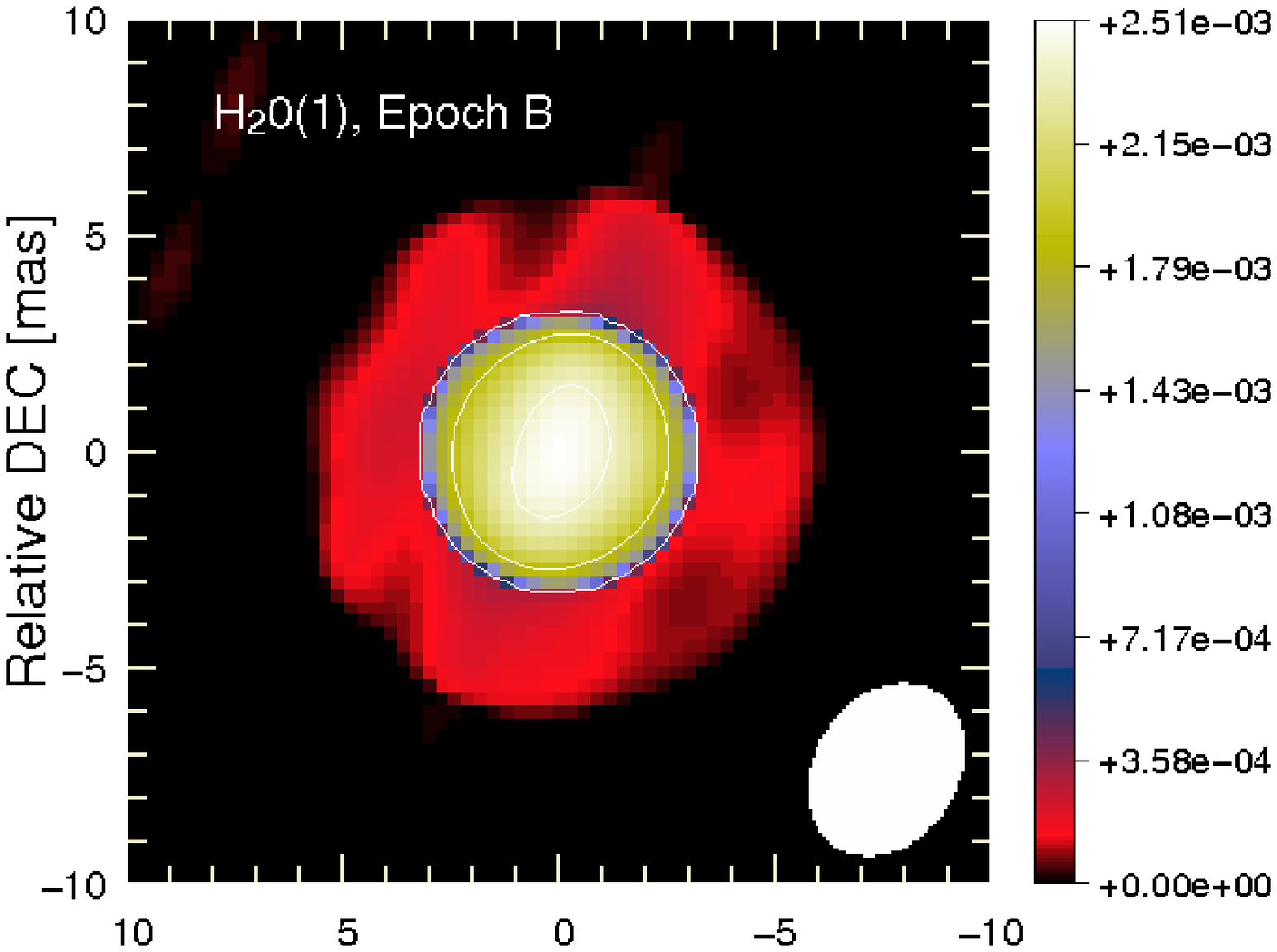}
 \includegraphics[width=3.5in]{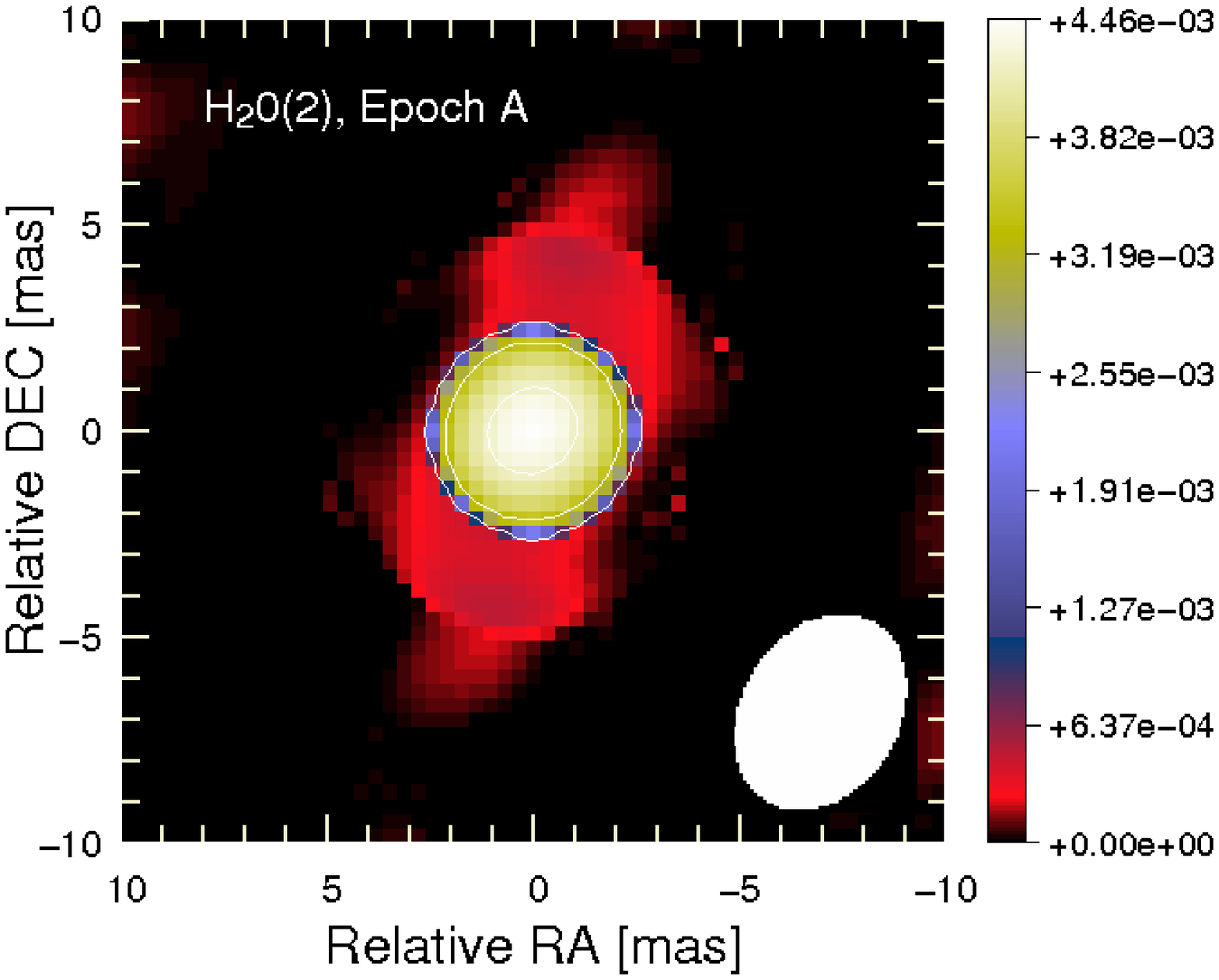}
  \hspace{0.2in} 
\includegraphics[width=3.5in]{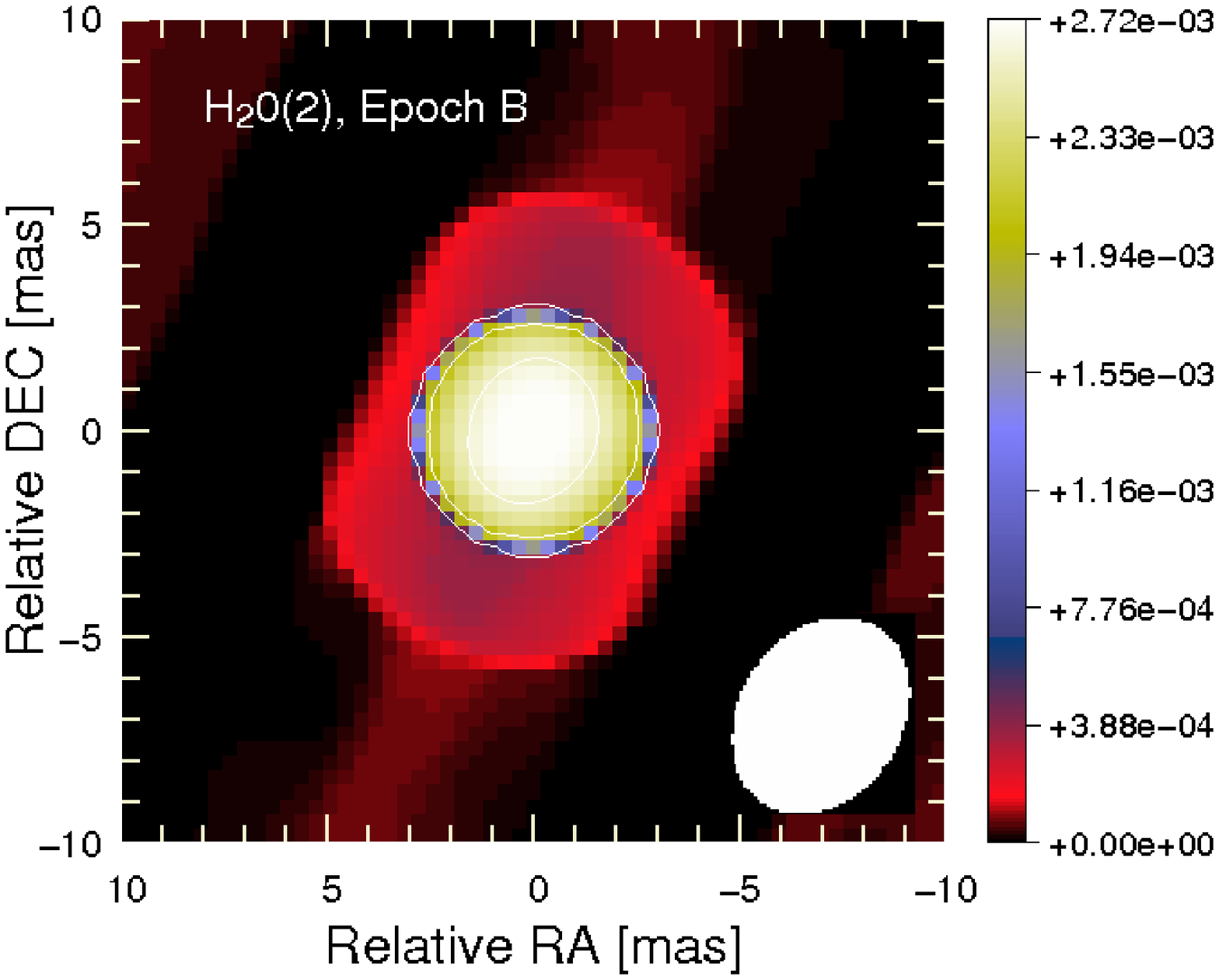}
\caption{Same as Fig.~\ref{images1} for the $\rm H_{2}O$ bands (bands 2 and 4).}
     \label{images2}
 \end{figure*}

\begin{figure*}[p]
\includegraphics[width=3.5in]{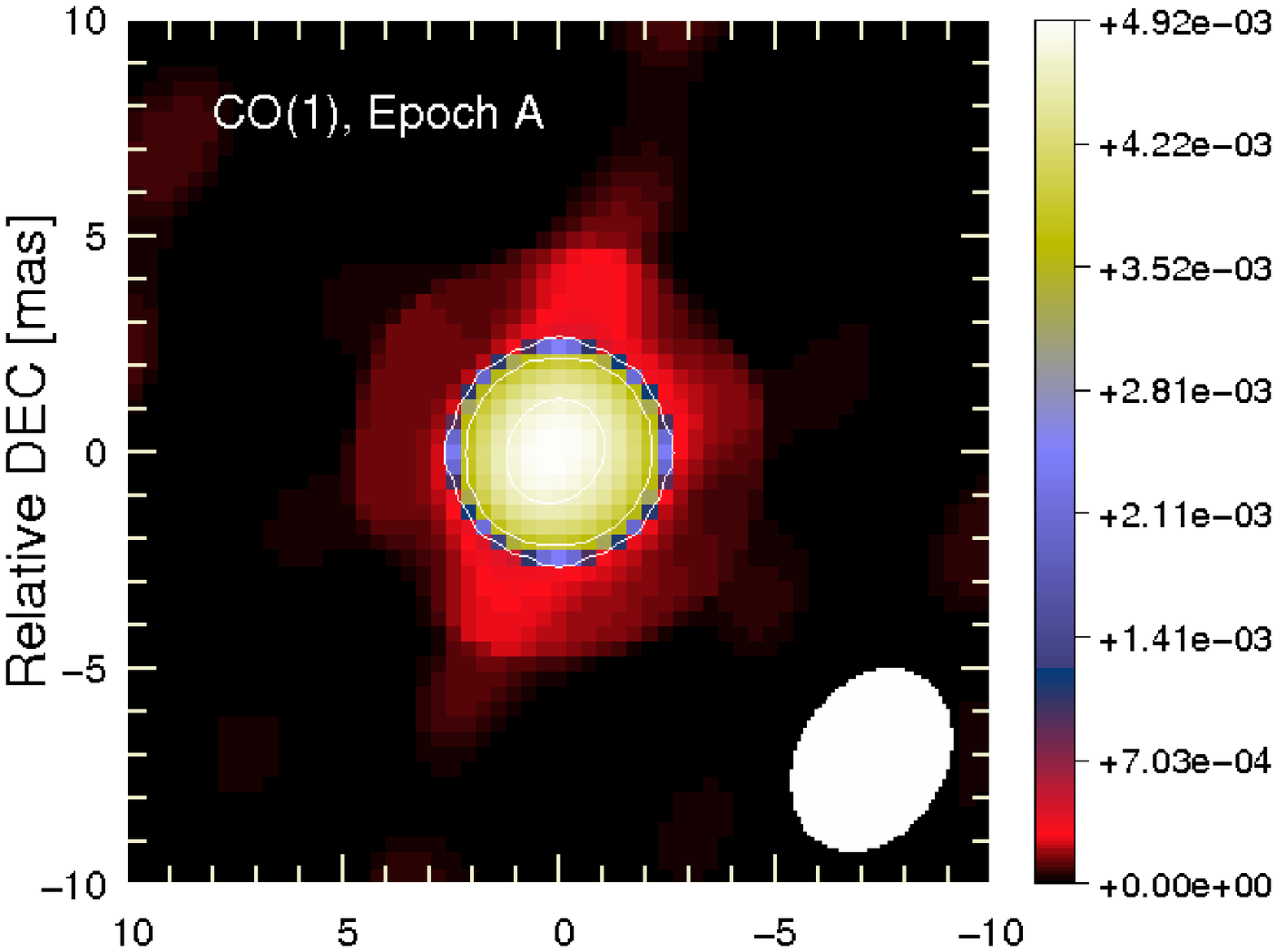}
\includegraphics[width=3.5in]{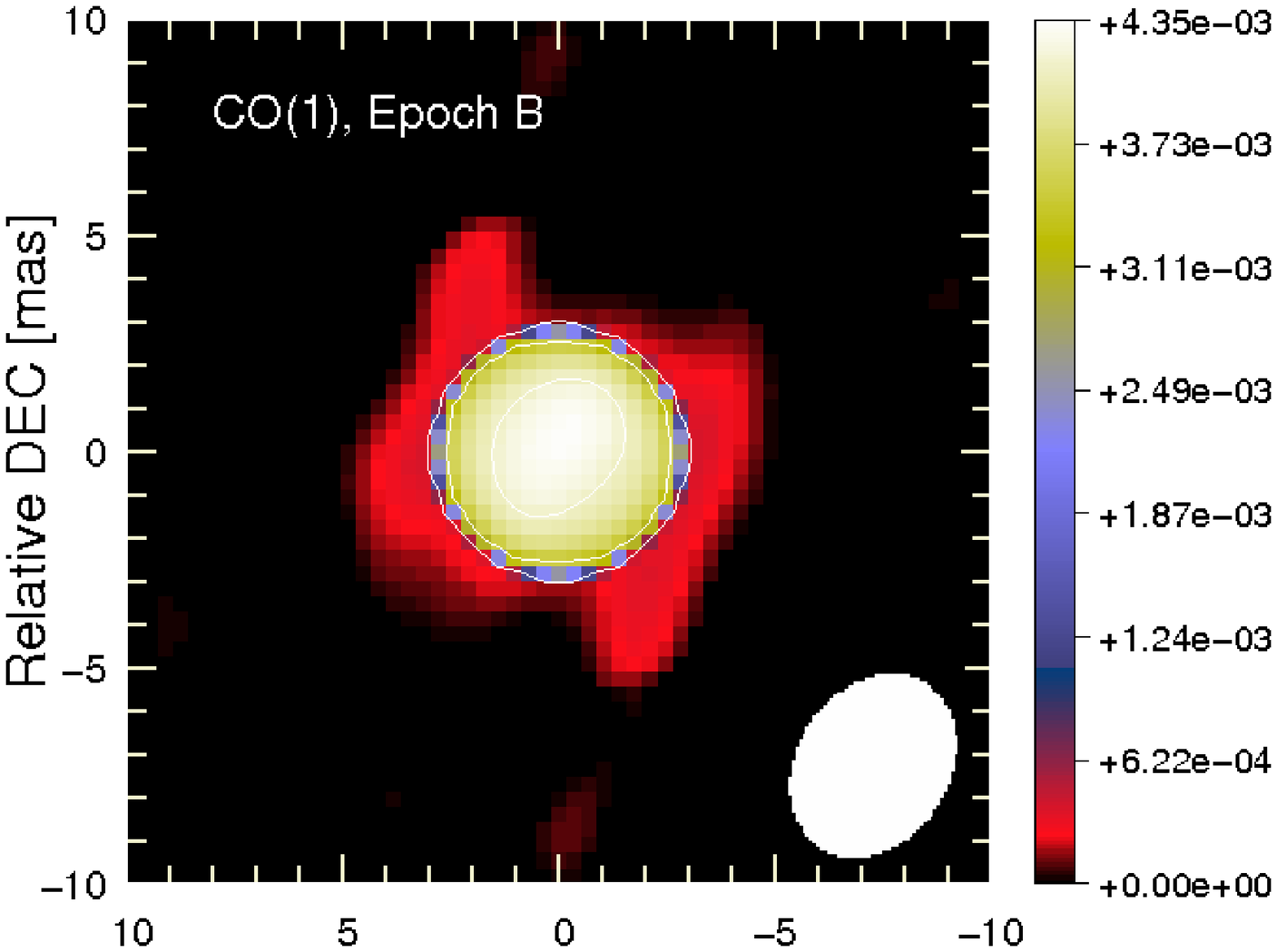}
 \includegraphics[width=3.5in]{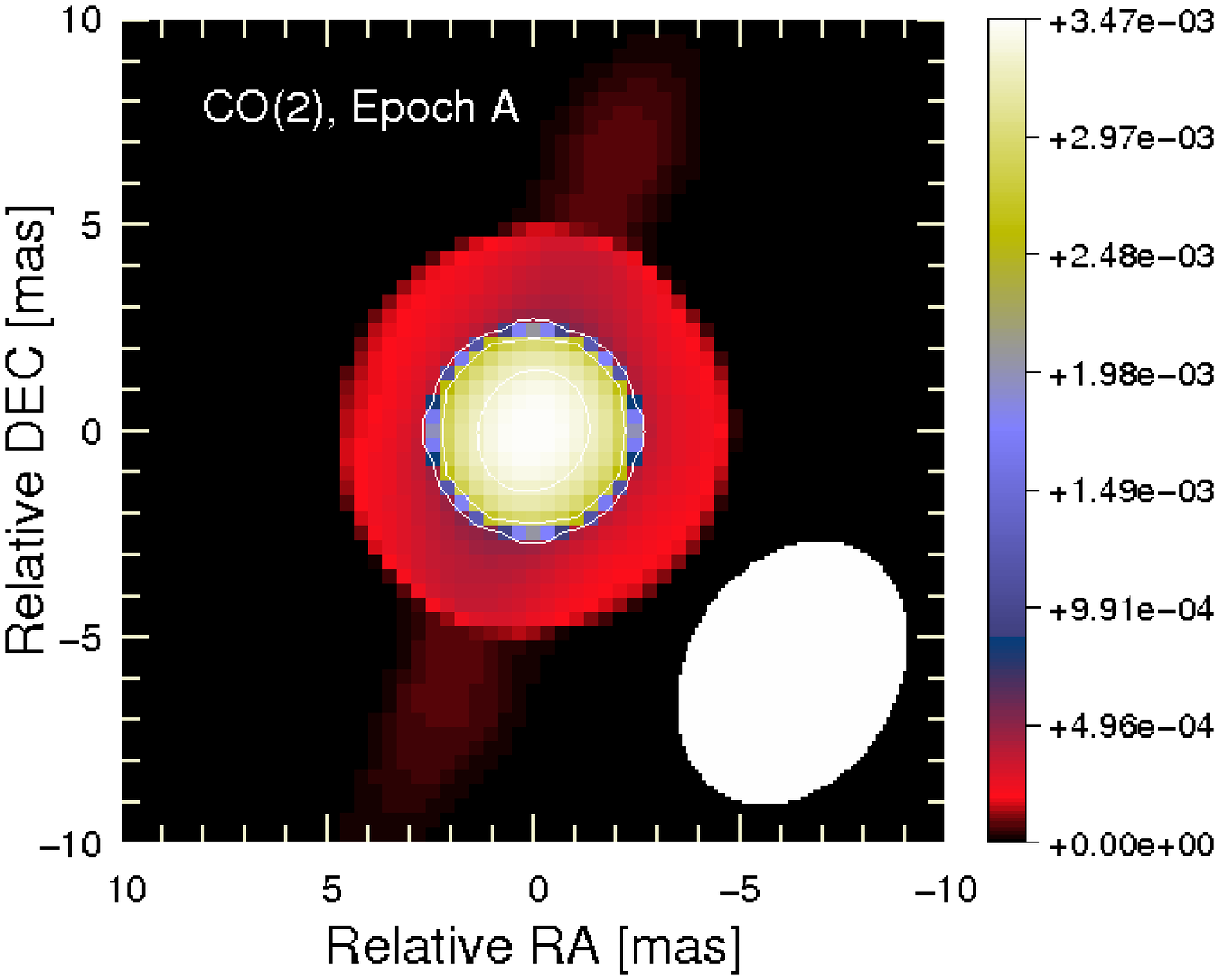}
  \hspace{0.2in} 
\includegraphics[width=3.5in]{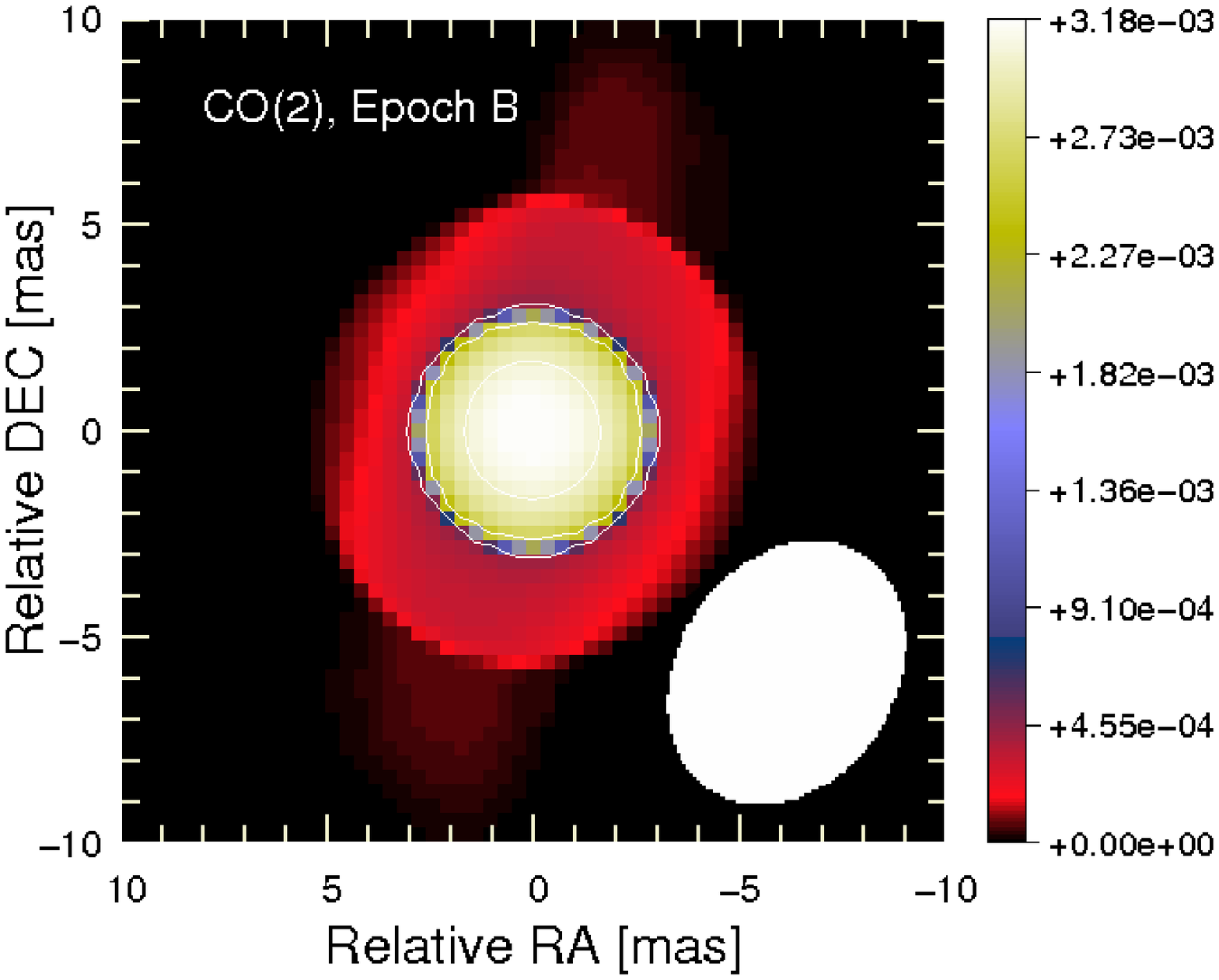}
\caption{Same as Fig.~\ref{images1} for the CO bands (bands 3 and 6).}
     \label{images3}
 \end{figure*}

Reconstructed images for the six spectral bands at the two epochs are presented in Figs.~\ref{images1}, \ref{images2} and \ref{images3}.

The fact that the uv coverages at the two observing periods are almost identical suggests that the morphology of X Hya really changed from epoch A to epoch B.
 At a $25\%$-intensity level of the reconstructed images shown in Fig.~\ref{images1} to ~\ref{images3},  the stellar morphology is disk-like in all spectral bands. If we use this intensity threshold to estimate the stellar diameter, the relative increase of this value between epochs A and B ranges from 13\% to 18\% depending on the spectral band (see Table~\ref{diam_images}). Considering the overlap of 1 $\sigma$-error bars, these results agree with the results of the UD modelling shown in Table~\ref{tab:DU}, except for band 1 for which the 25\%-intensity diameter variation is found to be weaker than the UD diameter variation.  A difference between the two diameter estimator ($\phi_{25\%}$ and  $\phi_{UD}$) values is expected because they take into account different intensity levels for the same given image. However, the relative variation of the two estimators should give close values if most of the spatial intensity distribution retains the same profile between the two epochs. For band 1, it seems to indicate that this is not the case. This could either be a real astrophysical effect or a consequence of noisy data (particularly for epoch B, right part of Fig.~\ref{band1}) in the reconstructed images (Fig.~\ref{images1}). The latter case could be explained by a combination of low visibilities and the effect of atmospheric turbulence, stronger at these wavelengths ($J$ band), on the accuracy of both $V^2$ and closure phase data.

\begin{table}[h!]
\centering
\begin{tabular}{cccc}
\hline
Spectral  & $ \phi_{25\%} $ (mas)  & $ \phi_{25\%} $ (mas) & $\Delta(\phi_{25\%}) / \phi_{25\%}$ \\  
band  & $\Phi =0.0$ & $\Phi =0.2$  &   \\
number  & (epoch A) &  (epoch B)  & \\ \hline
1 & 5.01 $\pm$ 0.15 \,\rm  & 5.67 $\pm$ 0.19 \,\rm  &  0.13 $\pm$ 0.04 \\ 
2 & 4.95 $\pm$ 0.16 \,\rm & 5.64 $\pm$ 0.17 \,\rm   & 0.14 $\pm$ 0.03   \\ 
3 & 4.76 $\pm$ 0.21 \,\rm & 5.57 $\pm$ 0.23 \,\rm  &  0.17 $\pm$ 0.05  \\  
4 & 4.73 $\pm$ 0.20 \,\rm & 5.57 $\pm$ 0.23 \,\rm  &  0.18 $\pm$ 0.05  \\  
5 & 4.84 $\pm$ 0.21 \,\rm & 5.57 $\pm$ 0.23 \,\rm  & 0.15 $\pm$ 0.05  \\  
6 & 4.90  $\pm$ 0.18 \,\rm &  5.57 $\pm$ 0.23 \,\rm  &  0.14 $\pm$ 0.04  \\ \hline
\end{tabular}
\caption{\label{diam_images} Stellar diameters estimated from the 25\% iso-intensity levels of Fig.~\ref{images1}, \ref{images2} and \ref{images3}. Diameter values were averaged over the PAs. Results are expressed as $\mu \pm 1 \sigma$, where $\mu$ is the average and $\sigma$ is the standard deviation computed over all the PAs. The fourth column represents the relative variation of stellar diameter between epochs A and B, relatively to epoch A.}
\end{table}

For the two epochs, the shape of the star varies quite significantly between pairs of conjugated spectral bands (1-5, 2-4, and 3-6). The first reason is the fact that each spectral band corresponds to a different number of spectral channels and consequently to a different number of $V^{2}$ and closure phase points (asymmetry information) for the image reconstruction. Secondly, the size of the interferometric beam, which is larger at longer wavelengths, makes it more difficult to resolve small features (including asymmetries) at longer wavelengths. Finally, as can already be seen from Fig.~\ref{graph_CODEX}, the size of the object as predicted by the radiative transfer calculations, does vary for all the bands. This explains why we can see different morphologies between bands 1 and 5 (continuum bands), 2 and 4 ($\rm H_{2}O$) and 3 and 6 (CO) for each epoch. 

We can distinguish two main components in the reconstructed images (Figs.~\ref{images1} to \ref{images3}):
\begin{itemize}
\item{the stellar disk, whose diameter variation is in agreement with previous estimations of geometrical models, and whose intensity is above 25\% of the maximum intensity level. Above this level, all images are centro-symmetric  apart from the Continuum (1) at Epoch B see the discussion above}.
\item{the close environment or atmosphere, whose intensity is below 25\% of the maximum intensity, that shows heterogeneous morphologies and whose characteristic angular size can reach up to 10 mas.}
\end{itemize}

We add a word of caution at this point. The 2014 Interferometric Imaging Beauty Contest \citep{2014SPIE.9146E..1QM}, which aimed at comparing image reconstruction methods, showed that reconstructed images based on the same dataset diverge for intensities lower than 25\% of the maximum relative intensity (see Fig.~5 of the quoted paper). Below this intensity level, it is therefore difficult to distinguish intrinsic astrophysical features from image reconstruction artefacts when one uses a single algorithm. Nevertheless in the following, we speculate on the possible nature of features and asymmetries if they prove to be real astrophysical properties of the star's morphology.

This hypothesis is supported by the fact that for most of the reconstructed images presented here, the two regularisation types and various hyperparameter values revealed the same features for intensities above 10\% of the maximum intensity ($\mu$ values typically ranged from 1e3 to 1e6 for the quadratic regularisation and from 1e1 to 1e3 for the TV regularisation). Exceptions are the CO(2) images at epoch A and the Cont(1) and $\rm H_{2}O(2)$ images at epoch B, which exhibit different environment shapes for the two regularisation methods even though they all produce good data fitting, such as the one presented in Table~\ref{tab:MIRA}. Nevertheless, for all spectral bands, the maximum size of the environment and the shape of the stellar disk remain stable with different regularisation methods.

\begin{figure*}
\includegraphics[width=3.5in]{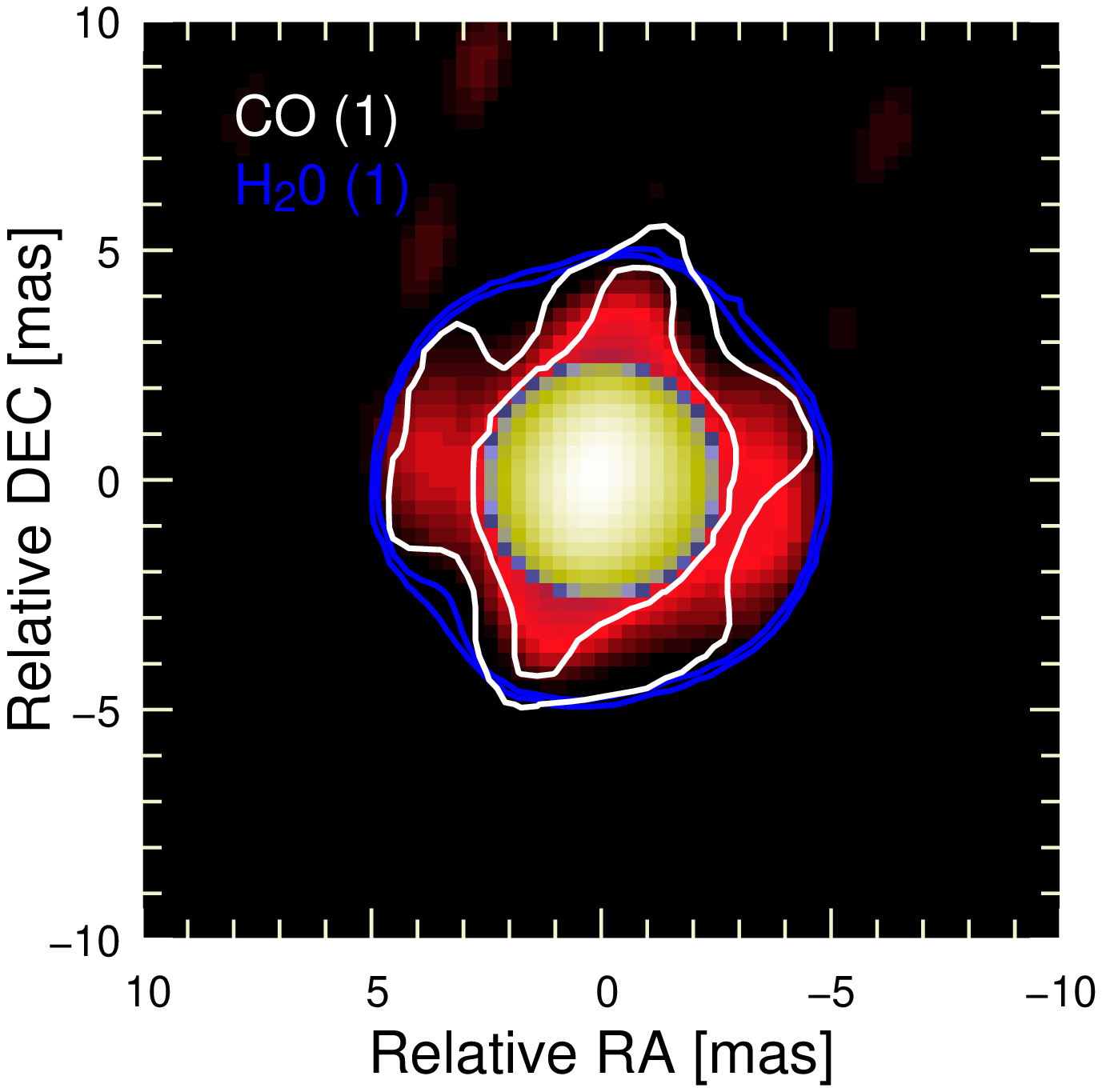}
 \includegraphics[width=3.5in]{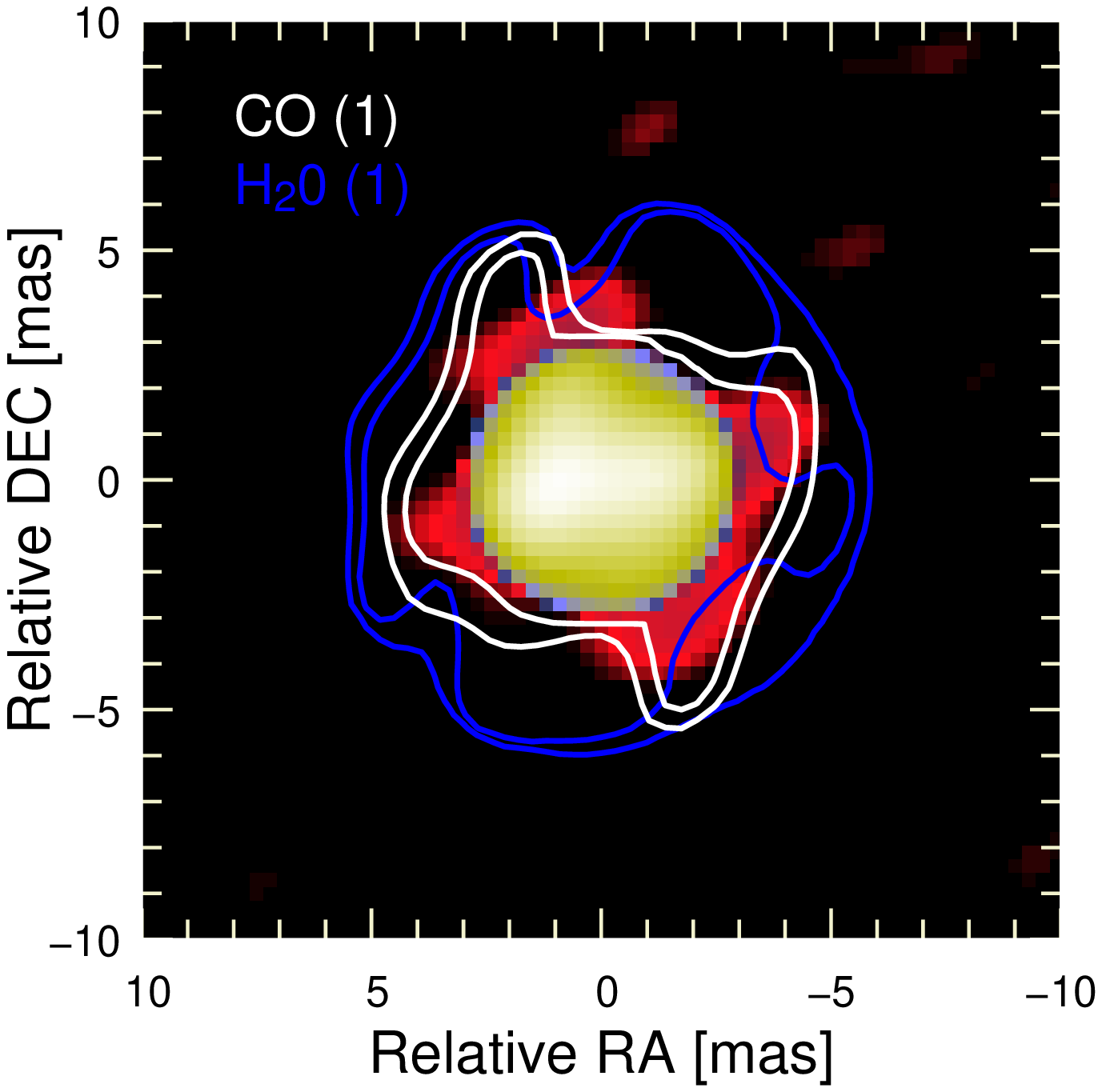}
 \includegraphics[width=3.5in]{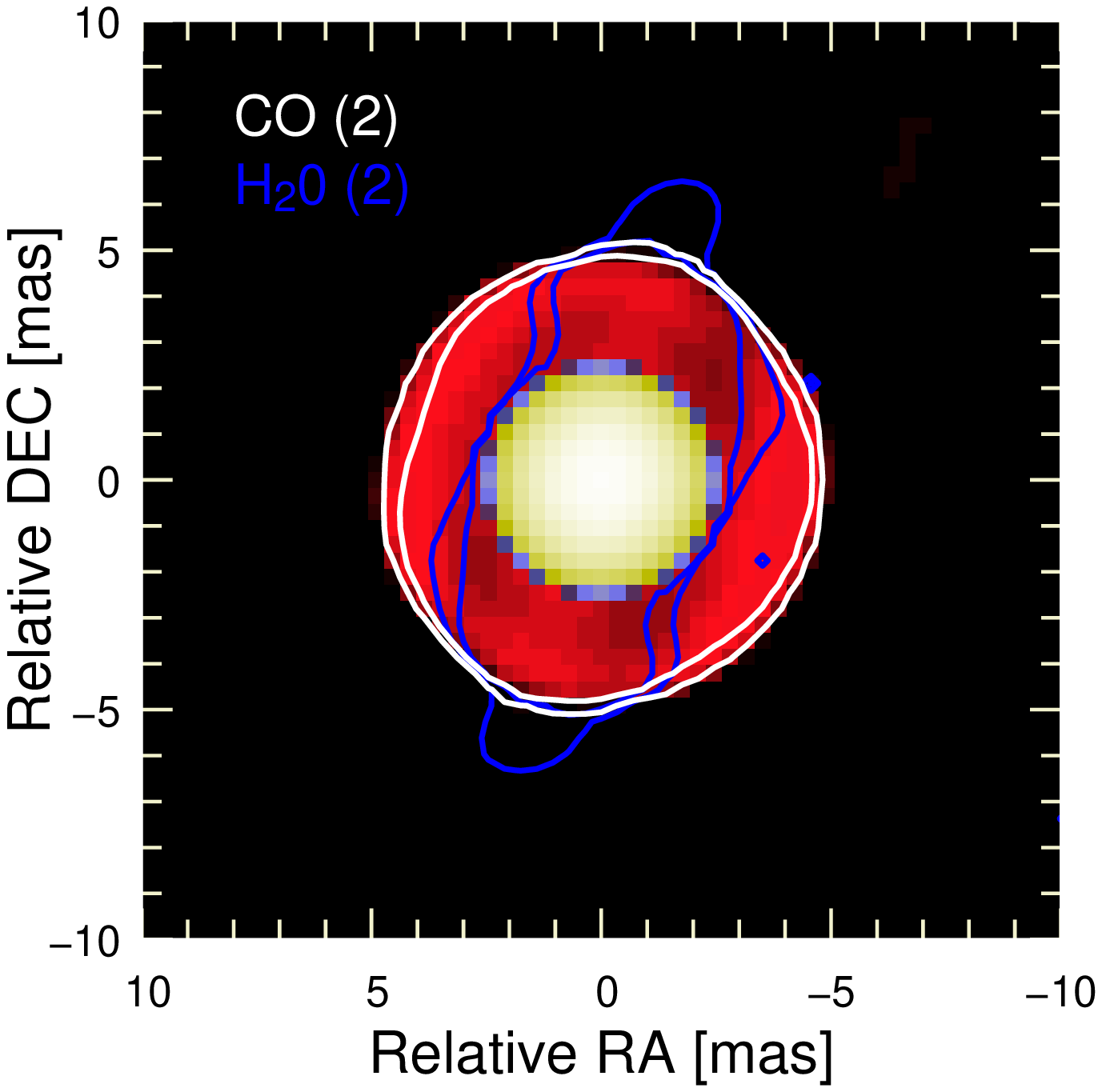}
  \hspace{0.2in} 
\includegraphics[width=3.5in]{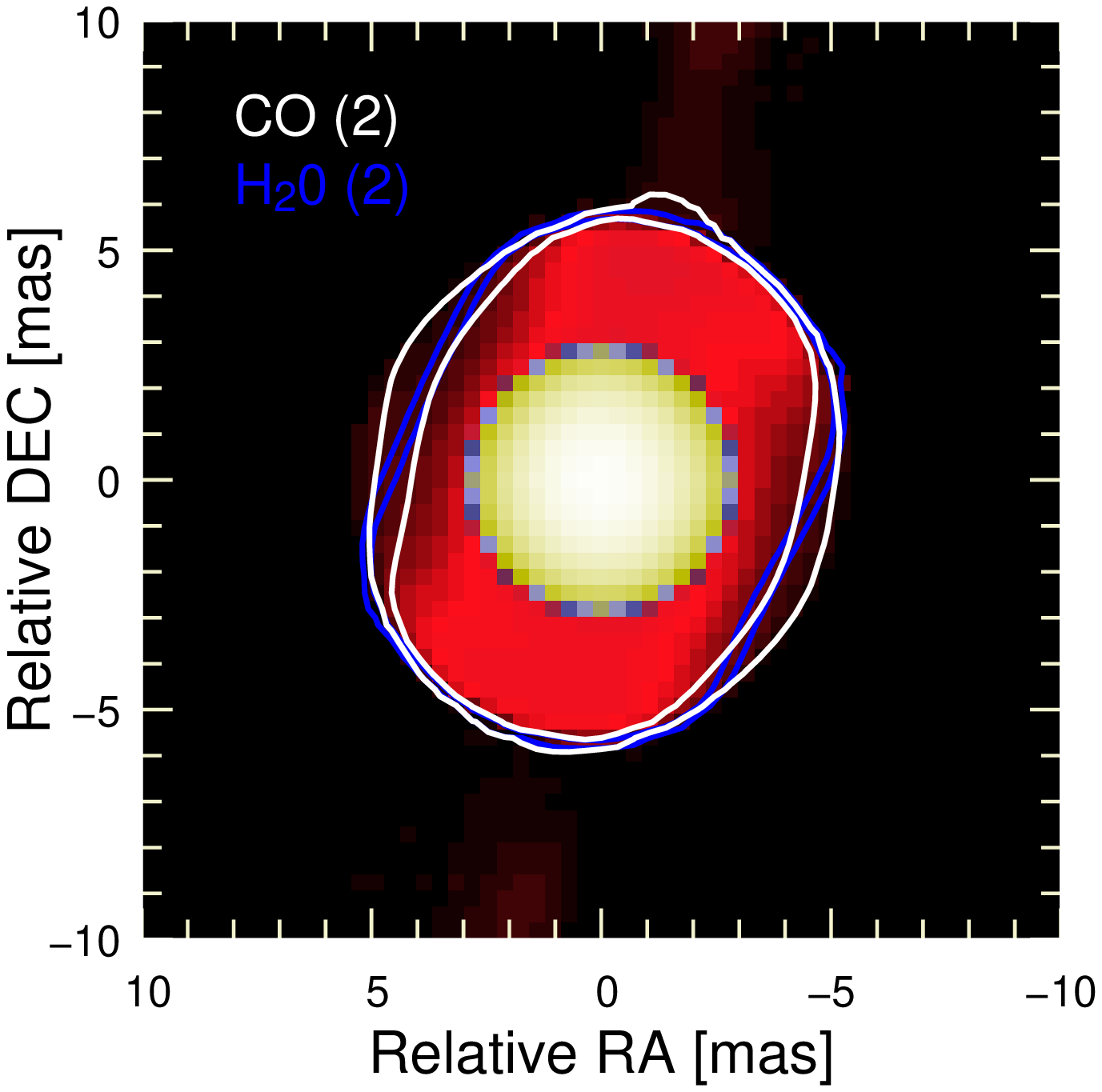}
\caption{Continuum images overplotted (Fig.~\ref{images1}) with $\rm H_{2}O$ (blue) and CO (white) intensity contours (from Figs.~\ref{images2} and \ref{images3}). Intensity contour levels represent  2\% and 5\% of the maximum intensity of each image and allow comparing the extension and shape of the environment between spectral bands for epochs A (left panels) and B (right epanels). The upper panels represent the Cont (1), $\rm H_{2}O(1)$ and CO(1) bands, whereas the lower panels represent the Cont (2), $\rm H_{2}O(2)$ and CO(2) bands.}
     \label{contour}
 \end{figure*}

For the two epochs, Fig.~\ref{contour} shows intensity contour lines of the CO and $\rm H_{2}O$ environments overlaid on the continuum images. Upper panels correspond to a smaller dirty beam and are better indicated to study the extension and the shape of the environment. For the two epochs, the azimuth-averaged extension of the environment is smaller in the continuum than in the $\rm H_{2}O(1)$ band, the environment seen at CO(1) being included between the two. However, due to the irregular shape of these structures, these relative extensions vary for given azimuths.

The impact of the uv coverage (represented as the dirty beam in the lower right corner of each image) on the environment's shape is well visible on some images. Particularly for Continuum(2) (epoch B) and CO(2) (epochs A and B) and for $\rm H_{2}O (2)$ at epoch B, we can see a global SE-NW elongation that is almost aligned with the dirty beam (which is larger at these wavelengths). If this elongation has no any obvious physical origin for a pulsating star like X Hya, it remains uncertain, however, whether it is entirely due to an artefact of the reconstruction.

This is not obvious for $\rm H_{2}O (2)$ at epoch A where the $V^2$ curve (Fig.~\ref{band4}) shows a clear departure from a simple radial symmetry. Moreover, the elongation is not exactly aligned with the dirty beam which potentially means that it is not an artefact. 


We now focus on the CO(1) and  $\rm H_{2}O (1)$ bands that show the most significant closure phase and irregular $V^2$ data: 
\begin{itemize} 
\item{$\rm H_{2}O (1)$: although the dataset contains more points and has larger error bars for Epoch B, it looks more complex than at Epoch A as can be seen from the non-regular $V^2$ curve and the scattering in the closure phase signal. Comparing the images at the two epochs, one can see the general morphology remained similar. It seems that the stellar surface and the close environment of the star expanded from Epoch A to Epoch B, and that some asymmetries appeared in the close environment at Epoch B (Fig.~\ref{images2}).}
\item{CO(1): on the two images, the close environment doesn't present the same structure. If the elongation of Epoch A is probably an effect of the uv coverage, it doesn't appear at Epoch B. Instead, two arms reproduce correctly the two parts of the second $V^2$ lobe.}
\end{itemize}


\section{Discussion}
\label{discuss}
The main goal of this observing campaign is to follow the evolution of asymmetries during the pulsation cycle of a Mira star. The degree of asymmetry as seen by the values of the closure phase signal increased from epoch A (maximum of the visual pulsation cycle) to epoch B (see Fig.~\ref{cpdata}), where the system star+ environment is also generally more extended angularly. We mention here that instead of the closure phase, an alternative estimator called the centrosymmetry parameter \citep[CSP, ][]{2014MNRAS.443.3550C} would be also interesting to use since it is designed to be more sensitive to asymmetries than the closure phase.

Compared to a similar imaging campaign of T Lep \citep{2009A&A...496L...1L}, it is noteworthy that we did not find any discontinuity in the intensity distribution of X Hya between the stellar surface to the close molecular environment. Moreover, CODEX models do not predict such a discontinuity in general.

The asymmetries in the close environment (typically 1.5 stellar radius) in $\rm H_{2}O$ and potentially CO molecular bands can be explained by an inhomogeneous distribution of circumstellar matter. 
We can therefore support the interpretation of previous medium-resolution AMBER observations of X Hya reported in \cite{2011A&A...532L...7W}, who suggested the presence of clumps in the water vapor band. Extrapolating from the $\rm H_{2}O(1)$ image at epoch B (Fig.~\ref{images2}), clumps would have a typical angular scale of $\sim$ 2 mas, representing a few percent of the total flux each. This conclusion agrees with the analysis of the closure phase signal of R Cnc reported in \cite{2011A&A...532L...7W}, which also suggests the presence of unresolved spot (up to $\sim$ 3 mas) contributing to up to $\sim$ 3\% of the total flux in the $\rm H_{2}O (2)$ band. It is noteworthy that this $\sim$ 2-3 mas clump size also represents the typical thickness of the atmosphere as modelled by CODEX series where the typical features of the stellar spectrum are formed.

Clumps in the atmosphere are indeed more likely to be present after the visual maximum phase, that is after the shock wave had time to penetrate the most external layers and possibly enhanced locally inhomogeneous distributions of matter.  On the other hand, recent progress in 3D radiation hydrodynamics simulations revealed that convection alone can account for asymmetrically extended features \citep[synthetic images in $J$, $H$ and $K$ bands are presented in Fig.~5 of][]{2014arXiv1410.3868C}.  It therefore remains possible that asymmetries also have a convective origin, whether they are formed directly by convective motions or whether convection set up density inhomogeneities and velocity fields on the scale of the photospheric radius, and these could be propagated into the environment by pulsation. A detailed comparison of intensity levels of the asymmetric features between these simulations and our multi-spectral observations has to be carried out in a future work.

%

\section{Conclusions}

We reported on AMBER low-resolution observations of the Mira star X Hya at two epochs.
Using advanced radiative transfer modelling tailored to studying dynamics in Mira star atmospheres, we successfully modelled the spectral variation of squared visibilities. Model-dependent images show asymmetries that are located in the immediate environment (at $\sim$ 1.5 stellar radius). As predicted by the CODEX models, no discontinuity is observed between the surface and the immediate environment of the star. The two best-fitting models provided an estimate of the fundamental parameters of X Hya: M = 1.1 M$_\odot$ (the mass is the same for the two models), L = 5300 $\pm$ 100L$_\odot$, R = 212  $\pm$ 3 R$_\odot$ and P = 319  $\pm$ 12 days. More epochs are needed to properly model the atmospheric dynamics and test further radiative transfer codes. More parameter space should be explored for the CODEX series as also concluded in \cite{2012A&A...538L...6H}.


We resolved asymmetric features on the Mira star \object{X Hya}:
   \begin{enumerate}
                \item As shown by the evolution of the closure phase signal (see Sect.~4) the object intensity distribution is more asymmetric at epoch B ($\Phi =0.2$) than at epoch A ($\Phi =0$) 
      \item Asymmetries may explain the discrepancy between the visual phase and the phase derived from the (1D) centro-symmetric dynamical models of pulsating atmospheres we used.
      \item The asymmetries located in the immediate environment could be inhomogeneous material enhanced by the shock wave passage or a feature of a convective pattern.
   \end{enumerate}

Further observing campaigns are necessary to confirm these conclusions on X Hya and on other Mira stars. A better uv coverage and more accurate closure phase measurements are needed to reconstruct model-independent images and unambiguously estimate the characteristics of asymmetries. Ideally, these new datasets should also make use of various imaging reconstruction algorithms to efficiently distinguish reconstruction artefacts from astrophysical asymmetries in the close environment. The hypothesis that asymmetries are due to a convective pattern could then be efficiently tested by comparing multi-band images with 3D radiation hydrodynamics simulations \citep[e.g. ][]{2014arXiv1410.3868C}.


\begin{acknowledgements}
XH thanks  Myriam Benisty and Gilles Duvert for their work on the reduction package amdlib, the \texttt{AMBER data reduction package} of the Jean-Marie Mariotti Center\footnote{Available at http://www.jmmc.fr/amberdrs}. For quick-look analysis, this work has made use of LITPro \citep{2008SPIE.7013E..1JT}. CNRS is acknowledged for having supported this work with allocation of 14 hours in Guaranteed Time Observations. Support for this work was provided by NASA through grant number HST-GO-12610.001-A from the Space Telescope Science Institute, which is operated by AURA, Inc., under NASA contract NAS 5-26555. 
\end{acknowledgements}


\begin{thebibliography}{}

\bibitem[Chelli et al.(2009)]{2009A&A...502..705C} Chelli, A., Utrera, O.~H., \& Duvert, G.\ 2009, \aap, 502, 705
\bibitem[Chiavassa \& Freytag(2014)]{2014arXiv1410.3868C} Chiavassa, A., \& Freytag, B.\ 2014, arXiv:1410.3868
\bibitem[Cruzal{\`e}bes et al.(2015)]{2015MNRAS.446.3277C} Cruzal{\`e}bes, P., Jorissen, A., Chiavassa, A., et al.\ 2015, \mnras, 446, 3277 
\bibitem[Cruzal{\`e}bes et al.(2014)]{2014MNRAS.443.3550C} Cruzal{\`e}bes, P., Jorissen, A., Rabbia, Y., et al.\ 2014, \mnras, 443, 3550 
\bibitem[Hillen et al.(2012)]{2012A&A...538L...6H} Hillen, M., Verhoelst, T., Degroote, P., Acke, B., \& van Winckel, H.\ 2012, \aap, 538, L6 
\bibitem[Haubois et al.(2009)]{2009A&A...508..923H} Haubois, X., Perrin, G., Lacour, S., et al.\ 2009, \aap, 508, 923 
\bibitem[Hinkle et al.(1984)]{1984ApJS...56....1H} Hinkle, K.~H., Scharlach, W.~W.~G., \& Hall, D.~N.~B.\ 1984, \apjs, 56, 1 
\bibitem[H{\"o}fner(2008)]{2008A&A...491L...1H} H{\"o}fner, S.\ 2008, \aap, 491, L1 
\bibitem[Ireland et al.(2008)]{2008MNRAS.391.1994I} Ireland, M.~J., Scholz, M., \& Wood, P.~R.\ 2008, \mnras, 391, 1994 
\bibitem[Ireland et al.(2011)]{2011MNRAS.418..114I} Ireland, M.~J., Scholz, M., \& Wood, P.~R.\ 2011, \mnras, 418, 114 
\bibitem[Kervella et al.(2000)]{2000SPIE.4006...31K} Kervella, P., Coud{\'e} du Foresto, V., Glindemann, A., \& Hofmann, R.\ 2000, \procspie, 4006, 31 
\bibitem[Lacour et al.(2009)]{2009ApJ...707..632L} Lacour, S., Thi{\'e}baut, E., Perrin, G., et al.\ 2009, \apj, 707, 632 
\bibitem[Lan{\c c}on \& Wood(2000)]{2000A&AS..146..217L} Lan{\c c}on, A., \& Wood, P.~R.\ 2000, \aaps, 146, 217 
\bibitem[Le Bertre(1993)]{1993A&AS...97..729L} Le Bertre, T.\ 1993, \aaps, 97, 729 
\bibitem[Le Bouquin et al.(2009)]{2009A&A...496L...1L} Le Bouquin, J.-B., Lacour, S., Renard, S., et al.\ 2009, \aap, 496, L1 
\bibitem[Mennesson et al.(2002)]{2002ApJ...579..446M} Mennesson, B., Perrin, G., Chagnon, G., et al.\ 2002, \apj, 579, 446 
\bibitem[Monnier et al.(2014)]{2014SPIE.9146E..1QM} Monnier, J.~D., Berger, J.-P., Le Bouquin, J.-B., et al.\ 2014, \procspie, 9146, 91461Q 
\bibitem[Perrin et al.(2004)]{2004A&A...426..279P} Perrin, G., Ridgway, S.~T., Mennesson, B., et al.\ 2004, \aap, 426, 279 
\bibitem[Petrov et al.(2007)]{2007A&A...464....1P} Petrov, R.~G., et al.\ 2007, \aap, 464, 1 
\bibitem[Ragland et al.(2006)]{2006ApJ...652..650R} Ragland, S., Traub, W.~A., Berger, J.-P., et al.\ 2006, \apj, 652, 650 
\bibitem[Ramstedt et al.(2008)]{2008A&A...487..645R} Ramstedt, S., Sch{\"o}ier, F.~L., Olofsson, H., \& Lundgren, A.~A.\ 2008, \aap, 487, 645 
\bibitem[Renard et al.(2011)]{2011A&A...533A..64R} Renard, S., Thi{\'e}baut, E., \& Malbet, F.\ 2011, \aap, 533, A64
\bibitem[Sahai et al.(2007)]{2007AJ....134.2200S} Sahai, R., Morris, M., S{\'a}nchez Contreras, C., \& Claussen, M.\ 2007, \aj, 134, 2200  
\bibitem[Tallon-Bosc et al.(2008)]{2008SPIE.7013E..1JT} Tallon-Bosc, I., Tallon, M., Thi{\'e}baut, E., et al.\ 2008, \procspie, 7013,  
\bibitem[Tatulli et al.(2007)]{2007A&A...464...29T} Tatulli, E., Millour, F., Chelli, A., et al.\ 2007, \aap, 464, 29 
\bibitem[Thi{\'e}baut(2008)]{2008SPIE.7013E..43T} Thi{\'e}baut, E.\ 2008, \procspie, 7013,  
\bibitem[Whitelock et al.(2008)]{2008MNRAS.386..313W} Whitelock, P.~A., Feast, M.~W., \& van Leeuwen, F.\ 2008, \mnras, 386, 313 
\bibitem[Wittkowski et al.(2008)]{2008A&A...479L..21W} Wittkowski, M., Boboltz, D.~A., Driebe, T., et al.\ 2008, \aap, 479, L21 
\bibitem[Wittkowski et al.(2011)]{2011A&A...532L...7W} Wittkowski, M., Boboltz, D.~A., Ireland, M., et al.\ 2011, \aap, 532, L7 

\end{thebibliography}
\end{document}